\documentclass[lettersize,journal]{IEEEtran}
\usepackage{amsmath,amsfonts}
\usepackage{algorithmic}
\usepackage{algorithm}
\usepackage{array}
\usepackage[caption=false,font=normalsize,labelfont=sf,textfont=sf]{subfig}
\usepackage{textcomp}
\usepackage{stfloats}
\usepackage{url}
\usepackage{verbatim}
\usepackage{graphicx}
\usepackage{cite}
\usepackage{orcidlink}
\usepackage{tabularx}
\usepackage{longtable}
\usepackage{multirow}
\usepackage{supertabular}
\usepackage{placeins}
\usepackage{tikz}
\usetikzlibrary{shapes,arrows, shadows}
\usetikzlibrary{shapes.geometric, positioning}
\usepackage{multirow}
\usepackage{array}
\usepackage{booktabs}
\usepackage{graphicx} 
\definecolor{colorgreen}{HTML}{005F69}

\begin{document}
\title{A Review on Deep Learning Autoencoder in the Design of Next-Generation Communication Systems}


\author{Omar Alnaseri, Laith Alzubaidi, Yassine Himeur, Mohammed Alaa Ala'anzy, Jens Timmermann and\\ Mohammed S. M. Gismalla
\thanks{Omar Alnaseri and Jens Timmermann (email: timmermann@dhbw-ravensburg.de) are with Department of Electrical Engineering, Baden-Wuerttemberg Cooperative State University, Friedrichshafen, Germany (Corresponding author: Omar Alnaseri email: omar.ahmed@daad-alumni.de).}
\thanks{Laith Alzubaidi is with School of Mechanical, Medical, and Process Engineering, Queensland University of Technology, Brisbane, 4000, QLD, Australia (email: l.alzubaidi@qut.edu.au)}
\thanks{Yassine Himeur is with College of Engineering and Information Technology, University of Dubai, Dubai, United Arab Emirates (email: yhimeur@ud.ac.ae).}
\thanks{Mohammed Ala'anzy is with Department of Compute Science, SDU University, Almaty, Kazakhstan (email: m.alanzy@ieee.org).}
\thanks{Mohammed S. M. Gismalla is with Department of Engineering Technology, South East Technological University, Waterford, Ireland (email: mohammed.gismalla@waltoninstitute.ie).}

}

\markboth{Proceedings of the IEEE, VOL. 00, NO. 00, July 2025}%
{Shell \MakeLowercase{\textit{et al.}}: A Sample Article Using IEEEtran.cls for IEEE Journals}


\maketitle

\begin{abstract}
Traditional mathematical models used in designing communication systems often fall short due to inherent simplifications, narrow scope, and computational limitations. In recent years, the incorporation of deep learning (DL) methodologies into communication systems has made significant progress in system design and performance optimization. Autoencoders (AEs) have become essential, enabling end-to-end learning that allows for the combined optimization of transmitters and receivers. Consequently, AEs offer a data-driven methodology capable of bridging the gap between theoretical models and real-world complexities. This paper presents a comprehensive survey of the application of AEs within communication systems, with a particular focus on their architectures, associated challenges, and future directions. We examine recent studies in the fields of wireless, optical, semantic, and quantum communication, providing a state-of-the-art review with critical analysis. This paper further examines the challenges encountered in the implementation of AEs, including the need for a solution for handling non-differentiable channels, computational complexity, convergence, data representation, and data training, including data scarcity. Moreover, a summary of the computational complexity calculation associated with AE-based systems is provided by conducting an analysis using the metric of floating-point operations per second (FLOPS). Finally, future directions are highlighted to realize the full potential of AEs in creating efficient, reliable, and high-performance communication systems. This survey aims to establish a roadmap for future research, emphasizing the transformative potential of AEs in the formulation of next-generation communication systems.

\end{abstract}


\section{Introduction}
\IEEEPARstart{I}{n} recent years, the field of communication systems has witnessed significant advancements, particularly with the integration of deep learning techniques. Among these, autoencoders (AEs) have emerged as a powerful tool for designing next-generation communication systems, due to their ability to provide end-to-end optimization systems, i.e., joint optimization of both the transmitter and receiver. This paper provides a comprehensive survey on the application of AE in communication systems, with a focus on their architectures, challenges, and future directions. The primary objective is to explore how AEs can address the limitations of current communication systems, which often rely on simplifying assumptions and may not capture the full complexity of real-world communication environments. 

\begin{table*}
\centering
\footnotesize
\caption{List of Acronyms}
\resizebox{\textwidth}{!}{%
\begin{tabular}{l|l|l|l|l|l}
\toprule
\textbf{Acronym} & \textbf{Full Form} & \textbf{Acronym} & \textbf{Full Form} & \textbf{Acronym} & \textbf{Full Form} \\
\midrule
ACL & Adjust Carrier Limitation & AE & Autoencoder & AEGNN & Autoencoder Graph Neural Network \\
\midrule
AEs & Autoencoders & ADC & Analogue-to-Digital Converters & ADL & Adaptive Deep Learning \\
\midrule
AICN & Additive Independent Cauchy Noise & AILN & Additive Independent Laplace Noise & ANNs & Artificial Neural Networks \\
\midrule
APQNN & Amplitude Phase Quantization with Neural Networks & AWGN & Additive White Gaussian Noise & BCC & Binary Cross-Entropy \\
\midrule
BER & Bit Error Rate & BLER & Block Error Rate & BMD & Bit-Metric Decoding \\
\midrule
BNN & Binary Neural Network & BPS & Blind Phase Search & BRNNs & Bidirectional Recurrent Neural Networks \\
\midrule
CCE & Categorical Cross-Entropy & CCDF & Complementary Cumulative Distribution Function & CD & Chromatic Dispersion \\
\midrule
CE & Cross-Entropy & CGAN & Conditional Generative Adversarial Network & CFAR & Constant False Alarm Rate \\
\midrule
CKF & Cubature Kalman Filter & CMA & Constant Modulus Algorithm & CNN & Convolutional Neural Network \\
\midrule
CPE & Carrier Phase Estimation & CR & Cognitive Radio & CRNs & Cognitive Radio Networks \\
\midrule
CSAE & Channel-Sensitive AE & CSI & Channel State Information & DBN & Deep Belief Network \\
\midrule
DL & Deep Learning & DMs & Diffusion Models & DNN & Deep Neural Network \\
\midrule
DMs & Diffusion Models & DRdA-CA & Dual-residual Denoising AE with Channel Attention & DSP & Digital Signal Processing \\
\midrule
DWDM & Dense Wavelength Division Multiplexing & EKL & Extended Kalman Filter & ELBO & Evidence Lower Bound \\
\midrule
ELU & Exponential Linear Unit & FCNN & Fully-connected Neural Network & FMCW & Frequency-modulated Continuous Wave \\
\midrule
FLOPS & Floating-Point Operations & FPGAs & Field Programmable Gate Arrays & FSO & Free-Space Optical \\
\midrule
GANs & Generative Adversarial Networks & GAs & Genetic Algorithms & GCS & Constellation Shaping \\
\midrule
GDR & Generalized Data Representation & GMI & Graphic Mutual Information & GRU & Gated Recurrent Units \\
\midrule
IMDD & Intensity Modulation with Direct Detection & ISI & Intersymbol Interference & JSCC & Joint Source-Channel Coding \\
\midrule
KLF & Kullback\textendash Leibler & KNL & Kerr Nonlinearity & LPN & Laser Phase Noise \\
\midrule
LRV & Latent Random Variables & LSTMs & Long Short-Term Memory & LSTM-AE & Long-Short-Term Memory-Autoencoder \\
\midrule
LTE & Long-Term Evolution & LLRs & Log-Likelihood Ratios & M-QAM & M-quadrature Amplitude Modulation \\
\midrule
MDC & Multi-Dimensional Constellations & MI & Mutual Information & MIMO & Multiple-Input and Multiple-Output \\
\midrule
ML & Maximum Likelihood & MLD & Maximum Likelihood Detection & mmWave & millimeter-Wave \\
\midrule
MSE & Mean Squared Error & NC-EA & Noncoherent Energy AE & OFDM & Orthogonal Frequency-Division Multiplexing \\
\midrule
OWC & Optical Wireless Communication & PAPR & Peak-to-Average Power Ratio & PCS & Probabilistic Constellation Shaping \\
\midrule
PDM & Polarization-Division Multiplexing & PHV & Packet Hot Vector & PN & Phase Noise \\
\midrule
PPM & Pulse Position Modulation & PS & Probabilistically-Shaped & QAM & Quadrature Amplitude Modulation \\
\midrule
QC-AE & Quantum-Classical Autoencoder & QoS & Quality-of-Service & QNNs & Quantum Neural Networks \\
\midrule
QPSK & Quadrature Phase-Shift Keying & RBF & Rayleigh Block Splitting & RF & Radio Frequency \\
\midrule
ReLu & Rectified Linear Unit & ResNet & Residual Neural Network & RIS & Reconfigurable Intelligent Surface \\
\midrule
RL & Reinforcement Learning & RNNs & Recurrent Neural Networks & RTNs & Radio Transformer Network \\
\midrule
SCCE & Sparse Categorical Cross-Entropy & SELU & Scaled Exponential Linear Unit & SER & Symbol Error Rate \\
\midrule
SIMO & Single-Input Multiple-Output & SM & Spatial Modulation & SNR & Signal-to-Noise Ratio \\
\midrule
SPSA & Simultaneous Perturbation Stochastic Approximation & SSFM & Split-Step Fourier Method & SVD & Singular Value Decomposition \\
\midrule
UOC & Underwater Optical Communication & VAEs & Variational AEs & VQ-VAEs & Vector-Quantized Variational AEs \\
\midrule
VLC & Visible Light Communication & WC & Wireless Communication & WiFi & Wireless Fidelity \\
\bottomrule
\end{tabular}
}%
\end{table*}

This review comprehensively investigates the application of AE in communication systems, drawing on recent studies published in reputed publishers, including ScienceDirect, Scopus, Springer, and IEEE Xplore, including 2025. The literature search used rigorous keywords such as \textit{"autoencoder in communication system"} and \textit{"end-to-end learning in communication system"} to identify relevant studies. By categorizing these works based on research issues such as system design, channel/signal processing, coding/modulation, computational complexity, and non-differentiable channels, we provide a structured overview of the current state of the field. Our analysis encompasses various communication domains, including wireless, optical, semantic, and quantum communication, to identify suitable AE approaches for each.
Identified AE approaches are analyzed based on model used, dataset/data type, best performance, and their limitation. Note that many existing studies on AE applications in communication systems rely on simplified channel models such as additive white Gaussian noise (AWGN) or basic fiber models. In addition, this review offers a critical analysis of current work and in-depth discussions on existing challenges and potential future research directions. To our knowledge, this is the first comprehensive survey of AE applications in communication systems, which can provide a valuable roadmap for future research.

This reset of this paper is organized as shown in Fig.~\ref{fig_stru}, where Section~\ref{sec2} discusses the challenges of designing next-generation communication systems using traditional approaches. Section~\ref{sec3} elaborates on the advantages of using AEs in communication systems. Section~\ref{sec4} describes the end-to-end AE learning process, exploring the AE-based communication architecture, including symbol-wise and bit-wise approaches, different loss functions used to optimize AE models, and the physical constraints that AEs must consider in communication systems. In addition, it delves into geometric and probabilistic constellation shaping techniques and (non)-differentiable channels and components, including how to determine differentiability, proposing solutions for non-differentiable scenarios. Finally, it details the calculation of the computational complexity of AE. Section~\ref{sec5} provides a comprehensive review of the literature on the use of AEs in communication systems in various domains, including wireless, optical fiber, free space optical, visible light, semantic, quantum and underwater communications. It also highlights the current challenges for which AE has been used. In addition to the state-of-the-art review, it provides a critical analysis of current approaches highlighting unsolved limitations. Based on that, Section~\ref{sec6} suggests future research areas, and finally Section~\ref{sec7} concludes the paper by summarizing the potential of AEs to revolutionize communication systems.

\begin{figure*}[ht!]

\definecolor{colorgreen}{HTML}{005F69}


\centering
\resizebox{1.8\columnwidth}{!}{
\begin{tikzpicture}[
  box/.style={rounded corners, draw=black, fill=colorgreen!40, minimum width=1.8cm, minimum height=0.5cm, drop shadow},
  container/.style={rounded corners, draw=black, inner sep=0.1cm, fill=gray!10},
  section/.style={rounded corners, draw=black, inner sep=0.1cm},
  node distance=0.4cm and 0.4cm,
  font=\small ]

\node[container] (main) at (0,0) {

  \begin{tikzpicture}[node distance=0.2cm]
    \node[align=center] (title2) at (0,1.2) {\textbf{Section II. Limitations of Mathematical Models}};
    \node[box] (M1) [align=center,below left=of title2, xshift=1cm] {Simplifications \\ and Narrow Scope};    
    \node[box] (M3) [align=center, right=of M1] {Inadequate Handling of\\ Nonlinearity and Complexity};
    \node[box] (M2) [align=center, right=of M3] {Validation Challenges and \\ Interdisciplinary Gaps};
  \end{tikzpicture}
};

\node[container] (main) at (-4,-2.4) {
  \begin{tikzpicture}[node distance=0.2cm]
    \node[align=center] (title3) at (0,1.2) {\textbf{Section III. Why AEs for Communication Systems?}};
    \node[box] (W1) [align=center, below left=of title3, xshift=3.2cm] {Transmitter/Receiver\\ joint optimization};    
    \node[box] (W2) [align=center, below=of W1] {Performance closer\\ to Shannon limit};
    \node[box] (W3) [align=center, right=of W1] {Robustness and adaptability\\ to varying channel conditions};
    \node[box] (W4) [align=center, below =of W3] {Fundamental end-to-end\\ communication framework};
  \end{tikzpicture}
};

\node[container] (main) at (-4,-4.5) {
  \begin{tikzpicture}[node distance=0.2cm]
    \node[align=center] (title4) at (0,1.2) {\textbf{Section VI. AE Architecture Fundamentals}};
    \node[box] (S1) [align=center, below=of title4] {Explaining the simple AE Architecture}; 
  \end{tikzpicture}
};

\node[container] (main) at (4,-3.1) {
  \begin{tikzpicture}[node distance=0.2cm]
    \node[align=center] (title4) at (0,1.2) {\textbf{Section V. AE based Communication Systems}};
    \node[box] (A1) [align=center, below=of title4] {AE Architecture approaches:\\ Regression, Symbol-/Bit-wise, Activation function,\\ loss function, physical constraints};    
    \node[box] (A2) [align=center, below=of A1] {AE Architecture Variant:\\ Classical AE, CAE, RAE, VAE, SAE, AAE};
    \node[box] (A3) [align=center, below left=of A2, xshift=3.8cm] {Geometric and Probabilistic\\ Constellation Shaping};
    \node[box] (A4) [align=center, right =of A3] {How to calculate\\ AE Complexity};
  \end{tikzpicture}
};

\node[container] (main5) at (0,-8.5) {
  \begin{tikzpicture}[node distance=0.3cm and 0.4cm]

    \node[section] (sec51) {
      \begin{tikzpicture}[node distance=0.2cm]
        \node[align=center] (title51) at (0,1.2) {\textbf{Domain-Specific Challenges/Solutions}};
        \node[box] (wc) [below=of title51, xshift=-0.7cm] {Wireless Communications};
        \node[box] (ofc) [below=of wc, xshift=0.35cm] {Optical Fiber Communications};
        \node[box] (owc) [below=of ofc, xshift=0.2cm] {Optical Wireless Communications};
        \node[box] (sc) [below=of owc, xshift=-0.5cm] {Semantic Communications};
        \node[box] (qc) [below=of sc] {Quantum Communications};
      \end{tikzpicture}
    };

    \node[section, right=of sec51, yshift=0.9cm] (sec52) {
      \begin{tikzpicture}[node distance=0.2cm and 0.3cm]
        \node[align=center] (title52) at (0,1.2) {\textbf{Research Issues}};
        \node[box] (r1) [below=of title52, xshift=-1.2cm] {System Design};
        \node[box] (r2) [align=center,below=of r1] {Channel/Signal\\ Processing};
        \node[box] (r3) [right=of r1] {Coding/Modulation};
        \node[box] (r4) [below=of r3] {Low Complexity};
      \end{tikzpicture}
    };

    \node[section, below=of sec52, yshift=0.2cm] (sec53) {
      \begin{tikzpicture}[node distance=0.1cm and 0.1cm]
        \node[align=center] (title53) at (0,0.2) {\textbf{AE Models}};
        \node[box] (m1) [below=of title53, xshift=-0.2cm] {Dense AE};
        \node[box] (m2) [below=of m1] {VAE};
        \node[box] (m3) [left=of m1] {Turbo AE};
        \node[box] (m4) [right=of m1] {GAN-AE};
        \node[box] (m5) [below=of m3] {Denoise AE};
        \node[box] (m5) [below=of m4] {...};
      \end{tikzpicture}
    };
    
    \node[section, right=of sec52, yshift=-0.75cm] (sec54) {
      \begin{tikzpicture}[node distance=0.2cm and 0.3cm]
        \node[align=center] (title54) at (0,1.2) {\textbf{Other Specific Challenges}};
        \node[box] (o1) [align=center, below=of title54] {Complexity and\\ Computational Constraints};
        \node[box] (o2) [below=of o1] {Training and Convergence};
        \node[box] (o3) [below=of o2] {Data Representation and Rate};
        \node[box] (o4) [below=of o3] {Data Scarcity};
        \node[box] (o5) [below=of o4] {Emerging HW Technologies};
      \end{tikzpicture}
    };
  \end{tikzpicture}
};

\node[container] (main) at (-4,-12.1) {
  \begin{tikzpicture}[node distance=0.2cm]
    \node[align=center] (title41) at (0,1.2) {\textbf{Section VII. (Non-) and Differentiable Channel}};
    \node[box] (D1) [align=center, below left=of title41, xshift=3.5cm] {How to determine\\ differentiability}; 
    \node[box] (D2) [align=center, right=of D1] {Solutions for\\ non-differentiable}; 
  \end{tikzpicture}
};

\node[container] (main) at (4,-12.2) {
  \begin{tikzpicture}[node distance=0.2cm]
    \node[align=center] (title42) at (0,1.2) {\textbf{Section VIII. Lessons Learned}};
    \node[box] (L1) [align=center, below left=of title42, xshift=2.5cm] {Regularization Strategies}; 
    \node[box] (L2) [align=center, below=of L1] {Loss Function Choices};  
    \node[box] (L3) [align=center, right=of L1] {Hyperparameter Tuning};  
    \node[box] (L4) [align=center, below=of L3] {Deployment Reliability}; 
  \end{tikzpicture}
};

\node[container] (main) at (0,-14.8) {
  \begin{tikzpicture}[node distance=0.2cm]
    \node[align=center] (title6) at (0,1.2) {\textbf{Section VX. Future Research Areas}};
    \node[box] (F1) [align=center, below=of title6] {Complexity and\\ Computational Constraints};    
    \node[box] (F2) [align=center, left=of F1] {Scalability and\\ High Dimensions};
    \node[box] (F3) [align=center, left=of F2] {Implementation\\ and HW Issues};
    \node[box] (F4) [align=center, right=of F1] {Non-differentiable channels\\ and DSP components};
    \node[box] (F5) [align=center, right=of F4, yshift=0.2cm] {Data and Training Issues};
    \node[box] (F8) [align=center, below=of F5] {Interpretability and Trust};
    \node[box] (F9) [align=center, below=of F8] {Hybrid models};
    
    \node[box] (F6) [align=center, below=of F3] {Advanced AE\\ architectures};
    
    \node[box] (F10) [align=center, below=of F4, xshift=0.2cm] {System-Level and\\ Practical Considerations};
    \node[box] (F11) [align=center, below=of F2, xshift=0.4cm] {Federated Learning for\\ Distributed AE Training};
    \node[box] (F12) [align=center, below=of F1, xshift=0.65cm] {Edge-AI Integration for\\ Low-Latency System};   
  \end{tikzpicture}
};

\node at (main5.north) [above=0.1cm] {\textbf{Section VI. AE-Based Solutions to Domain-Specific Challenges in Communication Systems}};

\end{tikzpicture}
}

    \caption{The structure of this paper}
    \label{fig_stru}
\end{figure*}

\section{Limitations of Mathematical Models}
\label{sec2}
Traditional mathematical models have long served as the basis for designing communication systems. However, their inherent assumptions and simplifications often do not address the complexities of next-generation networks. In the following, we categorize these limitations and highlight how autoencoder (AE)-based approaches offer promising alternatives, albeit with their own challenges.

\paragraph{Simplifications and Narrow Scope} Mathematical models frequently rely on idealized assumptions (e.g., linearity, Gaussian noise) that do not hold in real-world environments. For example, optical fiber models often ignore manufacturing defects or bending losses \cite{saleh2019fundamentals, senior2009optical, keiser2021fiber}, while wireless channel models oversimplify multipath effects in 5G mmWave systems \cite{sun2014mimo}. AE-based solutions bypass explicit modeling by learning channel characteristics directly from data, which allows for adaptive transceiver optimization (Section~\ref{IV_B}). 

\paragraph{Inadequate Handling of Nonlinearity and Complexity}
Nonlinear phenomena such as fiber Kerr effects and RF amplifier distortions are poorly captured by analytical models \cite{kaminow2010optical, mukherjee2006optical, marzetta2016fundamentals}. In wireless systems, understanding effects such as small-scale fading and Doppler shifts is key, especially in high-speed scenarios such as vehicle-to-vehicle networks\cite{tse2005fundamentals}. Likewise, in dense wavelength division multiplexing (DWDM) optical systems, ignoring nonlinearities such as four-wave mixing and self-phase modulation can result in crosstalk, spectral spreading, and signal degradation\cite{mukherjee2006optical}. Combining models with physical insight helps engineers better troubleshoot and optimize performance. AE-based solutions excel at modeling nonlinear relationships through deep neural networks (Section \ref{VI_C}). However, their computational complexity grows significantly with system dimensionality.

\paragraph{Validation Challenges} Testing models in real-world settings, especially for wireless networks in urban or suburban areas, is often resource-intensive. Validating small cell 5G deployments requires extensive trials in varying landscapes and interference levels\cite{heath2018foundations}. Similarly, in fiber optics, the complexity of signal behavior makes it difficult to validate models without experimental setups or detailed benchmarking against known references\cite{senior2009optical, keiser2021fiber, marcuse2013theory}. Data-driven AE reduce the dependency on closed-form models by training in empirical measurements (Section \ref{V_A_1}). However, mismatches between training and deployment environments can degrade performance (Section \ref{VI_D}).

\paragraph{Interdisciplinary Gaps} Mathematical models often operate in isolation, not integrating insights from fields such as materials science or electrical engineering \cite{snyder1983optical}. For example, developments in metamaterials could greatly impact antenna design. But if models do not account for such interdisciplinary advances, they risk lagging behind practical engineering progress \cite{david2012microwave}. AE-based solution can offer seamless integration with machine learning and signal processing techniques, which allows joint optimization of the physical and protocol layers (Section \ref{mimo}). However, such integration requires careful balance of modularity and end-to-end learning (Section Section \ref{VI_F}).

\section{Why AEs for Communication Systems?}
\label{sec3}
Although mathematical models provide valuable information, next-generation communication systems require a holistic approach that considers the above limitations. AEs are increasingly viewed as an elegant solution for next-generation communication systems primarily due to their end-to-end learning capability \cite{o2017introduction,cammerer2020trainable, raj2020design}. This paradigm allows for the \textit{joint optimization of the transmitter and receiver} based on data. It effectively treats the communication system as an AE aiming to reconstruct the input message at the receiver despite channel impairments. This bypasses the need for explicit, often complex, and idealized mathematical models of the channel and individual processing blocks, which can be particularly challenging in complex or dynamic environments. The end-to-end AE approach has the potential to \textit{achieve performance closer to the theoretical Shannon limit} by allowing the system to learn optimal joint encoding and decoding strategies directly from the data \cite{yu2017autoencoders}. The research highlights that end-to-end AE learning can outperform traditional human-engineered transceivers in various scenarios, even without prior knowledge of communication theory \cite{song2022benchmarking}.

Compared with traditional or hybrid approaches to AE, traditional block-based signal processing, while mature and often achieving near-optimal performance under well-defined conditions, suffers from suboptimal overall system performance due to separate optimization of individual blocks \cite{zhai2020design}. Traditional methods also rely heavily on explicit probabilistic models that may not adapt well to dynamic channel conditions or capture complex nonlinearities in real world environments \cite{li2025semantic}. AE-based systems, on the other hand, demonstrate \textit{robustness and adaptability to varying channel conditions} and can outperform traditional methods, especially in non-linear scenarios where the latter struggle \cite{song2022benchmarking, li2025semantic, lee2020autoencoder, huynh2023performance}. Furthermore, hybrid approaches have emerged that combine the adaptability of AEs with the efficiency of conventional demapping algorithms to address the drawbacks of purely AE-based systems, such as the high computational complexity and large model sizes associated with training and inference, especially on devices limited by resources \cite{ney2022hybrid}.  

AEs also compare with other machine learning models in their specific applications within communications. Although CNNs are used for tasks such as modulation classification and have shown improvements over traditional expert feature-based methods, AEs offer a more \textit{fundamental end-to-end communication framework} \cite{o2017introduction}.  GANs are noted for their potential in real-time channel modeling that can be relevant for training end-to-end systems by generating realistic channel impairments, which is relevant for AE training, particularly to address the gradient problem in channel backpropagation \cite{zhao2023end, vijayakumari2025survey}. 

AE-based end-to-end learning is an elegant candidate to address challenges in next-generation communication systems because of its ability to \textit{jointly optimize the entire communication link} and its \textit{data-driven nature}, which allows it to handle complex and unknown channel conditions where traditional methods struggle. A compilation of reasons that explain why end-to-end AE-based learning is considered an appropriate candidate includes \textit{overcoming the limitations of traditional modular designs} \cite{karanov2020end2, dorner2017deep, cammerer2018end}, \textit{handling complex and unknown channel models} \cite{wang2017deep, letizia2025deep}, \textit{adaptability and flexibility} \cite{wu2020deep, song2023autoencoders}, and \textit{simplifying design} \cite{o2017introduction, dorner2017deep}.

\section{AE Architecture Fundamentals}
\label{sec4}

\begin{figure}[htbp]
    \centering
    \includegraphics[width=1.0\linewidth]{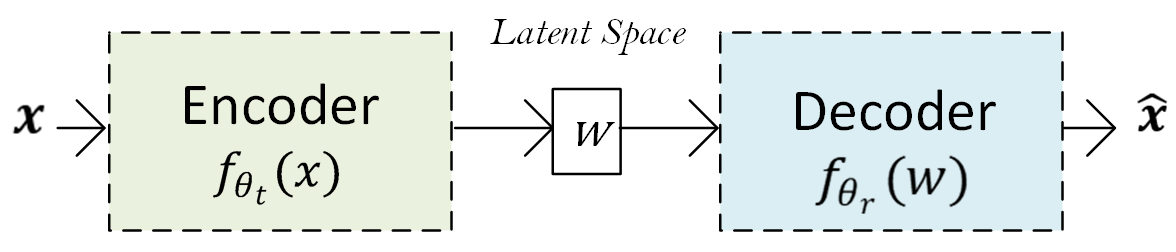}
    \caption{AE Architecture}
    \label{fig_AE}
\end{figure}

An end-to-end AE is a deep neural network (DNN) model that learns representations (encoding) for sets of data by attempting to reproduce its input in the output. It is an unsupervised machine learning algorithm. This means that it does not require labeled output or targets during training and instead tries to reconstruct inputs from their encoded representation in hidden layers. 
As shown in Fig.~\ref{fig_AE}, the architecture can be divided into three main parts: Encoder (mapping input $x$ to latent space $w$), Latent Space $w$ (bottleneck), and Decoder (mapping latent space $w$ to output $\hat{x}$).

\paragraph{Encoder} The encoder part of the AE takes input data, which can be any form of structured data, and is typically composed of a series of neural network layers with nonlinear activation functions such as rectified linear unit (ReLu) or exponential linear unit (ELU). Its goal is to transform the high-dimensional input into a lower-dimensional latent space representation that captures the most important features in an efficient manner.
\paragraph{Bottleneck/Latent Space} This part of AE represents compressed knowledge about the original data and is typically composed of one or more \textit{un-trainable} layers with fewer neurons than the input layer, hence creating a bottleneck effect where only important features are kept while less significant ones get discarded.
\paragraph{Decoder} The decoder part of an AE takes the compressed representation and reconstructs the original data. It mirrors the encoder architecture but in reverse, attempting to output a reconstruction that is as close as possible to its input (though not necessarily identical due to noise or other factors).

The entire process can be summarized as follows. Input $\mapsto$ Encoding $\mapsto$ Bottleneck/Latent Space $\mapsto$ Decoding $\mapsto$ Output. The AE learns optimal weights during training, which minimizes the reconstruction error between inputs and outputs using a loss function.
The key advantage of an end-to-end AE learning approach is that it allows for more efficient data processing or feature extraction since all these tasks are learned simultaneously. It also helps to capture complex patterns within the input data, which can be useful when dealing with high-dimensional datasets.

\section{AE-based Communication systems}
\label{IV_B}
\begin{figure*}[htbp]
    \centering
    \includegraphics[width=1.02\linewidth]{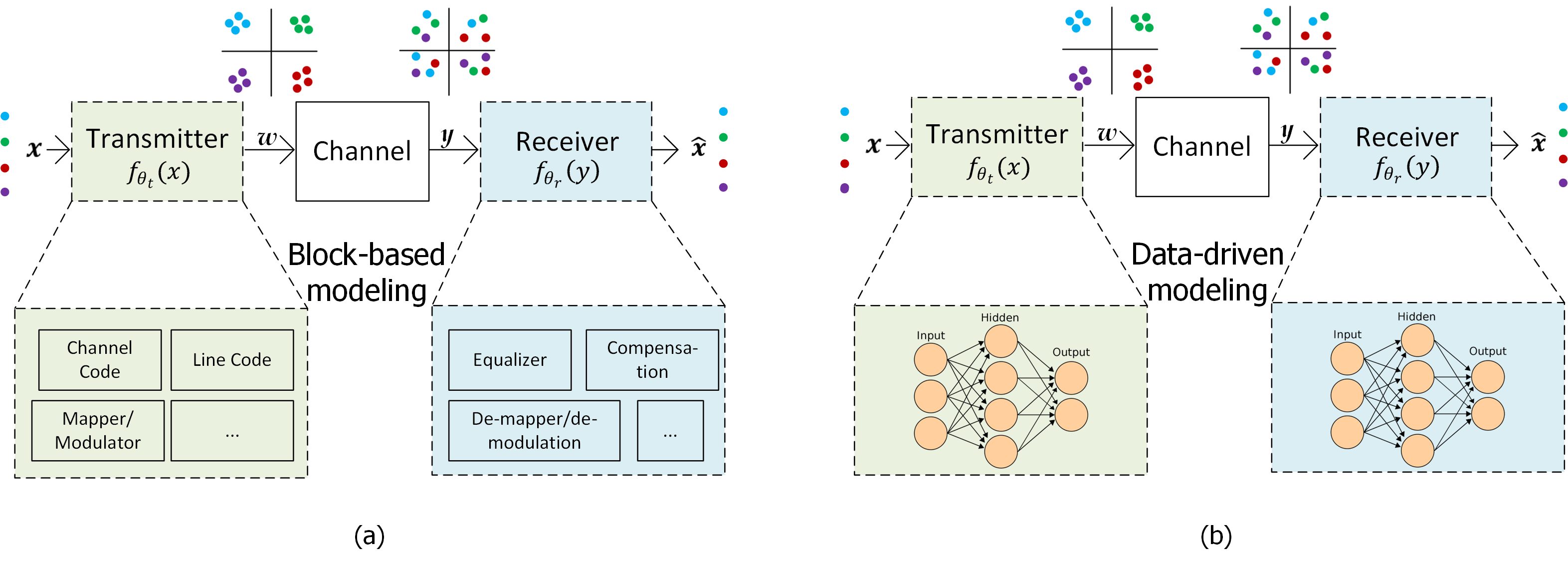}
    \caption{Communication System (a) Classical Block-based Model (b) AE Data-driven Model}
    \label{fig_AE_com}
\end{figure*}

Traditional communication systems employ a block-based approach, where separate transmitter and receiver signal processing blocks are cascaded as shown in Fig.~\ref{fig_AE_com} (a). This approach often leads to suboptimal performance due to two main challenges: (1) the difficulty of accurately modeling complex physical channels, as discussed in Section~\ref{sec2}, and (2) the need to develop specific algorithms to compensate for these effects based on the chosen mathematical model. A promising solution is to use DL-based approaches, inspired by AE architectures \cite{o2017introduction}. As shown in Fig.~\ref{fig_AE_com} (b), by taking advantage of the ability of deep learning models to learn robust latent representations, AE-based systems have the potential to optimize end-to-end communication performance. The latent space representation produced by the encoder $w$ is passed through a channel model, where it undergoes channel impairments and produces $y$. Thus, instead of relying on a chain of separate transmitter and receiver signal processing blocks, a data-driven model based on AEs architecture can effectively capture the characteristics of the communication channel, enabling more efficient and reliable transmission. 

The main goal of a communication system, acting as an AE, is to find a robust latent space representation $w$ of the input data in the transmitter to counteract any impairments introduced by the channel. In particular, through an end-to-end training, the AE learns to effectively reconstruct the input data at the receiver, considering the presence of channel impairments. During the training process, the encoder component of the AE acquires resilient symbol sequence representations for all incoming data. Once training is complete, both the transmitter and the receiver can function independently, using their respective layer dimensions and weights. Next, we delve into the various aspects of employing AEs in communication systems, encompassing the two AE approaches, activation functions, loss functions, physical constraints, and AE architectural variants.

\subsection{AE Architecture Approaches}
In the context of communication systems, there are two variant ways of dealing with the communication problem, either as a regression problem or a categorical problem \cite{Erpek2019}. 
A regression problem in the context of machine learning and communication systems typically involves predicting continuous or numerical data based on the input data. By setting the output layer of the AE to a regression layer (such as a linear layer), the AE can be trained to reconstruct a continuous target value, which can be used effectively in communication systems. In contrast, in classification problems, the predicting data is a categorical or discrete class. End-to-end AE learning for communication systems using a classification problem involves training the AE to recognize and classify different constellation diagrams such as QPSK, 16-QAM, and 64-QAM. This can be done by setting the output layer of the AE to a classification layer (such as the sigmoid activation layer) and compute the cross-entropy. 
\begin{figure}[htbp]
    \centering
    \includegraphics[width=1.0\linewidth]{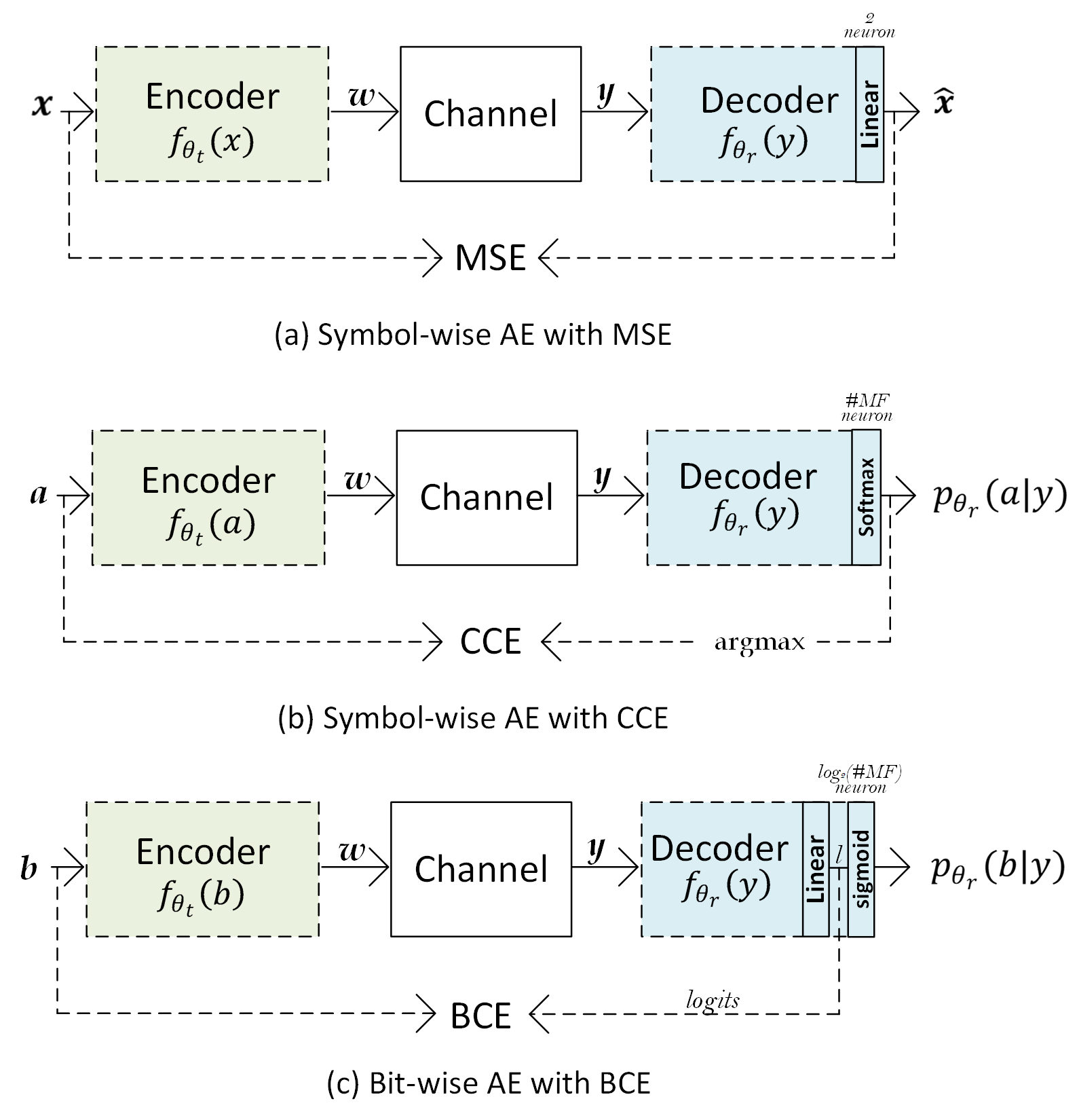}
    \caption{Symbol and Bit-wise AE learning based communication system}
    \label{fig_AE_com_bit}
\end{figure}

Fig.~\ref{fig_AE_com_bit} shows three different end-to-end learning architectures based on AE. Regression problems can be used in AEs for end-to-end learning of communication systems. By setting the output layer of the AE to a linear regression layer as shown in Fig.~\ref{fig_AE_com_bit} (a), the AE can be trained to predict continuous values for new inputs\cite{lau2022machine, alnaseri2024Noisy}. This results in complex output data consisting of both in-phase (I) and quadrature (Q) components. This data is then fed into a demapper for symbol detection and subsequently a hard decision circuit to recover the original bit sequence.   
The AE encoder processes complex baseband symbols ($x \in \mathbb{C}$) from modulation formats such as quadrature phase-shift keying (QPSK), or M-quadrature amplitude modulation ($M$-QAM). The encoder, parametrized by $\theta_{t}$, generates resilient latent representations ($w \in \mathbb{C}$) by mapping $f_{\theta_t}:x \mapsto w$. These representations are then passed through a channel model, where they undergo impairments. The decoder, parameterized by $\theta_{r}$, aims to reconstruct the original symbol from the channel-corrupted output ($y \in \mathbb{C}$) by mapping $f_{\theta_r}:y\mapsto \hat{x}$, where $\hat{x} \in \mathbb{C}$. As it tries to predict the data of a numerical symbol, the Mean Squared Error (MSE) is a suitable loss function; see Section~\ref{LF}. Since NNs cannot directly process complex data, complex input data $N$ must be converted to $2N$ real-valued representation for both the encoder and decoder by implementing a $\mathbb{C}2\mathbb{R}$ layer. The resulting real-valued output is then converted back to complex data using a $\mathbb{R}2\mathbb{C}$ layer. 

Symbol-wise AEs and Bit-wise AEs are two different approaches to designing a communication system. Their primary difference lies in how they handle the mapping of bits to symbols and bit-labeling to effectively determine the learned constellation. This may directly impact system performance characteristics such as bit error rate (BER), symbol error rate (SER), and computational complexity. In both symbol-wise AE and bit-wise AE, the input to the encoder AE can be pre-processed to one hot encoded.

As shown in Fig.~\ref{fig_AE_com_bit} (b), a symbol-wise approach based on minimizing categorical cross-entropy (CCE) can be fed with a message $a \in \{0,1,...,M-1\}$, $M$ denotes the number of different symbols as $M = 2^m$ where $m$ is the number of bits per symbol. Thus, the encoder then processes the mapping to complex data as $f_{\theta_t}:\{0,1,...,M-1\} \mapsto w$, i.e. $f_{\theta_t}:a \mapsto w$. Another key distinction lies in the output layer, the Softmax activation layer, which is used to obtain a probability vector of length $M$ which corresponds to the probabilities of possible $M$ messages, $p_{\theta_r}(a|y)$ \cite{cammerer2020trainable}. This represents the estimated posterior distribution of the transmitted signal given the received signal. The constellation obtained through symbol-wise training is not optimized for bit-labeled. Bit-labeled indicates that each constellation point must be uniquely mapped to a specific bit vector. To enable joint constellation optimization and bit labeling, a bit-wise AE can be used, as shown in Fig.~\ref{fig_AE_com_bit} (c), the bit vector $b$ can be fed to the encoder to perform the mapping $f_{\theta_t}:b \mapsto w$, where the mapping of bits into symbols is integrated within the encoder. The AE output is used directly without applying an activation function. These output values are called logits ($l$) which is a raw data and represents un-normalized probabilities of the data belonging to a specific class. Logits are normally suitable for classification tasks. Therefore, the loss function of binary cross-entropy (BCC) is suitable for such an approach. The Sigmoid function outputs a probability for each bit independently, which is applied to the decoder logits $l$ . This allows for the direct computation of the estimated posterior probabilities $p_{\theta_r}(b|y)$ for each bit, facilitating the use of log-likelihood ratios (LLRs) in the analysis \cite{cammerer2020trainable}. The BCC minimization function can then be used to optimize AE, more details are given in Section~\ref{LF}. 

Although symbol-wise AEs suit low-complexity applications, bit-wise AEs are preferred for high-performance systems where joint optimization is critical, albeit at increased computational cost. The decision should weigh trade-offs between the complexity of training, the performance of the system, and the feasibility of the hardware.

\subsection{Activation Functions}
\label{AF}
In the context of using AE for end-to-end learning in communication systems. The ReLU is widely used in hidden layers of encoder and decoder parts of the AE model. The ELU activation function is a non-saturating function that accelerates training while mitigating the risk of gradient explosion or vanishing gradient problems. The existing literature commonly employs ReLU and/or ELU activation functions in the hidden layer of AE-based communication systems. For example, \cite{goutay2021end} initially employed ReLU in their AE model but later switched to ELU for better results. Therefore, it is highly recommended to investigate both in order to determine the optimal choice. Furthermore, further research is needed to determine the combination use of both, for example, one for the encoder and the other for the decoder. However, the LeakyReLU activation function allows for a small, non-zero gradient when the input is negative, which helps mitigate the "dying ReLU" problem often encountered with standard ReLU activations. This ensures a small gradient when the unit is inactive, which helps to keep the learning process alive \cite{mohamed2022lstm}. 
Tanh activation has zero-centered output, which can make optimization easier. It is also useful in hidden layers to model complex relationships. In \cite{aoudia2018end}, Tanh activation is used to extract the feature. In \cite{asif2020ofdm}, it is used in all the 1D convolutional layers. The scaled exponential linear unit (SELU) can induce self-normalization, which means that the activation automatically converges to zero mean and unit variance. This helps maintain the stability of the model during training, which is crucial for AEs that require consistent encoding and decoding \cite{sakketou2019invariance}. It is suited for the hidden layer of AEs to leverage its self-normalization, which can lead to faster convergence. More research is required to investigate the potential benefits of alternative activation functions, such as Swish, for better gradient flow in the hidden layers. Finally, Softmax and Sigmoid activation layers are typically used in the output layer for classification tasks. The first can be used for symbol detection, while the second can be used for binary detection.

\subsection{Loss Functions}
\label{LF}
\subsubsection{Mean Squared Error (MSE)}
It serves as a loss function during the training phase, which is determined by computing the mean squared deviations between the predicted values and the actual ground truth values \cite{dulhare2020machine}. The general expression of MSE is 
\begin{equation}
\label{eq:M1}
\mathcal{L}_{MSE} = \frac{1}{N} \sum^{N}_{i=1} |X_{estimate} - X_{true}|^{2},
\end{equation}
where $N$ is the number of samples, $X_{estimate}$ and $X_{true}$ are the estimate and ground-truth values. As AE consists of two parts: an encoder and decoder, the encoder mapping can be expressed as $w=f(x;\theta_{t})$, which produces the latent representation $w$ (see Fig.~\ref{fig_AE_com_bit}(a)). The decoder mapping, on the other hand, can be expressed as $\tilde{x}=f(y;\theta_{r})$, which reproduces a replica of the original input values of $x$. The $\theta_{t}$ and $\theta_{r}$ are the training parameters of the encoder and decoder parts, respectively. The channel layer of the AE architecture is differentiable but not learnable. As a consequence, the MSE of the proposed AE is defined as 
\begin{equation}
\label{eq:M2}
\mathcal{L}_{MSE}(\theta) = \frac{1}{N} \sum^{N}_{i=1} |f(x;\theta) - x|^{2},
\end{equation}
where $\theta=\{\theta_{t},\theta_{r}\}$ are the learning parameters of the AE model. 

Gradient-based learning \cite{roberts2022principles, montavon2012neural} is used to find optimal parameters $\theta=\{\theta_{t},\theta_{r}\}$ that minimize the loss function $\mathcal{L}_{MSE}(\theta)$ as 
\begin{equation}
\label{eq:G1}
\theta = \text{argmin} (\mathcal{L}_{MSE}(D_{set}; \theta)),
\end{equation}
where $\mathcal{L}_{MSE}(D_{set}; \theta)$ is the loss associated with $D_{set}$ set of data. This is achieved iteratively by updating $\theta$ using gradient descent as 
\begin{equation}
\label{eq:G2}
\theta_{i} = \theta_{i-1} - \beta \Delta \mathcal{L}_{MSE}(D_{set}; \theta_{i-1}),
\end{equation}
where $\beta$ represents the learning rate, and the stochastic gradient Descent, such as Adam optimizer, enhances convergence. 

\subsubsection{Cross-Entropy (CE)}
CE is a widely used metric for training classification models as it is differentiable, which is crucial for gradient-based optimization algorithms\cite{Goodfellow2016, Bishop2006}. It is a quantitative metric in information theory that is used to measure the divergence between two probability distributions. This means that it measures the dissimilarity between two probability distributions. In the context of neural networks, one distribution is the output of the predicted probability distribution by the network, and the other is the true (target) distribution. Categorical cross-entropy (CCE) and sparse categorical cross-entropy (SCCE) are two additional variations of the Cross-Entropy. While both are used for multiclass classification, the latter is optimized for datasets with integer encoded labels, offering computational efficiency benefits. One challenge of using CE is that it can be sensitive to class imbalance, where one class has significantly more samples than the other. This can lead to biased models.

Given a predicted probability distribution $p$ and a true distribution $q$, the cross-entropy is defined as:
\begin{equation}
\label{eq:CE1}
H_{CE}(p, q) = -\sum_i q_{i} log(p_{i}),
\end{equation}
where $i$ iterates over all possible classes, $p_{i}$ is the learned probability of class $i$, and $q_{i}$ is the true
posterior distribution of class $i$.In the context of AE end-to-end learning, the learned probability distribution is defined $p_{\theta_r}(a|y)$ as shown in Fig.~\ref{fig_AE_com_bit} (b). 
The encoder with the learning parameter $\theta_t$ is trained to output the symbol distribution $f_{\theta_t}(a)$ based on the input data $a$.
The optimization of the AE model is then promised by minimizing CCE \cite{Krizhevsky2012} as 

\begin{equation}
\label{eq:CE2}
H_{CCE}(p, q) = -\sum_{a,y} p_{\theta_t}(a|y) log(p_{\theta_r}(a|y)).
\end{equation}

In the case of binary classification, where $q$ is a one-hot vector, the equation simplifies to:
\begin{equation}
\label{eq:CE3}
H_{BCE}(p, q) = -z log(p) - (1-z) log(1-p), 
\end{equation}
where $z$ is the true label (0 or 1), and $p$ is the predicted probability of the positive class. This is known as BCE \cite{LeCun2015}. 


\begin{table}[]
\caption{Loss Functions Comparison}
\label{tab1}
\centering
\begin{tabular}{p{0.7cm}p{2cm}p{2cm}p{2cm}}
\toprule
\textbf{Loss Func.} &
  \textbf{Advantages} &
  \textbf{Disadvantages} &
  \textbf{Use Cases} \\
  \midrule
MSE &  averages squared error between predictions and actual values & Sensitive to outliers and Less suitable for classification tasks &
  Regression tasks \\ \midrule
BCE & quantifies prediction errors against actual distributions, differentiable & Sensitive to class imbalance and assumes independence of classes & classification tasks \\ \midrule
CCE &  Same as BCE &  Same as BCE & Multi-class classification\\ \midrule
SCCE & optimized for large datasets with numeric labels &
  Same as binary CE & same as CCE but with integer labels\\
  \bottomrule
\end{tabular}
\end{table}

\subsection{Physical constraints}
AEs offer significant value due to their ability to learn latent representations under various constraints. By tailoring the training process, AEs can be designed to capture specific desired properties within the latent space. In the context of communication systems, this enables generating a signal that acknowledges physical limitations. Therefore, some constraints can be introduced to generate robust signal based on specific physical limitations, such as:
\begin{itemize}
    \item Average power, energy, amplitude constraints,
    \item Bandwidth constraint,
    \item Peak-to-average power ratio (PAPR) constraint, and
    \item Adjust carrier limitation (ACL) constraint.
\end{itemize}

The constraints can be realized in several ways. One way is to incorporate a regularization layer into the trainable network, within the AE encoder part. This layer can effectively maintain the level of encoded signals by amplitude $|w_i| \leq 1 \ \forall w_i$, average power $\mathop{\mathbb{E}}[ |w_i|^2 ] \leq 1 \ \forall w_i$,  or energy $||w||_2 ^2 \leq n $, where $n$ is channel use \cite{o2017introduction,morocho2020learning}.  
Some constraints such as reducing PAPR and ACL can be overcome by introducing an additional loss function that penalizes the AE encoder to promote a restriction on the transmit signal such as \cite{duque2024autoencoders}.
\begin{equation}
\label{eq:pen}
\mathcal{L}_{total} = \alpha_{AE} \mathcal{L}_{AE} + \alpha_{add} \mathcal{L}_{add}, 
\end{equation}
where $\mathcal{L}_{AE}$ is the AE loss function like MSE or CE, and $\mathcal{L}_{add}$ is the added penalties to realize the constraints. $\alpha_{AE}$ and $\alpha_{add}$ are the weights in penalties that help determine how much importance is given to minimizing the loss of the learning process. This penalty is applied exclusively to the encoder weights\cite{goutay2021end, aoudia2022waveform}.

\subsection{AE Architecture Variant}
\label{IV_C}
Based on the existing literature, the following AE architectures have been used in communication systems:
\subsubsection{Classical AEs}
They typically use fully connected layers (also known as dense layers)  \cite{alzubaidi2021review}. This is the simplest form of an AE and is designed for general-purpose dimensionality reduction or feature learning tasks. In this architecture, both the encoder and the decoder are composed of fully connected layers, where each neuron in a layer is connected to every neuron in the subsequent layer. This means that the AE maps the input directly to a lower-dimensional latent space and back to the original space.
\subsubsection{Convolutional AEs}
They are based on convolutional neural network (CNN), the convolutional AEs utilize convolutional layers instead of fully connected layers. This architecture is more suitable for handling large, high-dimensional data by efficiently learning spatial patterns. Several CNN architectures have been proposed, such as AlexNet, ZefNet, ResNet, etc., more details are available under \cite{alzubaidi2021review}.
\subsubsection{Recurrent AEs}
These RAEs use recurrent neural networks (RNNs), such as long short-term memory (LSTMs) or gated recurrent units (GRUs), to handle sequential data. The main idea is to leverage the sequential processing capabilities of RNNs to reconstruct ordered or temporal data the same way traditional AE work with static data \cite{berahmand2024autoencoders}.
\subsubsection{Variational AEs} 
They are a generative model that learns the input data distribution and can generate new data samples \cite{berahmand2024autoencoders}. Unlike classical AEs that learn a deterministic mapping from input to latent space, variational AEs (VAEs) learn a probabilistic latent space. This makes them more robust for generating realistic and diverse outputs. The encoder in VAEs outputs two vectors: the mean $\mu$ and the standard deviation $\sigma$ of a latent variable distribution. Instead of encoding the input to a fixed latent vector, VAEs assume that the latent space follows a Gaussian distribution. In addition to the reconstruction loss function that measures how well the reconstructed data match the input, the Kullback-Leibler (KL) divergence loss is used to measure how close the learned latent distribution is to the standard Gaussian distribution:
\subsubsection{Sparse AEs}
These SAEs impose sparsity constraints on hidden layers, ensuring that only a few neurons are activated at a time. This forces the AE to learn efficient and compact representations of the input data, even if the dimensionality of the latent space is not explicitly reduced. they are typically enforced by adding a regularization term such as penalty to the loss function. This encourages the network to activate only a subset of neurons.
\subsubsection{Adversarial AEs}
These AAEs utilize generative adversarial networks (GANs) to regularize the learned latent space, making it match a specific prior distribution \cite{berahmand2024autoencoders}. The goal is to achieve better generative capabilities, producing more realistic and diverse data compared to traditional AEs. Along with the reconstruction loss function, adversarial loss is used to ensure that latent space follows a target prior distribution. Compared to VAEs which use a probabilistic approach with KL divergence, AEs use adversarial training through a GAN framework.

Future research should investigate in more depth the following AE architectures for the area of communication systems \cite{berahmand2024autoencoders}: Bayesian AEs, Diffusion AEs, Contractive AEs, Denoising AEs, Semi‑supervised AEs, and Masked AEs   

\subsection{Geometric and Probabilistic Constellation Shaping}
Geometric and probabilistic constellation shaping are two techniques used to optimize the performance of communication systems \cite{b13,stark2019joint,b15,b16,b17}. Although they share some similarities, there are also significant differences between them. Table~\ref{tab2} presents the main differences between them. 

GCS aims to optimize the constellation shape for better performance by optimizing the spatial distribution of modulation symbols in a two-dimensional plane. That is, it involves designing a specific geometric shape for the modulation format to achieve better performance in terms of maximizing the signal-to-noise ratio (SNR) or minimizing the BER. This is typically done by defining a set of rules or constraints that define the geometry of the constellation and then optimizing the modulation format accordingly. GCS can be achieved through various methods, such as mathematical optimization techniques, graph theory-based approaches, or machine learning algorithms. 

On the other hand, probabilistic constellation shaping (PCS) involves controlling the probability distribution of constellation points. That is, it considers the geometric distribution and introduces a probabilistic aspect to it. Control both symbol distance and their occurrence probabilities in order to optimize system performance further. The aim is typically two-fold: to enhance spectral efficiency while simultaneously improving the SNR for better BER. This is typically done by optimizing the modulation format to achieve better performance in terms of data rate, reach, and noise tolerance. Probabilistic shaping can be achieved through various methods such as optimization algorithms, machine learning techniques, or signal processing approaches. 

\begin{figure}[htbp]
    \centering
    \includegraphics[width=1.0\linewidth]{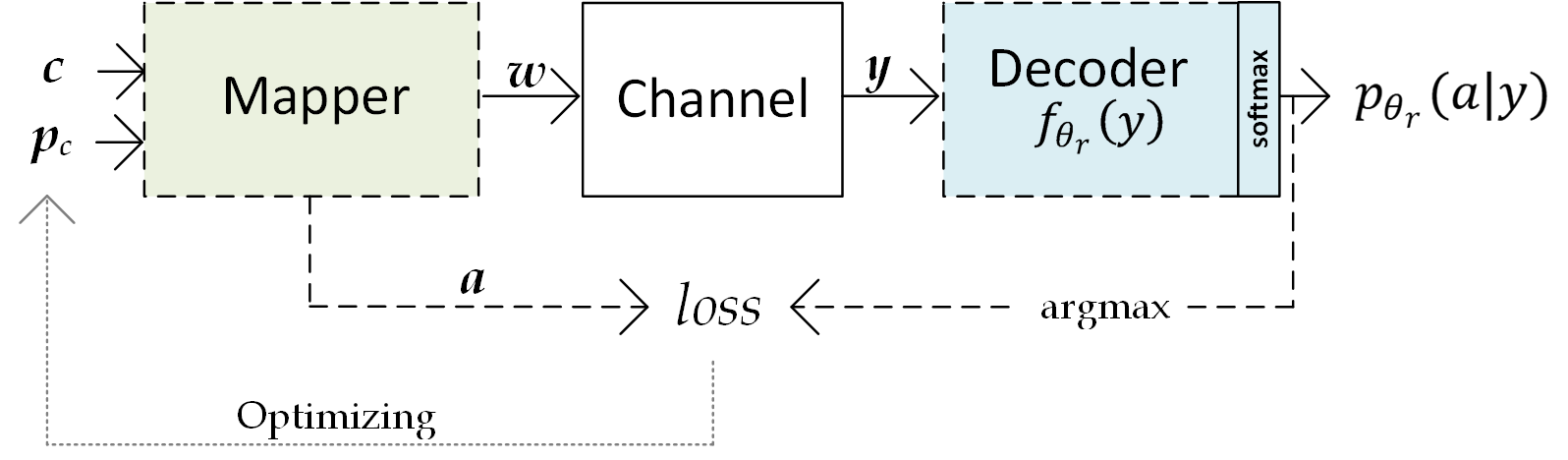}
    \caption{Geometric ($c$) and probabilistic ($p_c$) constellation shaping}
    \label{fig_ps_gs}
\end{figure}

Another key difference is the level of complexity involved in each approach. GCS can be simpler to implement, but may have limitations in terms of flexibility and adaptability to changing channel conditions. On the other hand, PCS can be more complex to implement and optimize due to its dual consideration, optimizing both spatial and temporal aspects (occurrence probabilities), and it requires advanced mathematical techniques and signal processing algorithms. In the context of end-to-end learning based on AE in the communication system, joint geometric and probabilistic shaping can optimize the constellation for enhanced performance, such as maximization of mutual information \cite{aref2022end, neskorniuk2022model, neskorniuk2022memory}. Fig~\ref{fig_ps_gs} illustrates the two inputs to the mapper on the transmitter sides: the constellation points $c$ and their corresponding probabilities $p_c$. Optimizing $c$, GCS is achieved, while optimizing $p_c$ enables probabilistic constellation shaping.

\begin{table*}
\caption{Geometric Constellation Shaping vs. Probabilistic Constellation Shaping}
\label{tab2}
\centering
\begin{tabular}{lll} 
\toprule
\textbf{Aspect}  & \textbf{Geometric Constellation Shaping} & \textbf{Probabilistic Constellation Shaping} \\ \midrule
\textbf{Objective}                 & \begin{tabular}[c]{@{}l@{}}optimize the spatial distribution of \\modulation symbols in a 2D plane\end{tabular} & Control the probability distribution of constellation points \\ \midrule
\textbf{Performance Metrics}       & maximize SNR, minimize BER   & Enhance spectral efficiency, improve SNR for better BER \\ \midrule
\textbf{Methods}                   & \begin{tabular}[c]{@{}l@{}}Mathematical optimization, graph theory, \\machine learning\end{tabular}             & \begin{tabular}[c]{@{}l@{}}optimization algorithms, machine learning, \\signal processing\end{tabular} \\ \midrule
\textbf{Constellation Geometry}    & optimized for specific geometries & \begin{tabular}[c]{@{}l@{}}Not optimized for specific geometries, \\focuses on probability distribution\end{tabular} \\ \midrule
\textbf{Adaptability}  & Less adaptable to changing channel conditions & More adaptable to different channel conditions \\ \midrule
\textbf{Learning and optimization} & \begin{tabular}[c]{@{}l@{}}Relies on pre-defined geometric shapes \\or optimization techniques\end{tabular}     & \begin{tabular}[c]{@{}l@{}}Uses learning algorithms to determine \\optimal probability distribution based on channel conditions\end{tabular}\\ \midrule
\textbf{Implementation Complexity} & Simpler to implement but less flexible  & \begin{tabular}[c]{@{}l@{}}More complex due to dual consideration \\of spatial and temporal aspects\end{tabular} \\ \midrule
\textbf{Flexibility}   & Limited flexibility and adaptability  & High flexibility and adaptability to various scenarios \\ \bottomrule
\end{tabular}
\end{table*}

\subsection{AE Complexity}
This section briefly outlines the calculation methodology of the AE complexity model. An examination of the literature on the complexity of AE is presented in Section \ref{Comp_review}. Moreover, Section \ref{VI_C} offers a discussion of future developments concerning AE complexity and computational constraints.

Computational complexity is essential for both encoder and decoder components of AE-based systems. A practical approach involves counting floating-point operations (FLOPs) \cite{hunger2005floating}, which is commonly used to assess the computational performance of algorithms and hardware in machine learning. It measures the computational performance of algorithms and hardware, which is more practical for machine learning algorithms. Additionally, counting FLOPs for algorithm complexity has different advantages such as efficiency analysis of the algorithm, hardware benchmarking, and optimizing the algorithm. In the context of AEs, FLOPs must be calculated for both the encoder and the decoder. For each layer, the number of FLOPs for matrix multiplication, bias addition, and nonlinear activation functions should be considered as follows:
\subsubsection{Matrix Multiplication}
The AE encoder receives a one-hot vector $1_l \in \{0,1\}^L$ of length $L$. This input is then multiplied by a weight matrix $W \in R^{N \times L}$, where $N$ is the number of neurons in the layer. 
Following \cite{hunger2005floating}, the computational complexity of matrix multiplication between matrices $X$ and $Y$ with sizes ($a \times b$) and ($b \times c$) respectively, is given by ($2abc - ac$). Therefore, in the context of an AE layer with $N$ neurons, the number of FLOPs required is $N(2L-1)$. 
\subsubsection{Bias Addition} 
Each neuron is associated with bias addition represented by a vector $B \in R^L$, where $L$ is the number of neurons in a layer. Adding the bias term requires one FLOP per neuron. Therefore, the total number of FLOPs for both matrix multiplication and bias addition can be calculated as $N(2L-1) + L$. 
\subsubsection{Nonlinear Activation Function}
Table~\ref{Tab_F} provides a breakdown of the FLOPs counted for various activation functions in a layer with $N$ neurons.

In summary, formula $N(2L-1) + L + F_{act}$ can be used to calculate the total number of FLOPs, where $F_{act}$ the FLOPs associated with the chosen activation function based on Table ~\ref{Tab_F}.

\begin{table}[h]
\centering
\caption{Number of FLOPs of Activation Functions}
\label{Tab_F}
\begin{tabular}{ll|ll}
\toprule
\textbf{Activation Functions} & \textbf{FLOPs} & \textbf{Activation Functions} & \textbf{FLOPs} \\
\midrule
ReLU & $N$ & ELU & $8N$ \\ 
Leak ReLU & $2N$ & Tanh & $13N$ \\ 
SELU & $9N$ & Swish & $8N$ \\
Softmax & $7N - 1$ & Sigmoid & $7N$ \\ 
\bottomrule
\end{tabular}
\end{table}

\section{AE-Based Solutions to Domain-Specific Challenges in Communication Systems and their Limitations}
\label{sec5}
This section provides a detailed analysis of communication challenges and AE roles in various domains, including wireless, optical fiber, free-space optical, visible light, semantic, quantum and underwater communications. First, it highlights the general limitations and challenges in using AE in communication systems, Then for each area, it outlines key difficulties within each area, such as complexity, channel variability, interference, data and training issues for machine learning, system and hardware limitations, security, performance, and interpretability, while also discussing potential solutions of using deep learning AE architectures as promising approaches. Moreover, it also explores various limitations encountered when implementing deep learning AEs in these specific areas of communication systems. 

\subsection{Wireless Communication} 
Wireless communication systems face a wide array of challenges, spanning from fundamental issues like channel impairments and interference to specific challenges in wireless sub-field systems such as multiple-input multiple-output (MIMO), OFDM, integrated sensing and communication (ISAC), and CR technology. Figure~\ref{fig_challenge_1} shows an overview of typical challenges of wireless communication systems, where the problem is addressed by AE-based architecture.

\begin{figure}[ht!]
    \definecolor{leo-blue}{RGB}{64,115,158}
\definecolor{geo-orange}{RGB}{218,124,48}
\tikzset{box/.style={rectangle, draw=black, rounded corners=2pt, thick},
    param/.style={box, font=\large, minimum width=4cm, minimum height=0.6cm, text width=7.5cm, align=left, drop shadow},
    header/.style={box, font=\Large, text=black, minimum width=5cm, minimum height=0.8cm, drop shadow},
    icon/.style={circle, draw, thin, minimum size=6mm}}

\centering
\resizebox{1.0\columnwidth}{!}{
\begin{tikzpicture}[y=-0.8cm]

\node[header, fill=gray!30] at (0,-0.4) {Wireless Communications};
\node[header, text=white, fill=leo-blue] at (4cm,0.8) {Solutions};
\node[header, text=white, fill=geo-orange] at (-4cm,0.8) {Challenges};

\node[param, fill=geo-orange!30] at (-4,3) {\textbf{Channel Impairments and Variability}: lack of exact channel models, mismatch between model and real conditions, noise/quantization robustness, channel coding/equalization};

\node[param, fill=geo-orange!30] at (-4,6) {\textbf{Interference Management}: dynamic and unknown interference, co-channel effects, trade-off between suppression and performance};
\node[param, fill=geo-orange!30] at (-4,8.7) {\textbf{MIMO Systems:} curse of dimensionality, decentralized processing schemes, signal detection, nonlinear/interference distortion, CSI-efficient feedback, ZIC};
\node[param, fill=geo-orange!30] at (-4,11.1) {\textbf{OFDM Systems}: inherent waveform characteristics, PAPR, ACLR, synchronization, phase noise};
\node[param, fill=geo-orange!30] at (-4,13.2) {\textbf{ISAC Systems}: JCAS, difficulties in detecting multiple radar targets, balance sensing and communication};

\node[param, fill=leo-blue!30] at (4,3) {Model-free AEs, stochastic approximation algorithms, GAN-based channel modeling, Turbo AE, VAE equalizer, denoising AE};

\node[param, fill=leo-blue!30] at (4,6) {Interference-aware training, joint transmitter–receiver AE design, adaptive deep learning};

\node[param, fill=leo-blue!30] at (4,8.7) {Bit-wise AE, BWAE-MIMO, SWAE-MIMO, VAE-based, AE-based architecture, PRVNet};
\node[param, fill=leo-blue!30] at (4,11.1) {AE-driven waveform design methodologies, AE-based GCS, MC-AE};
\node[param, fill=leo-blue!30] at (4,13.2) {AE-based Joint constellation shaping, AE-based for permutation-invariant encoding};

\draw[] (-0.15,3) -- (0.15,3);
\draw[] (-0.15,6) -- (0.15,6);
\draw[] (-0.15,8.7) -- (0.15,8.7);
\draw[] (-0.15,11.1) -- (0.15,11.1);
\draw[] (-0.15,13.2) -- (0.15,13.2);

\end{tikzpicture}
}
    \caption{Wireless communication challenges and solutions}
    \label{fig_challenge_1}
\end{figure}

\subsubsection{Communication System Design}
\label{V_A_1}
Table~\ref{state_system} lists the recent literature that focuses on the design of AE-based communication systems. Real-world conditions introduce significant channel impairments that degrade signal quality in wireless communication systems. These include extreme variability such as noise and distortion, which complicate the development of comprehensive and accurate channel models \cite{zou2021channel}. A central challenge in this context is the \textit{unavailability of an exact channel model}, which hinders the optimization of the AE parameters \cite{asif2020ofdm, zou2021channel}. Moreover, traditional modeling approaches often fail to capture complex relationships in channel behavior and are inadequate to accurately represent real world conditions \cite{Erpek2019}. Training systems with mismatched channel models can also lead to significant performance degradation \cite{che2022trainable}.

\begin{table*}[t!]
\caption{State-of-the-art comparison of AE-based models in system design}
\label{state_system}
\centering
\resizebox{\textwidth}{!}{%
\begin{tabular}{p{0.4cm}p{4.5cm}p{2.5cm}p{2.5cm}p{4cm}p{3.5cm}}
\toprule
\textbf{Ref.} & \textbf{Main Contribution} & \textbf{Model(s) Used} & \textbf{Dataset/Data Type} & \textbf{Best Performance Achieved} & \textbf{Limitation} \\ \midrule
\cite{asif2020ofdm} & Proposes stochastic approximation for AE training without exact channel model & Stochastic Approximation AE & OFDM, Fading Channels & Maintains robustness under channel uncertainty & Complexity in implementation \\ \midrule
\cite{Erpek2019} & Uses GANs for real-time channel modeling with AEs & GAN-based AE & Real-world channels & Effective channel representation and interference handling & Needs high-quality training data \\ \midrule
\cite{morocho2020learning} & End-to-end AE-based full communication system & End-to-End AE & Single-antenna, impaired channels & Adaptable, all-in-one AE framework & No direct comparison to baselines \\ \midrule
\cite{zou2021channel} & Compares model-assumed vs model-free channel AEs & GAN-CAE, RL-CAE & AWGN, Rayleigh Channels & Comparable BLER, robust under complex channels & Model-free requires more training \\ \midrule
\cite{che2022trainable} & Highlights AE performance loss from channel-model mismatch & Trainable Channel AE & Synthetic channels & Insight into model mismatch sensitivity & Lack of practical mitigation strategy \\ \midrule
\cite{xu2019performance} & Evaluates model-free AEs under various channel models & Dense AE & AWGN, Rayleigh & Performance stable under different impairments & Training time increases with channel complexity \\ \midrule
\cite{aoudia2019model} & Introduces end-to-end AE without differentiable channel model & Dense AE & AWGN, RBF, fiber-optic & Comparable to model-based AE, hardware-implementable & Assumes accurate feedback for training \\ \midrule
\cite{davaslioglu2022autoencoder} & Develops interference-aware AE with randomized smoothing & Interference-aware AE & Dynamic interference, jamming & Robust to unknown interference, improved fidelity & Complexity in balancing suppression and accuracy \\ \midrule
\cite{choubey2022autoencoder} & Highlights AE difficulty in adapting to jamming and dynamic interference & End-to-End AE & Adversarial channel conditions & Realistic interference mitigation examples & No general framework proposed \\ \midrule
\cite{wu2020deep} & Adaptive AE for dynamic interference management in beyond-5G systems & ADL AE & Multi-user Gaussian interference channels & 2 dB improvement in BER over conventional schemes & Performance highly dependent on interference estimation and higher complexity \\ \midrule
\cite{jiang2019turbo} & Proposes TurboAE with near-optimal AWGN performance & TurboAE & AWGN, Non-AWGN & Exceeds LDPC/Turbo in mid-range block length & High SNR training requires large memory \\ \midrule
\cite{jiang2020joint} & Integrates modulation with TurboAE (TurboAE-MOD) & TurboAE-MOD & Non-Gaussian channels & More robust than traditional modulation & Training requires differentiable channels \\ \midrule
\cite{balevi2020autoencoder} & AE-based Turbo coding for 1-bit quantized systems & ConvAE + Turbo Code & AWGN, Quantized Inputs & Enables 16-QAM, lowest BER for 1-bit RX & Suboptimal AE training leads to local minima \\ \midrule
\cite{van2020deep} & Energy-based AE for multiuser noncoherent systems & NC-EA & Multicarrier SIMO fading channels & BLER=$10^{-3}$ with 8 dB gain (single-user), 6.5 dB (multi-user) & Performance depends on accurate sub-carrier power allocation \\ \midrule
\cite{caciularu2020unsupervised} & VAE equalizer for blind decoding w/o pilots & VAE Equalizer & Linear/Nonlinear ISI & BER $10^{-3}$ at 6 dB, pilot-free & High sensitivity to hyperparameters \\ \midrule
\cite{zhang2021design} & Bi-GRU AE optimized for ISI channels & Bi-GRU AE & ISI, AWGN & Better than LDPC at low SNR & Needs performance boost at high SNR \\ \midrule
\cite{saidutta2021joint} & VAE-based JSCC for image and Gaussian sources & VAE with MoE & Gaussian, Laplace, Image Data & SDR near Shannon limit, robust to mismatch & JSCC lacks modularity and standards \\ \midrule
\cite{letizia2021capacity} & Mutual info-based AE loss for capacity-approaching codes & MI-Regularized AE & Unknown Channels & Better mutual info and BLER vs cross-entropy & Not scalable to high dimensions \\ \midrule
\cite{khan2019robust} & AE + DAE for robust latent image transmission & Stacked AE/DAE & Latent Vectors (Image) & Improved CA, MSE resilience to quantization & Assumes known latent vector, ideal conditions \\ \midrule
\cite{ke2021real} & LSTM DAE for modulation classification on edge devices & LSTM Denoising AE & RadioML2018.01A, PSD Data & 96.56\% accuracy, real-time on Raspberry Pi 4 & Limited advanced modulation coverage \\ \midrule
\cite{ali2017automatic} & AE with non-negativity for improved mod classification & AE + Constraint & 4th-order cumulants & 100\% accuracy at 10-15 dB SNR & Limited to low-dim features \\ \midrule
\cite{zhu2017modulation} & Stacked DAE for modulation classification under noise & Stacked Denoising AE & Synthetic modulated signals & Higher accuracy than conventional classifiers & Lacks real-world validation \\ \midrule
\cite{rajapaksha2020low} & Low-complexity AE for coded comms at low SNR & Low-Complexity AE & AWGN & Better BER than 16-QAM in low SNR & Needs validation vs modern codes \\ \midrule
\cite{rajapaksha1911low} & Early low-complexity AE showing BER gains & Dense (ReLU) & Binary, AWGN & Beats 16-QAM in ideal cases & Idealized assumptions only \\ \midrule
\bottomrule
\end{tabular}
}
\end{table*}
Therefore, AEs have been explored for designing entire communication systems. Morocho et al. \cite{morocho2020learning} showcased the versatility of AEs by developing an end-to-end communication system that replaced traditional transmitter and receiver tasks. Their approach optimized system operations and was effectively adapted to channel impairments. However, the core challenge being addressed is the sub-optimality of conventional block-based designs. To address the challenge of unavailability of an exact channel model, AEs can be designed to dynamically adjust to measured channel characteristics, both \textit{model-assumed} and \textit{model-free channel} AEs have been proposed as viable methods for mitigating channel impairments and improving performance \cite{xu2019performance, zou2021channel, aoudia2019model}. A recent innovation involves an end-to-end learning algorithm based on AE that eliminates the need for a \textit{differentiable channel model}, yet achieves performance comparable to traditional model-based training across various channel conditions \cite{aoudia2019model}. Furthermore, \textit{stochastic approximation algorithms} offer a solution to the absence of exact channel models by enabling robust learning under uncertainty \cite{asif2020ofdm}. Additionally, deep learning AE architectures that use \textit{generative adversarial networks (GANs)}, facilitate real-time channel modeling and adaptation \cite{Erpek2019}. Both model-assumed and model-free channel AEs are primarily presented as solutions designed to address channel impairments in communication systems. However, when comparing the two approaches, some characteristics could be seen as relative limitations. First, model-assumed AEs are noted to outperform model-free AEs in simpler environments \cite{zou2021channel, aoudia2019model}. This implies that model-free AEs might have limitations in achieving peak performance when the channel conditions are well-behaved and can be accurately modeled. Second, model-free AEs require more training time compared to model-assumed AEs, although they can achieve comparable performance. This longer training time is a limitation in terms of efficiency and practical deployment speed.

\textit{Interference} presents a major challenge in wireless communication systems, particularly in densely populated spectral environments, where isolating the desired signal becomes increasingly difficult. In multi-user multiple access channels (MAC), inter-user interference further complicates reliable communication, which requires robust interference management strategies. End-to-end communication frameworks that use AEs inherently capture a wide range of impairments, including interference, through \textit{joint transmitter–receiver optimization} \cite{Erpek2019}. Moreover, AE-based systems that incorporate \textit{interference-aware training techniques}, such as randomized smoothing demonstrate a strong ability to suppress unknown and dynamically varying interference sources, thereby improving resilience and communication fidelity \cite{davaslioglu2022autoencoder}. Adapting AE-based communication systems to dynamic and diverse interference scenarios is non-trivial, as achieving effective suppression without compromising system performance remains a significant hurdle. Several open challenges and limitations are identified when using an AE-based system to address interference impairments. First, adapting AE communications to various interference scenarios, including co-channel interference, is considered complex \cite{davaslioglu2022autoencoder, choubey2022autoencoder}. This complexity extends to handling unknown and dynamic interference conditions, such as jamming effects. Second, achieving effective interference suppression while maintaining communication performance is highlighted as a significant challenge when using AEs in the presence of interference \cite{davaslioglu2022autoencoder}. There is also the challenge of balancing accuracy and robustness in these conditions. Lastly, there is a need for modeling interference in addition to channel effects. This implies that simply capturing the channel may not inherently handle complex interference scenarios without specific design considerations. 

Wu et al. \cite{wu2020deep} proposed a novel \textit{adaptive deep learning} (ADL) based AE to address the challenge of dynamic interference in multi-user Gaussian interference channels, which is particularly challenging under high interference conditions where standard constellations are not optimal. It demonstrated significant performance improvements over conventional n-psk and n-QAM modulation schemes under high interference conditions. However, it considered adapting modulation and decoding strategies to specific channel conditions and interference levels.  

The importance of AEs for \textit{channel coding and decoding} comes from their potential to redefine communication system design through end-to-end optimization. Therefore, Jiang et al. \cite{jiang2019turbo} proposed Turbo AE (TurboAE), which achieved state-of-the-art performance in conventional AWGN channels for moderate block lengths and exceeded traditional codes such as the LDPC and Turbo codes. The design of TurboAE automates channel coding through deep learning using novel training algorithms and structures for the encoder and decoder. However, a limitation noted is that performance at larger block lengths requires extensive training memory, which remains challenging due to the rarity of error events. Jiang et al. \cite{jiang2020joint} also introduced TurboAE-MOD, which combines channel coding and modulation within an end-to-end AE framework. This approach exhibited enhanced reliability and robustness, especially under non-Gaussian channel conditions, compared to conventional modulation techniques. The challenges in this approach include the requirement for \textit{differentiable channel models} for backpropagation during training, although model-free learning algorithms that do not require differentiable channels have also been proposed and implemented on hardware \cite{aoudia2019model, zou2021channel}. Furthermore, Balevi et al. \cite{balevi2020autoencoder} proposed a system combining turbo codes with AE for one-bit quantized receivers. This deep learning-based coding scheme was designed for AWGN channels under one-bit quantization constraints with the aim of enhancing error correction performance. The proposed method outperformed conventional turbo codes and enabled operational one-bit quantization for higher modulation orders like 16-QAM. However, the research also acknowledges challenges such as the unattainability of perfect AE training, which leads to \textit{suboptimal training}, and the inherent difficulty in training deep networks, often resulting in local minima rather than global optima. 

Van et al. \cite{van2020deep} introduced a single user \textit{non-coherent energy AE} (NC-EA) system, which uses DNN for encoding and decoding. This system operates by allowing the decoder to use energy combined from all receive antennas, efficiently recovering transmitted data without requiring CSI estimation. Novel DNN structures were developed for both uplink and downlink NC-EA multiple access schemes. The simulation results demonstrated the superiority of the proposed NC-EA schemes over baseline methods in several key performance indicators, including reliability, flexibility, and complexity. The NC-EA system's ability to avoid CSI estimation simplifies the receiver design, which can be particularly attractive for applications requiring reliable, low latency, and low complexity connectivity, such as machine-type communications (MTCs). While It offers significant advantages, the need for effective strategies to handle the \textit{mismatch between training models and real-world channel conditions} is imperative.

Another application of AEs is \textit{channel equalization}. Caciularu et al. \cite{caciularu2020unsupervised} introduced \textit{VAE equalizers} specifically for the purpose of blind equalization. It aims to reconstruct transmitted symbols with unknown channel parameters, including impulse response and noise variance, without relying on pilot symbols. VAE equalizers have shown significant improvements in error rates for reconstructed symbols compared to existing blind equalization methods and enable faster channel acquisition. Critical analysis notes that VAE equalizer performance can be significantly influenced by \textit{hyperparameters}, and while effective for linear impairments, extending them to handle nonlinearities remains an area for future research. Zhang et al. \cite{zhang2021design} introduced a Bi-GRU-based AE tailored for ISI channels that offers flexible coding rates and outperforms convolutional and LDPC codes in low SNR regimes. Nevertheless it requires further enhancements at high SNR. 

AE have also been applied to \textit{joint source-channel coding} (JSCC), where the goal is to encode the source data directly into symbols for transmission over a noisy channel to optimize for end-to-end distortion. Saidutta et al. \cite{saidutta2021joint} developed a \textit{VAE-based JSCC scheme}, which achieved near-optimal SDR for Gaussian sources while maintaining robustness across different channel conditions. The approach also demonstrated superior performance on image datasets compared to traditional methods. Saidutta et al. \cite{saidutta2021joint} developed a VAE-based JSCC scheme, which demonstrated effective performance over analog noise channels and achieved performance comparable to or better than existing methods for Gaussian sources over AWGN channels. A key finding is that this methodology generalizes to other source distributions, including Laplace and image datasets, which shows consistent improvements and demonstrating robustness to changes in channel conditions. Despite these successes, JSCC faces challenges such as \textit{residual noise accumulation} over multiple hops, a \textit{lack of coding standards} for learned methods, loss of modularity that requires separate designs for different source types, and issues with \textit{non-differentiable components} in standard codecs that hinder training. 

Another significant area is the design of codes that approach \textit{theoretical channel capacity} using novel loss functions. Letizia et al. \cite{letizia2021capacity} proposed an \textit{AE loss function that incorporates mutual information} to design capacity-approaching codes. This loss function is inspired by the information bottleneck method and channel capacity. The aim was to design capacity-approaching codes under power constraints without prior channel knowledge by jointly maximizing mutual information and minimizing cross-entropy. It showed improved decoding schemes and rates and that traditional cross-entropy loss functions can lead to overfitting and suboptimal signal representation. The MI regularization term can help to control the information stored in the latent representation. However, this research faced critical challenges that the AE struggled to scale with\textit{ high code dimensions}, which risk local minima or failure to converge to zero-error codes. Furthermore, the mutual information estimation methods used, such as the MINE block, can be unreliable with a large number of samples. The experiments carried out were limited to a 3-bit system. This suggested a need for further experimentation and hyperparameter tuning for in-depth analysis and generalization to higher bit scenarios.

\textit{Robustness against noise and quantization} is crucial for practical systems. Khan et al. \cite{khan2019robust} combined basic AE and \textit{denoising AE} to improve robustness in transmitting latent image representations under noise and quantization, which demonstrated significant improvement in image reconstruction quality and classification accuracy (CA). Nonuniform quantization of the parameters of the AE was found to significantly reduce MSE and improve CA, which closely matches the performance without quantization. Conversely, fewer quantization bits led to increased MSE and decreased CA. Critical analysis notes the vulnerability of digital transmission using conventional AEs to quantization and channel noise. Limitations included scenarios in which the original transmitted latent vectors were assumed available, neglecting real impairments, and disregarding the effects of digitization and quantization of latent variables and parameters in practical systems. 

In the realm of \textit{modulation classification}, Ke et al. \cite{ke2021real} designed a \textit{ LSTM-based denoising AE} for radio technology and modulation classification. It achieved high classification accuracy, reaching 96.56\% at high SNR on the RadioML2018.01A dataset. A key contribution was the development of a compact neural network architecture, specifically designed for real-time classification on low-cost computational platforms like the Raspberry Pi 4. The proposed LSTM Denoising AE aimed to enhance classification accuracy while simultaneously reducing computational complexity. It was noted that existing spectrum monitoring techniques often struggle with achieving the required accuracy and computational efficiency on low-cost platforms. Similarly, Ali et al. \cite{ali2017automatic} introduced an \textit{AE with non-negative constraints}, which achieved 100\% classification accuracy at high SNR (10$-$15 dB) and demonstrated robustness under fading conditions. This approach outperformed traditional sparse AEs in both reconstruction error and classification accuracy. Furthermore, Zhu et al. \cite{zhu2017modulation} presented a stacked denoising AE for rapid and accurate modulation classification. The proposed approach outperforms traditional methods, especially in noisy environments. For the classification task, despite achieving high accuracy and efficiency on low-cost platforms. The difficulty in distinguishing certain modulation schemes with similar spectral properties persists.

Researchers have focused on developing \textit{low-complexity AE architectures} for communication systems. Rajapaksha et al. \cite{rajapaksha1911low, rajapaksha2020low} developed a low-complexity AE that aimed to reduce model dimensionality and training complexity. It outperformed convolutional coding, particularly within the range of 0-5 dB $E_b/N_0$ in the low to moderate SNR ranges and also achieved better BER than the 16-QAM half rate with hard decision decoding. However, this performance analysis was based on an ideal communication system assuming perfect synchronization, and further research is needed to evaluate its performance in non-ideal conditions and against more advanced channel codes like LDPC and polar codes. 

\subsubsection{MIMO Systems}
\label{mimo}
Table~\ref{state_mimo} lists the state-of-the-art comparison of AE-based models for MIMO systems.

\begin{table*}[t!]
\caption{State-of-the-art comparison of AE-based models for MIMO systems}
\label{state_mimo}
\centering
\resizebox{\textwidth}{!}{%
\begin{tabular}{p{0.4cm}p{4.5cm}p{2.5cm}p{2.5cm}p{4cm}p{3.5cm}}
\toprule
\textbf{Ref.} & \textbf{Main Contribution} & \textbf{Model(s) Used} & \textbf{Dataset/Data Type} & \textbf{Best Performance Achieved} & \textbf{Limitation} \\ \midrule
\cite{dorner2021bit}&Curse of dimensionality&Bit-wise AE, dense (ReLU)&Complex Gaussian&APP receiver for k=8&Scalability limitations for $k>8$ \\ \midrule
\cite{bui2023deep} & Introduced AE-based MIMO systems with improved BER & BWAE-MIMO, SWAE-MIMO & Simulated MIMO systems & Achieved BER comparable to MIMO-ML with RTN detector & Limited evaluation on specific antenna configurations \\ \midrule
\cite{zhao2021variational} & VAE-based deep learning for signal detection with reduced runtime & VAE & MIMO-OFDM-IM systems & BER performance comparable to traditional detectors with lower runtime & Requires offline training with simulated datasets \\ \midrule
\cite{tao2020autoencoder} & AE-based intelligent hybrid beamforming for mmWave systems & AE-HB & mmWave MIMO systems & Over 3 dB BER improvement over linear beamforming & Limited evaluation on specific mmWave configurations \\ \midrule
\cite{cherif2023autoencoder} & end-to-end learning approach for uplink distortion compensation & end-to-end AE & Multi-user MIMO uplink channels & Near-linear SER performance under Rayleigh fading channels & Focused on high-power amplifier distortions \\ \midrule
\cite{sahay2023defending} & Mitigation of adversarial attacks on power allocation models & Denoising AE & Adversarial attack simulations for massive MIMO & Retained accurate performance across threat models & Limited application outside adversarial contexts \\ \midrule
\cite{le2022ris}&RIS-assisted MIMO reliability&Fully connected (ReLU)&200k symbols, AWGN&Lowest BER across SNR&Single deterministic channel realisation \\ \midrule
\cite{song2022benchmarking}&Avoid interference&Dense Layer, geometric shaping&Rayleigh fading, AWGN&Outperforms Alamouti&Error floor at high SNR \\ \midrule
\cite{jang2019deep} & Addressed feedback delay and errors in CSI feedback & Deep AE & Frequency division duplex (FDD) MIMO systems & Improved normalized MSE and robustness to feedback errors & Focused on specific feedback errors and delay cases \\ \midrule
\cite{hussien2023prvnet} & Efficient CSI feedback with robustness to noise & PRVNet (VAE-based) & Indoor and outdoor FDD-MIMO scenarios & NMSE: $-27.7$ dB (indoor), $-13.9$ dB (outdoor) at CR=1/4 & Limited to feedback channels with specific compression ratios \\ \midrule
\cite{ravula2021deep} & Joint quantization and compression for CSI feedback & AE with entropy bottleneck layer & Indoor (5.3 GHz) and outdoor (300 MHz) CSI feedback & NMSE: $-24$ dB (indoor), $-13$ dB (outdoor) at BPP=3.0 & Requires specific tuning for optimal entropy bottleneck performance \\ \midrule
\cite{zhang2023deep} & Interference-aware constellations with adaptive shaping & DAE for Z-interference channels & Simulated ZIC scenarios & BER reduction: up to 2 orders of magnitude under imperfect CSI & High complexity for dynamic interference adaptation \\ \midrule
\cite{shin2024autoencoder} & Improved relay quantization algorithms with BER enhancements & APQNN & Simulated relay systems with multiple antennas & BER $\approx 10^{-3}$ at SNR=12 dB with $q=8$ bits & Increased computational complexity due to neural network-based relay \\ \midrule
\bottomrule
\end{tabular}
}
\end{table*}

The integration of deep learning AE architectures into MIMO systems introduces significant challenges, primarily due to \textit{the curse of dimensionality}. As the number of antennas and the bit representation in vector constellations increases, AE architectures experience exponential growth in complexity, which leads to computational and training difficulties.  To address this issue, D{\"o}rner et al. \cite{dorner2021bit} introduced a \textit{bit-wise AE} designed to achieve shaping gains in MIMO systems introduced a bit-wise AE. However, the system is limited to constellations with up to 8 bits due to training difficulties for larger configurations. Complexity increases exponentially with setups exceeding 8 bits, which led to decreased mutual information performance.  This curse of dimensionality presents a significant barrier for AE architecture, especially with binary detection and decoding. 

Expanding on AE-based approaches for \textit{signal detection}, Bui et al. \cite{bui2023deep} introduced two innovative systems, \textit{BWAE-MIMO and SWAE-MIMO}, designed to leverage AEs for improved signal detection performance in MIMO systems. Notably, the BWAE-MIMO system, equipped with a radio transformation network (RTN) detector, demonstrated BER performance comparable to the maximum likelihood detection (MLD) approach when evaluated in 2x2 MIMO configurations. However, several critical considerations temper the practical application of AE-based MIMO systems. Scaling AE architectures to accommodate larger numbers of bits per vector symbol presents significant challenges due to the curse of dimensionality and associated training difficulties. Furthermore, implementing AE systems in real-world scenarios is complicated by the requirement for \textit{differentiable channel models} (or effective workarounds) and the persistent challenge of ensuring robustness and adaptability across \textit{varying channel conditions} not perfectly matched during training.

In a similar vein, Zhao et al. \cite{zhao2021variational} developed a \textit{VAE-based signal detection} architecture, specifically adapted to MIMO-OFDM-IM systems. It achieved BER performance that was not only close to existing detection algorithms but also featured reduced runtime, thus underscored the efficiency of deep learning-driven solutions. Nevertheless, this approach also faces challenges. While VAEs have shown promise in areas such as blind channel equalization and handling noisy latent codes, their performance can degrade in high-resolution or \textit{high-dimensional scenarios}. Achieving optimal results often requires meticulous hyperparameter tuning, and ensuring robustness against complex noise scenarios remains an area requiring further study. 

In the domain of millimetre-wave (mmWave) communications, Tao et al. \cite{tao2020autoencoder} introduced an innovative AE-based hybrid beamforming method (AE-HB). By effectively mapping the traditionally non-convex hybrid beamforming optimization problem onto an AE neural network, their approach transformed it into a more tractable neural network training process. This method demonstrated a significant 3 dB improvement in BER compared to conventional linear beamforming systems. However,  AE-based transceivers trained on specific channel models can experience performance degradation when there are discrepancies between the training model and the actual transmission link. Furthermore, end-to-end AE systems often require \textit{differentiated channel} models for backpropagation, which can hinder development unless workarounds are used.

On the other hand, Cherif et al. \cite{cherif2023autoencoder} developed an AE end-to-end learning approach aimed at mitigating nonlinear distortions, such as those introduced by high-power amplifiers in massive MIMO systems. The proposed approach achieved near-linear SER performance at high SNRs with 16-QAM modulation, which significantly outperformed traditional distortion mitigation techniques. However, achieving optimal performance often requires a careful selection of network parameters and \textit{hyperparameters}, which can be a complex process. Learned constellations optimized for specific receiver algorithms may not achieve optimal performance with different receiver types. In addition, \textit{generalizability} across different types and sources of nonlinearity needs thorough investigation. Massive MIMO systems present another layer of complexity \cite{Hangyang2023massive}, particularly with \textit{decentralized processing schemes}. These systems are constrained by interconnect bandwidth, and current compression techniques often fail to reduce data transfer rates to acceptable levels. As a result, overall efficiency and scalability are significantly impaired. This motivates the need for innovative solutions to mitigate data throughput demands.
Sahay et al. \cite{sahay2023defending} investigated the vulnerability of DNNs used for power allocation in massive MIMO networks to adversarial attacks. They developed a DAE specifically for mitigating adversarial attacks in massive MIMO power allocation networks. It effectively filters adversarial perturbations before inputting data into the deep neural networks for power allocation. The research demonstrated the DAE's effectiveness without requiring retraining of the AE deep learning models and showed that it outperforms existing benchmarks for regression-based adversarial attack mitigation. Crucially, the DAE retains accurate performance in the absence of attacks and operates with low computational overhead, while addressing the high computational overhead identified as a limitation in traditional optimization-based methods. 

Optimization of communication reliability in reconfigurable intelligent surface (RIS)-assisted MIMO systems is explored in \cite{le2022ris}. It replaced the transceiver and the RIS model with three fully-connected neural network (FCNN) models and jointly trained the proposed system to minimize BER. The findings indicated significant gains in BER performance compared to benchmarks and demonstrated robustness even when perfect channel information is unavailable. Critically, it is clear that the need for a framework robust to imperfect channel information and the increased complexity with efficient signal detection/processing .
Furthermore, Song et al. \cite{song2022benchmarking} analyzed how AE for \textit{interference channels} learns to avoid interference in a rotated reference frame, with AWGN, open-loop MIMO, closed-loop MIMO, and MIMO broadcast interference channels. The study demonstrated that end-to-end AE learning can outperform traditional human-engineered transceivers in memoryless MIMO and MU systems without requiring prior knowledge of communication-theoretic principles. However, the study identifies potential pitfalls in interpreting learned communication schemes, particularly regarding interference management and the AE learning of interference avoidance in a rotated reference frame. Additionally, the AEs considered have a fixed data rate, which limits their adaptability for rate adaptation, and the neural network parameters used may not be fully optimal.

Jang et al. \cite{jang2019deep}, Hussien et al. \cite{hussien2023prvnet}, and Ravula et al. \cite{ravula2021deep} each addressed the challenge of \textit{CSI-efficient feedback} in MIMO systems through innovative deep learning approaches. The first proposed a deep AE-based method that significantly reduced normalized MSE and demonstrated robust performance under feedback errors and delays, which ensured reliable CSI reconstruction. The latter introduced PRVNet, a VAE-based model optimized for high-compression scenarios. PRVNet achieved exceptional NMSE values of $-27.7$ dB in indoor environments and $-13.9$ dB in outdoor settings in a compression ratio of 1/4, which outperformed many existing methods. Similarly, the latter proposed an entropy bottleneck layer within an AE-based feedback framework, which enabled joint optimization of bit rate and distortion. This approach achieved NMSE values of $-24$ dB for indoor channels and $-13$ dB for outdoor channels at specific bit-per-pixel (BPP) values, which demonstrated superior reconstruction quality over traditional methods. Collectively, these studies highlighted the potential of AE and VAE-based techniques to improve CSI feedback efficiency and accuracy in diverse wireless environments. However, assuming that the CSI information is available at the transmitter may not always be the case. Furthermore, more studies are required on the impact of AE-based compression on CSI accuracy.

Furthermore, Zhang et al. \cite{zhang2023deep} proposed a deep AE-based architecture for \textit{Z interference channels}. They proposed system introduced interference-aware constellations that adapt their shape based on interference intensity with the explicit goal of minimizing the BER. They addressed the challenge of the two-user Z-interference channel (ZIC) with finite-alphabet inputs, and the conventional finite-alphabet approaches in this context are their reliance on predefined constellations that do not adapt to interference. The structure was also extended to work effectively with imperfect CSI, directly addressing the challenges posed by estimation and quantization errors. A notable feature of their design was the inclusion of an I/Q power allocation layer, which enables the generation of nonuniform constellation shapes, thereby enhancing robustness against interference. The simulation results indicated performance improvements of up to two orders of magnitude compared to standard methods. In another innovative effort, Shin et al. \cite{shin2024autoencoder} introduced the AE framework based on amplitude phase quantization with neural networks (APQNN) for relay systems, which achieved significant BER improvements. Their approach demonstrated performance close to \(10^{-3}\) BER at high SNR levels, which emphasized its effectiveness in optimizing the relay system.

\begin{table*}[ht!]
\caption{State-of-the-art comparison of AE-based models for OFDM systems}
\label{OFDm}
\centering
\resizebox{\textwidth}{!}{%
\begin{tabular}{p{0.4cm}p{4.0cm}p{2.5cm}p{2.5cm}p{4cm}p{3.5cm}}
\toprule
\textbf{Ref.} & \textbf{Main Contribution} & \textbf{Model(s) Used} & \textbf{Dataset/Data Type} & \textbf{Best Performance Achieved} & \textbf{Limitation} \\ \midrule
\cite{stark2019joint}&optimize discrete distributions&Tx: Dense (ReLU, linear), Rx: Dense (ReLU)&AWGN, Rayleigh fading&Close to capacity on AWGN&Finding optimal marginal distribution is challenging \\ \midrule
\cite{goutay2021end}&PAPR reduction&ResNet blocks (sep. conv, ReLU)&Simulated, AWGN&1.4 dB difference at CCDF $10^{-3}$&BCE computation complexity \\ \midrule
\cite{aoudia2022waveform}&Joint Tx/Rx, constellation, labelling&ResNet blocks (sep. conv, ReLU)&Simulated, AWGN, multipath&ACLR $-$29.95 dB, PAPR 5.60 dB&NN architecture complexity \\ \midrule
\cite{aoudia2021end}&Pilot signal reduction&ResNet blocks (sep. conv, ReLU)&Realistic wireless, time/freq-selective&BER close to perfect knowledge&Practical implementation challenges \\ \midrule
\cite{felix2018ofdm}&Joint Tx/Rx optimisation&Dense (nonlinear), seq. detector&Complex-valued, WiFi-inspired&Improved block error rate (BLER)&Complexity of analysing AE-based system \\ \midrule
\cite{marasinghe2024constellation}&Mitigate hardware impairments&SC-FDE BMD&AWGN, Wiener phase noise&0.3 dB BLER improvement&Limited to SC-FDE, AWGN channel \\ \midrule
\cite{xu2022turbo} & flexible MC-AE for any channel coding & Turbo-style Multi-Carrier AE & OFDM Systems & Achieves frequency diversity gain and surpasses OFDM-IM & High computational cost due to multi-stage learning \\ \midrule
\bottomrule
\end{tabular}
}
\end{table*}
\subsubsection{OFDM Systems}
Conventional OFDM systems face significant performance and efficiency limitations due to \textit{inherent waveform characteristics} such as high peak-to-average power ratio (PAPR) and low adjacent channel leakage ratio (ACLR). High PAPR leads to nonlinear distortions in power amplifiers that degrade system performance \cite{stark2019joint, goutay2021end, alnaseri2025papr}. Similarly, poor ACLR performance signifies inadequate spectral containment, which can result in increased interference with adjacent frequency bands and reduced spectral efficiency. Table~\ref{OFDm} lists the state-of-the-art comparison of AE-based models for OFDM systems.

The use of \textit{AE-driven waveform design methodologies} that account for PAPR and ACLR constraints during training has been proposed \cite{goutay2021end, aoudia2022waveform}. They focused on jointly optimizing a high-dimensional modulation scheme while satisfying the PAPR and ACLR constraints without compromising spectral containment or inducing power inefficiencies. The proposed AE used ResNet blocks with separable convolutional layers and the ReLU activation function, along with a separable two-dimensional (2D) convolution output layer. However, such approaches involve some limitations, such as difficulty in defining target values for PAPR and ACLR, complexity of computing true average energy, and limitation in accurately reflecting and optimizing for real-world channel conditions. In the same direction, Aoudia et al. \cite{aoudia2021end} proposed a learning-based method for the joint design of transmit and receive filters, constellation geometry, and bit labeling, along with a neural network detector, evaluated on AWGN and 3GPP multipath channels. Their approach aimed to maximize the achievable information rate while satisfying ACLR and PAPR constraints. 

Another critical aspect addressed by AE-based systems is \textit{synchronization} in OFDM systems. Felix et al. \cite{felix2018ofdm} investigated the integration of an AE into an OFDM system with a CP. This was shown to mitigate synchronization issues and simplify equalization over multipath channels. Furthermore, their work demonstrated that the AE inherently learns to manage certain hardware impairments, such as non-linear amplifiers, without requiring explicit compensation algorithms. 

\textit{Phase noise} (PN) is another significant hardware impairment, which particularly impacts performance at higher frequencies like sub-THz as frequency and bandwidth increase. Coupled with the need for low PAPR in sub-THz due to amplifier limitations, constellation shaping becomes crucial for mitigating these issues. Marasinghe et al. \cite{marasinghe2024constellation} addressed the challenges of phase noise and PAPR in sub-THz communications through geometric constellation shaping. They proposed optimizing constellation shapes to be robust to PN while maintaining low PAPR under practical conditions, using an augmented Lagrangian approach for waveform optimization. Their results indicate a trade-off between PAPR and BLER performance, and stricter PAPR constraints can lead to reduced maximum amplitude and increased constellation point density. This work highlights the potential of tailored constellations to replace conventional QAM or APSK formats. The study mentions using filtered Gaussian techniques to generate phase noise samples. 

Although these learning-based approaches show promise in PAPR/ACLR reduction, pilotless transmissions, and handle hardware impairments, several challenges and areas remain, such as complexity and generalization. The complexity of the used architecture, such as the use of ResNet, which often incurs high computational complexity in the training process. The generalization is that systems that are trained under a specific channel model often encounter performance deterioration when utilized in real-world channels, primarily due to the divergences that exist between the training model and the actual channel conditions. 

Furthermore, Xu et al. \cite{xu2022turbo} introduced a \textit{turbo-style multi-carrier AE} (MC-AE) for OFDM systems. The MC-AE design allows for flexible DNN integration into channel coded systems and is capable of handling unknown channel models. The decoder within this framework is modified to produce reliable soft-bit decisions to achieve beneficial iteration gain through the exchange of extrinsic information with the channel decoder. Simulation results indicate substantial performance improvements of MC-AE compared to traditional methods, including conventional OFDM and OFDM-based index modulation in channel coded systems. However, there are limitations concerning the choice of channel coding schemes and parameters in deep learning integration. While the MC-AE demonstrates adaptability to unknown channel models, the interplay between the learned AE components and traditional channel coding elements requires careful consideration regarding parameter selection. 

\subsubsection{ISAC Systems}
ISAC systems face critical challenges in \textit{jointly optimizing communication and sensing (JCAS)} performance, primarily due to the inherently conflicting design requirements for each functionality. Communication systems typically prioritize constellations optimized for data throughput and robustness, whereas sensing applications demand constellations that enhance detection accuracy and resolution. This divergence results in a fundamental trade-off between the two objectives \cite{Geiger2025Joint}. Table~\ref{OFDm} lists the state-of-the-art comparison of AE-based models for ISAC systems.

\begin{table*}[h!]
\centering
\caption{Comparison of Studies on AE-based ISAC systems}
\label{isac_comparison}
\resizebox{\textwidth}{!}{%
\begin{tabular}{p{0.4cm}p{4cm}p{3cm}p{3cm}p{3.5cm}p{3.5cm}}
\toprule
\textbf{Ref.} & \textbf{Main Contribution} & \textbf{Model(s) Used} & \textbf{Dataset/Data Type} & \textbf{Best Performance} & \textbf{Limitation} \\ \midrule
\cite{Geiger2025Joint}& Joint GSC and PSC for ISAC & OFDM-ISAC, AWGN & Simulated-based experiments & Increase GMI by 0.16 bit/symbol while reducing kurtosis constraint by 0.19& hyperparameters fine-tuning, discrepancies between desired and achieved kurtosis levels \\ \midrule
\cite{chen2019generalized}&Improved data transmission&Dense (ReLU), symbol-wise&Simulated, stochastic AWGN&BLER below $10{-3}$ at SNR $>$ 0 dB&Evaluated only in AWGN \\ \midrule
\cite{muth2023autoencoder}&Multi-target detection and localisation&Fully connected (ELU)&Simulated, Rayleigh fading&BMI 2.94 bits, RMSE 0.04 radians&Performance degrades with increased snapshots \\ \midrule
\end{tabular}
}
\end{table*}

To balance these competing demands, constellation shaping has emerged as a pivotal strategy. However, achieving optimal performance requires careful consideration of both geometric and probabilistic constellation design, each of which offers distinct advantages under different operational constraints. \textit{Geometric shaping} tends to be more effective under strict sensing constraints, while \textit{probabilistic shaping} performs better when such constraints are relaxed. A notable challenge remains in optimizing detection probability while simultaneously satisfying performance and regulatory constraints, such as maintaining a fixed false alarm rate \cite{Geiger2025Joint, chen2019generalized}. 
Another approach to \textit{balance sensing and communication} performance is the use of \textit{Joint constellation shaping using AE frameworks} that has shown significant promise to enable simultaneous optimization of communication reliability and sensing accuracy \cite{Geiger2025Joint}. These frameworks leverage novel loss functions that integrate both communication and sensing performance metrics to ensure balanced optimization. The \textit{discrepancy between analytical models and real-world sensing performance} further complicates this optimization process, as a result it often leads to inconsistencies between simulated and empirical detection probabilities \cite{Geiger2025Joint}. 
On the other hand, AE-based systems have been extended to support \textit{permutation-invariant encoding strategies}, such as counting-based encoding schemes, which improve radar target detection in JCAS scenarios \cite{muth2023autoencoder}. However, this permutation invariance of targets during detection poses training challenges for AEs. Therefore, modern ISAC systems, particularly those that implement JCAS, encounter \textit{difficulties in detecting multiple radar targets} due to the inherent permutation invariance of the targets. This property introduces challenges for training AEs, as traditional loss functions and network architectures are not naturally suited for such scenarios \cite{chen2019generalized, muth2023autoencoder}.

In RF signal classification, AEs have been used in the field of human sensing where AEs have been proposed to extract a compressed knowledge representation of the original input in RF sensing tasks \cite{nirmal2021deep}. On the other hand, Kompella et al. \cite{kompella2024augmenting} proposed using a Vector-Quantized Variational Autoencoder (VQ-VAE) to augment training data for RF signal classification. They generate high-fidelity synthetic RF signals using the VQ-VAE. Experimental results demonstrate that augmenting datasets with this synthetic data significantly improves classification performance. However, the VQ-VAE equalizer requires additional computation, and it struggles to scale with high code dimensions.

\subsubsection{CR Technology} 
Deep AEs improve CRN by improving spectrum sensing and interference mitigation, allowing efficient and adaptive communication in dynamic environments. For spectrum sensing, AEs extract essential features from signals, detect anomalies, and identify underused spectrum bands, with Denoising AEs proving effective in reconstructing clean signals under noisy conditions. In interference mitigation, AEs learn to suppress interference, adapt modulation schemes, and reconstruct signals in multi-user scenarios, which showed their ability to predict interference levels and dynamically adjust communication parameters. By automating feature extraction, providing real-time adaptability, and handling complex multidimensional data, AEs optimize spectrum use and ensure reliable, interference-aware communication in CRNs. Table~\ref{crn_comparison} lists the state-of-the-art comparison of AE-based models for CR technology.

Abdalzaher et al. \cite{abdalzaher2021deep} proposed a deep AE-based framework called (DAE-TRUST) for detecting malicious nodes in CR-enabled IoT networks. The DAE-TRUST model employed a deep AE to identify anomalies in input data in ensuring spectrum access without collisions while protecting primary user rights by filtering out malicious reports. The effectiveness of the model was evaluated across six different environments, incorporating realistic conditions, including shadow fading measurements. A lack of prior research on AE-based trust models for security against jamming attacks in CR-assisted IoT networks puts this work as an early exploration in this specific area. However, challenges include the difficulties in integrating ML techniques into existing infrastructures without requiring additional hardware, the impact of large-scale fading on performance, and the complexity introduced by false reports affecting node reputation. 

Liu et al. \cite{liu2020data} introduced a stack hybrid AE framework for industrial CRNs. The proposed method is a semi-supervised classification approach that combines SHAE and a semi-supervised linear discriminant for effective feature extraction and classification. It effectively utilized CWD time-frequency analysis for feature extraction from wireless signals. The SHAE method outperformed AE, DAE, and SAE in classification accuracy, especially at low SNR. However, a significant challenge lies in the development of algorithms that generalize across various radio access scenarios and the need for solutions applicable across multiple radio standards. 

Teganya et al. \cite{teganya2021deep} developed a deep-completion AE for estimating radio occupancy maps, which learned spatial propagation patterns with fewer measurements compared to traditional methods. The approach addressed the challenge of accurately estimating radio maps due to complex electromagnetic wave propagation, which represents a significant area of interest in the field of cognitive radio networks. A key finding is that this method can significantly reduce the number of measurements required for accurate map estimation compared to existing schemes. However, this approach is sensitive to interference, based on the assumption that the measurements come from a single known transmitter. Finite-precision arithmetic can introduce significant errors in map estimates. Furthermore, a high computational complexity arises when predictions are needed in many locations.

Almazrouei et al. \cite{almazrouei2019using} demonstrated the efficacy of convolutional denoising AEs (CDAE) in denoising radio signals, which focused on denoising preambles from IEEE 802.11 protocols using simulated data. It achieved 96\% accuracy at SNR = 0 dB and with robust performance under varying channel conditions. The use of CDAE for denoising radio signals is emphasized as a technique that can enhance classification performance within a machine learning pipeline. Critically, the complexity of applying denoising AE in a classification pipeline remains a challenge. In addition, there is a need for solutions that are applicable across multiple radio standards. 

\begin{table*}[ht!]
\centering
\caption{Comparison of Studies on Deep Learning AEs for CR Technology}
\label{crn_comparison}
\resizebox{\textwidth}{!}{%
\begin{tabular}{p{0.4cm}p{4cm}p{3cm}p{3cm}p{3.5cm}p{3.5cm}}
\toprule
\textbf{Ref.} & \textbf{Main Contribution} & \textbf{Model(s) Used} & \textbf{Dataset/Data Type} & \textbf{Best Performance} & \textbf{Limitation} \\ \midrule
\cite{abdalzaher2021deep} & Malicious node detection in CR-assisted IoT & DAE-TRUST & IoT network data & 100\% accuracy in line-of-sight scenarios & Reduced accuracy in challenging environments \\ \midrule
\cite{liu2020data} & Signal classification in ICRNs & Stack Hybrid AE & Industrial wireless signals & 97.75\% accuracy in high SNR & Lower performance in highly uncertain conditions \\ \midrule
\cite{teganya2021deep} & Efficient radio occupancy map estimation & Deep Completion AE & Radio map data & High accuracy with minimal data & Generalization to unseen environments \\ \midrule
\cite{almazrouei2019using} & Radio signal denoising & Convolutional Denoising AE & IEEE 802.11 signal data & 96\% accuracy at 0 dB SNR & Performance drops at very low SNR \\ \midrule
\end{tabular}
}
\end{table*}

\subsection{Optical Fiber Communications}
Optical fiber communication presents unique challenges, especially with increasing data rates and the need to overcome physical limitations like chromatic dispersion and nonlinearity, alongside complexities in signal processing such as equalization and carrier recovery in transceiver design. Figure~\ref{fig_challenge_2} illustrates these identified challenges and also presents the solutions of AE-based methodologies. 

\begin{figure}[ht!]
    \definecolor{leo-blue}{HTML}{b67d77}
\definecolor{geo-orange}{HTML}{7d77b6}
\tikzset{box/.style={rectangle, draw=black, rounded corners=2pt, thick},
    param/.style={box, font=\large, minimum width=4cm, minimum height=0.6cm, text width=7.5cm, align=left, drop shadow},
    header/.style={box, font=\Large, text=black, minimum width=5cm, minimum height=0.8cm, drop shadow},
    icon/.style={circle, draw, thin, minimum size=6mm}}

\centering
\resizebox{1.0\columnwidth}{!}{
\begin{tikzpicture}[y=-0.8cm]

\node[header, fill=gray!30] at (0,-0.4) {Optical Fiber Communications};
\node[header, text=white, fill=leo-blue] at (4cm,1) {Solutions};
\node[header, text=white, fill=geo-orange] at (-4cm,1) {Challenges};

\node[param, fill=geo-orange!30] at (-4,2.7) {\textbf{Dispersion and Nonlinearity:} CD, ISI, Kerr nonlinearity, long memory optical channel};

\node[param, fill=geo-orange!30] at (-4,4.5) {\textbf{Equalization and Carrier Recovery:} robust
equalization and carrier recovery};
\node[param, fill=geo-orange!30] at (-4,6) {\textbf{Constellation Shaping:} non-Gray labeling, dedicated shaping encoder};

\node[param, fill=leo-blue!30] at (4,2.7) {Joint design transmitter and receiver, RNN, Low-complexity offline learning,  ... active challenge area };


\node[param, fill=leo-blue!30] at (4,4.5) {(VAE, VQ-VAE, VAE-LE) Equalizers, ... active challenge area};
\node[param, fill=leo-blue!30] at (4,6) {Joint GS and PS through
AE, differentiable bit-level PS ... active challenge area};

\draw[] (-0.15,2.7) -- (0.15,2.7);
\draw[] (-0.15,4.5) -- (0.15,4.5);
\draw[] (-0.15,6) -- (0.15,6);

\end{tikzpicture}
}
    \caption{Optical fiber communication challenges and solutions}
    \label{fig_challenge_2}
\end{figure}

\subsubsection{Dispersion and Nonlinearity}
Optical fiber communication systems face significant performance degradation due to \textit{chromatic dispersion} and \textit{intersymbol interference} (ISI) effects over fiber links \cite{karanov2018end, karanov2020optical}. These impairments are particularly problematic in long-haul fiber scenarios, where achieving robustness against dispersion variations remains a critical challenge \cite{karanov2025towards}. Furthermore, \textit{long memory of optical fibers} complicates accurate modeling of interference effects across symbol sequences, which poses difficulties for both DSP techniques and end-to-end learning frameworks \cite{karanov2018end, lauinger2023improving}. Additional challenges emerge from the nonlinear behavior of the fiber, such as \textit{Kerr nonlinearity}, which further restricts system capacity and complicates signal recovery \cite{uhlemann2020deep}.

To mitigate these impairments, Uhlemann et al. \cite{uhlemann2020deep} explored the use of AE deep learning for the optimization of coherent optical communication systems, particularly under the influence of CD and Kerr nonlinearity. The fiber channel model was implemented using the split-step Fourier method (SSFM). The findings indicated that the AE could learn compensation and achieve high spectral efficiency even under challenging conditions. However, AE struggles to compensate for impairments at \textit{high input powers}, and the training process may not consistently converge to a suitable solution. The \textit{interpretability} of the AE learned behavior may also limit its performance. Furthermore, the lack of a closed-form solution for certain fiber models restricts the AE compensation capabilities, and training complexity remains a practical limitation, which requires careful adjustment of the AE structure. 

AEs have facilitated the development of end-to-end optimization frameworks that jointly design transmitter and receiver components, bypassing the need for explicit mathematical models of complex fiber optic channels \cite{karanov2020optical}. These frameworks have shown notable gains in performance and reach, especially when adapted to intensity modulation/direct detection (IMDD) systems. \textit{Recurrent neural network (RNN)} architectures, including bidirectional and feedforward variants, have also been shown to be effective in capturing and compensating for channel memory effects, making them particularly suitable for dispersive optical environments \cite{karanov2020end, karanov2020end2}. A differentiable link has been modeled that accounts for CD and memory effects in the optical fiber channel. The FFNN AE is noted as being unable to compensate for interference outside of the symbol block. In addition, diminishing returns in performance gains are observed as system nonlinearities and noise increase. However, \textit{high complexity} and large model size of ANNs limit the effectiveness of AE architectures. Furthermore, such transceivers learned on specific model assumptions often perform poorly when applied to actual transmission links due to discrepancies, necessitating optimization of ANN parameters using experimental data. On the other side, finding the optimal symbol distribution for fiber channels remains an open problem, and the complexity of optimizing probabilistically shaped coherent links is a challenge. Furthermore, there is also a lack of exploration into the effects of various analog or digital impairments beyond those explicitly modeled.

In the same direction, \textit{low complexity offline learning approaches} have demonstrated improved adaptability of NN-based transceivers to dispersion variations in highly dispersive IMDD links \cite{karanov2025towards}. However, a primary difficulty lies in the need to develop \textit{efficient optimization procedures} that remain robust across varying link conditions without relying on additional on-device learning, which can introduce prohibitive computational overhead.

Table~\ref{tab:optical} lists the state-of-the-art comparison of AE-based models for optical fiber communication. 
\begin{table*}[t!]
    \centering
    \caption{Comparison of AE-based models in optical fiber communications}
    \label{tab:optical}
    \resizebox{\textwidth}{!}{%
    \begin{tabular}{p{0.4cm} p{3.5cm} p{2.5cm} p{3cm} p{3.5cm} p{3.5cm}}
    \toprule
    \textbf{Ref.} & \textbf{Main Contribution} & \textbf{Model(s) Used} & \textbf{Dataset/Data Type} & \textbf{Best Performance} & \textbf{Limitation} \\ \midrule
    \cite{aref2022end} & Accurate gradient estimation & Sigmoid/Softmax & Constellation points, AWGN & Near-optimal MI/GMI & Complexity for complex channels, large constellations \\ \midrule  
    \cite{Rode2023} & optimize constellation shaping & Bitwise & Synthetic datasets, Wiener phase noise & Improves 64-ary performance & Lack of detailed complexity analysis \\ \midrule
    \cite{karanov2020end} & optimize input probability distributions & Distance-agnostic AE & Experimental data, Generative model & Outperforms uniform distributions & Optimisation on specific model assumptions \\ \midrule
    \cite{karanov2018end} & optimize transceiver chain & Fully-connected (ReLU) & Labelled input-output pairs & 42 Gb/s at $<$ 6.7\% HD-FEC & Sensitivity to distance variations \\ \midrule
    \cite{karanov2020optical} & Dispersion-induced interference & SBRNN, FFNN & CD, memory effects & Transmission below 6.7\% HD-FEC & Trained on specific channel models \\ \midrule
    \cite{uhlemann2020deep} & Adaptation to nonlinear fiber links & Dense (ELU) & AWGN, CD, KNL & High spectral efficiency gain & Training complexity \\ \midrule
    \cite{karanov2020end2} & Dispersion-induced interference & FFNN, BRNN & Random messages, CD & Operation beyond 70 km & FFNN limitations \\ \midrule
    \cite{lauinger2023improving} & Bootstrapping of blind equalizers & VAEflex & Linear frequency domain, AWGN, PMD & Highest BMI & Higher percentage of failed runs \\ \midrule
    \cite{lauinger2022blind} & Carrier recovery, channel estimation & VAE, 1D Conv (ELU) & Dual-polarisation, AWGN, ISI & VAE-based outperform CMA & Convergence issues at high symbol rates \\ \midrule
    \cite{song2023blind} & Slow convergence of traditional methods & VQ-VAE & Unshaped 64-QAM, PDM & Frequency-domain VQ-VAE comparable to time-domain, outperforms CMA & ELBO computation challenges \\ \midrule
    \cite{jones2018deep} & Design constellations for nonlinearities & Dense (SELU) & Nonlinear fiber channel, ASE noise & 0.13 bit/4D gain & Model agnostic to embedded channel \\ \midrule
    \cite{jones2019end} & optimize geometric constellations & Dense (ReLU) & fiber channel with WDM & Up to 0.2 bits/symbol GMI gain & Implementation penalty \\ \bottomrule
    \cite{rode2022geometric} & Laser phase noise & Bit-wise (ReLU) & Simulated symbols, Wiener phase noise & 0.1 bit/symbol gain & May not consistently outperform \\ \midrule
    \cite{rode2023optimized} & Differentiable CPE & Fully-connected (ReLU) & Simulated signals, Wiener phase noise & $\mu$ = 4 provides best performance & Additional computational complexity \\ \midrule   
    \bottomrule
     \end{tabular}
    }
\end{table*}

\subsubsection{Equalization and Carrier Recovery}
Achieving robust equalization and carrier recovery in coherent optical fiber communication systems remains a critical and complex challenge. Traditional algorithms such as the adaptive zero-forcing (ZF) approach suffer from convergence issues during startup, while independent signal analysis (ISA) methods are prone to noise-induced numerical instabilities \cite{rode2025machine, vijayakumari2025survey}. On the other hand, bootstrapping blind equalization algorithms, especially in systems using adaptive equalizers, has proven difficult, with commonly employed techniques such as the constant modulus algorithm (CMA) and its variants failing to converge at critical operating points \cite{lauinger2023improving}. These issues are further exacerbated in systems that use higher-order modulation formats and PCS, where equalizers face significant difficulty maintaining reliable performance \cite{lauinger2023improving, lauinger2022blind}. Conventional decision-directed algorithms also struggle in low SNR regimes, while the convergence of CMA becomes notably slower for complex modulation schemes \cite{song2023blind}. Furthermore, designing blind receivers for PCS-modulated systems poses additional complexity, particularly in time-varying channels where ISI and noise further degrade performance \cite{lauinger2022blind, Rode2023}. Despite extensive research, there remains a lack of effective blind maximum likelihood (ML)-based equalizers suitable for PCS formats and dynamic conditions \cite{lauinger2022blind}.

In response to these challenges, \textit{pilot-less VAE equalizers} demonstrated substantial gains in channel estimation accuracy, which can eliminate the need for overhead while maintaining high performance \cite{rode2025machine}. Frequency-domain blind equalization schemes and \textit{vector quantized VAE (VQ-VAE) equalizers} have also been proposed to deliver lower complexity without sacrificing accuracy \cite{song2023blind}. However, VQ-VAE equalizers require additional computation to reconstruct channel observations from decoded symbols. In addition, training VQ-VAE equalizers may face difficulties in computing the ELBO for certain channels. 

On the other hand, \textit{VAE-based linear equalizers (VAE-LE)} have been explored for blind equalization and channel estimation. Lauinger et al. \cite{lauinger2022blind} investigated adaptive blind equalizers based on variational inference for carrier recovery in optical communications, specifically leveraging VAEs for channel estimation. A dispersive linear optical dual-polarization channel is employed, along with an AWGN and ISI time-varying channel. They demonstrated that VAEs can outperform traditional methods such as CMA in fixed and time-varying channels. In the same direction, they addressed the improvement in the performance of blind adaptive equalizer bootstrapping, specifically addressing the challenges faced by algorithms such as CMA during the start-up at critical working points \cite{lauinger2023improving}. Their work demonstrated that novel VAE-based equalizers could achieve reliable convergence at these critical points for both standard and PCS formats, unlike CMA and its variants which often failed. The demonstration used a fiber channel using a linear frequency domain channel matrix and assuming a time-invariant channel during the bootstrapping phase AWGN and first-order PMD. Critically, this work mainly focused on linear impairments and assuming that potential nonlinearities were negligible or compensated separately. The extension of VAE-LE towards nonlinearities remains an area for future research. Furthermore, the performance of the VAE-LE can be significantly influenced by \textit{hyperparameters} such as the symbol rate and the filter length, which might lead to performance penalties under certain conditions. 

Alternatively, Song et al. \cite{song2023blind} presented a novel frequency-domain blind equalization scheme using VQ-VAEs. It demonstrated that this approach can achieve performance similar to that of time-domain methods while reducing computational complexity, particularly in long-memory polarization division multiplexing (PDM) channels. This implies that the VQ-VAE equalizer can overcome the slow convergence of traditional blind equalization methods, such as CMA, especially for high-order modulation formats. However, there are challenges in the training process, as the ELBO computation for VAE equalizer training may face difficulties for certain channels. Additionally, the VQ-VAE equalizer requires additional computation to reconstruct channel observations from decoded symbols.

\subsubsection{Constellation Shaping}
Constellation shaping plays an essential role in modern optical fiber communication systems. It offers the potential to enhance mutual information (MI), spectral efficiency, and transmission reach. However, both geometric shaping (GS) and probabilistic shaping (PS) present non-trivial design and implementation challenges. In GS, \textit{non-Gray labeling} incurs implementation penalties and complicates receiver design, while PS typically requires a \textit{dedicated shaping encoder}, which poses difficulties for efficient and parallelized implementation \cite{jones2019end}. The overall complexity of shaping, including its iterative demapping requirements and reliance on non-binary FEC, adds further constraints to practical deployment. A key challenge in both shaping paradigms lies in jointly optimizing the constellation shape and system parameters to adapt to real-world impairments. These include nonlinear channel effects, fiber dispersion, and hardware imperfections, all of which must be considered during optimization to achieve shaping gains in practical fiber-optic channels \cite{jones2018deep}. 

To address these multifaceted challenges, AE-based solutions have emerged as promising solutions. These models enable \textit{joint geometric and probabilistic shaping through AE end-to-end learning frameworks} that adaptively optimize constellation layouts and symbol distributions for nonlinear and dispersive fiber channels \cite{jones2018deep, jones2019end, neskorniuk2022model}. In probabilistic shaping with AE also poses some challenges, such as it requires a \textit{shaping encoder}, which can complicate parallelized implementation, \textit{Sampling uniformly} from constellation points complicates the process for channels with memory, such as optical fiber, and difficulties in estimating gradients of cost functions related to constellation probabilities. Innovative training techniques now enable AEs to sample from constellation distributions without approximation to improve both shaping accuracy and training stability \cite{aref2022end}.

Furthermore, a series of recent works have proposed \textit{differentiated bit-level probabilistic shaping (BPS)} modules that enable flexible and optimized constellation generation within deep learning pipelines \cite{rode2022geometric, Rode2023}. These approaches outperform conventional shaping methods by achieving significant gains in MI and transmission reach, particularly under nonlinear channel conditions. However, there are open challenges in geometric shaping with AE such as it incurs an implementation penalty due to \textit{non-Gray labeling} in conventional BICM systems and it requires \textit{iterative demapping} or \textit{non-binary FEC}. The works in \cite{rode2023optimized} investigated the optimization of both geometric and probabilistic constellation shaping within Wiener phase noise channels. The core objective is to leverage deep learning AE to design constellations and associated transceiver functions that are robust to phase noise by integrating CPE directly into the end-to-end optimization process. This was particularly explored using the BPS algorithm, and also variations of the Viterbi \& Viterbi (V\&V) algorithm. However, this integration required the development of a \textit{differentiated implementation of BPS/CPE}. The standard BPS algorithm is inherently non-differentiable due to its reliance on operations like arg min, which prevents the direct application of gradient descent necessary for end-to-end deep learning optimization. 

A recent work in \cite{chimmalgi2025end} introduced a new approach for PCS in communication systems by leveraging automatic differentiation and importance sampling. The importance sampling in this work is used to draw symbols from a simplified distribution while continuously updating the approximation to the actual constellation probabilities. This method leads to a more stable and straightforward optimization process, as the gradients of the cost functions with respect to the constellation point probabilities become exact, removing potential errors caused by manual gradient correction. However, the proposed method is sensitive to how it is started and requires careful setting of various parameters. Additionally, it demands precise adjustments of hyperparameters to avoid getting trapped in local minima, which means that poor choices in setup could lead to suboptimal performance. Furthermore, relying on standard automatic differentiation without additional correction steps makes the system simpler, but it also carries the risk that the optimization process may not explore the entire solution space effectively.

While joint geometric and probabilistic shaping through AE-based systems provides a transformative direction for the design of next-generation fiber optic networks capable of supporting high-throughput, low-latency, and energy-efficient communications, the \textit{complexity of jointly learning probabilistic and geometric shaping} poses a significant hurdle in optimizing the physical layer. Therefore, further exploration is needed in the joint learning of probabilistic and geometric shaping, particularly in determining the optimal placement of constellation points. Generally, learned constellations are often tailored for specific receiver algorithms, making it challenging to achieve optimal performance universally.

\subsection{Optical Wireless Communications (OWC)}
AEs are being actively explored to address specific challenges within various optical wireless communication systems, including free-space optical (FSO), visible light communication (VLC), underwater optical communication, where those systems face challenges related to the unique properties of light as a transmission medium, including atmospheric effects and intensity/nonnegativity constraints, as well as general challenges in applying deep learning techniques. as shown in Fig.~\ref{fig_challenge_3}.

\begin{figure}[ht!]
    \definecolor{leo-blue}{HTML}{3d92ad}
\definecolor{geo-orange}{HTML}{ad3d5a}
\tikzset{box/.style={rectangle, draw=black, rounded corners=2pt, thick},
    param/.style={box, font=\large, minimum width=4cm, minimum height=0.6cm, text width=7.5cm, align=left, drop shadow},
    header/.style={box, font=\Large, text=black, minimum width=5cm, minimum height=0.8cm, drop shadow},
    icon/.style={circle, draw, thin, minimum size=6mm}}

\centering
\resizebox{1.0\columnwidth}{!}{
\begin{tikzpicture}[y=-0.8cm]

\node[header, fill=gray!30] at (0,-0.4) {Optical Wireless Communication};
\node[header, text=white, fill=leo-blue] at (4cm,0.8) {Solutions};
\node[header, text=white, fill=geo-orange] at (-4cm,0.8) {Challenges};

\node[param, fill=geo-orange!30] at (-4,2.4) {\textbf{Channel Characteristics and Modeling:} non-negativity, intensity modulation constraints, variable channel conditions};

\node[param, fill=geo-orange!30] at (-4,4.2) {\textbf{FSO Specific Challenges:} coding efficiency, degradation in strong turbulence};
\node[param, fill=geo-orange!30] at (-4,6) {\textbf{Underwater Optical Communication:} severe channel attenuation, hardware design and signal processing};

\node[param, fill=leo-blue!30] at (4,2.4) {AE-based OWC systems, sparse-representation AE, ... active challenge area };

\node[param, fill=leo-blue!30] at (4,4.2) {AE-based transceiver for FSO, convolutional AEs ... active challenge area};

\node[param, fill=leo-blue!30] at (4,6) {AE-based model tailored to underwater optical communication ... active challenge area};


\draw[] (-0.15,2.4) -- (0.15,2.4);
\draw[] (-0.15,4.2) -- (0.15,4.2);
\draw[] (-0.15,6) -- (0.15,6);

\end{tikzpicture}
}
    \caption{Optical wireless communication challenges and solutions}
    \label{fig_challenge_3}
\end{figure}

Table~\ref{tab:optical_wireless} lists the state-of-the-art comparison of AE-based models for optical wireless communication. 
\begin{table*}[t!]
    \centering
    \caption{Comparison of AE-based models in optical wireless communication}
    \label{tab:optical_wireless}
    \resizebox{\textwidth}{!}{%
    \begin{tabular}{p{0.4cm} p{3.5cm} p{2.5cm} p{3cm} p{3.5cm} p{3.5cm}}
    \toprule
    \textbf{Ref.} & \textbf{Main Contribution} & \textbf{Model(s) Used} & \textbf{Dataset/Data Type} & \textbf{Best Performance} & \textbf{Limitation} \\ \midrule
    
    \cite{zhai2020design} & Channel attenuation & Fully connected & Underwater optical, AWGN & Best 16QAM performance & Sensitivity to channel conditions \\ \midrule
    \cite{huynh2023performance}&optimize constellation for near Shannon capacity&Dense (ReLU), softmax&Shared communication, AWGN& Improves SER over TDMA/FDMA& computational complexity \\ \midrule
    \cite{mohamed2022lstm} & PAPR reduction & LSTM-AE & VLC DCO-OFDM, AWGN & 4.2 dB at CCDF $10^{-3}$ & Limited PAPR improvement \\ \midrule
    \cite{chimmalgi2025end} & new cost functions and a weight learning method & NN-layers, COCOB Optimizer & AWGN, 256-QAM & MI differences at maximum around 0.01 bits per symbol & Sensitive to hyperparameters, Local Minima Risk \\ \midrule
    \cite{kim2022autoencoding} & Scalable transceiver & AEGNN & AWGN, Multicolor VLC & Comparable SER to traditional methods & Relies on channel model assumptions \\ \midrule
    \cite{liu2020autoencoder} & Long-distance atmospheric turbulence & Dense (ReLU) & AWGN, multipath fading & Outperforms conventional QAM & Suboptimal compared to global solution \\ \midrule
    \cite{soltani2018autoencoder} & Design techniques for optical channels & Dense (linear) & Simulated messages, AWGN & 1 dB inferior to soft-decision decoder & Fails when both transmitters send same symbols \\ \midrule
    \cite{si2020model} & Efficient transceiver design & SR-AE, DNN & Poisson Channel & DNN outperforms existing designs & SR-AE performance loss in some scenarios \\ \midrule
    \cite{zhang2023autoencoder} & Atmospheric turbulence & Symbol differentiation & Random messages, Rytov variance & 12 dB improvement over PPM & ML receiver performance worse than DNN \\ \midrule
    \cite{cao2023end} & BER degradation in unknown CSI & 1D Conv (ELU) & Random bits, Gamma-Gamma & BER $10^{-5}$ at SNR 20.5 dB & Degrades with increasing turbulence \\ \midrule
    \bottomrule
     \end{tabular}
    }
\end{table*}

\subsubsection{Channel Characteristics and Modeling}
OWC systems face distinct challenges arising from the physical nature of the optical channel, including \textit{nonnegativity} and \textit{intensity modulation constraints}, as well as \textit{variable channel conditions} that differ markedly from traditional radio frequency (RF) models \cite{liu2020autoencoder, soltani2018autoencoder}. Furthermore, developing effective communication techniques that do not rely heavily on existing analytical channel models remains a pressing challenge, as these models often fail to capture the \textit{complexity of real-world OWC channels} \cite{soltani2018autoencoder}. These difficulties are particularly pronounced in Poisson channels, which are commonly used to model photon-limited optical intensity channels. 

To address these challenges, \textit{AE-based OWC systems} have been shown to outperform conventional QAM formats in terms of BER, particularly under multipath fading conditions \cite{liu2020autoencoder}. They also present a viable solution to the modeling and design challenges posed by Poisson channels. AE end-to-end learning techniques tailored specifically for these channels enable transceiver architectures to learn the unique statistical characteristics of photon-limited communication without requiring precise analytical models \cite{huynh2023performance}. Recent proposals demonstrate that \textit{sparse-representation AE (SR-AE)} can match or exceed the performance of traditional transceivers while offering reduced computational complexity \cite{si2020model}. They addressed the challenge of designing a transceiver over the Poisson channel, which often relies on approximations and numerical searches. The AE architecture may not fully meet the requirements for non-negative and intensity constraints. 

Another open issue is the unclear impact of varying the modulation order on overall system performance. Both high and low-order modulation schemes can hinder communication efficiency under different OWC conditions. This requires \textit{adaptive modulation strategies} that can dynamically respond to channel variability \cite{liu2020autoencoder}. Traditional techniques such as Quadrature Amplitude Modulation (QAM) are often not suitable for OWC environments, especially in the presence of multipath fading, which leads to signal distortion and degraded performance \cite{liu2020autoencoder}. 

\subsubsection{VLC Systems}
In the area of VLC, Mohamed et al. \cite{mohamed2022lstm} aimed to address the problem of high PAPR by proposing a long-short-term memory-AE (LSTM-AE) model. This model used for the first time the LSTM-AE architecture with a fully connected layer using the LeakyReLU activation function. It shows the ability to maintain BER while effectively reducing PAPR with its superiority over existing techniques. However, increasing the hyperparameter beyond 0.3 does not significantly improve the PAPR values. Additionally, while the LSTM-AE has fewer parameters compared to a Dense AE, its processing time was slightly longer than that of the Dense-AE, which is a consideration in real-time applications.

On the other hand, Kim et al. \cite{kim2022autoencoding} developed a versatile AE framework known as the autoencoding graph neural network (AEGNN) to address limitations in existing AE models and improve scalability of the transceiver design, particularly in VLC. The work addressed fixed AE computational structures that lack flexibility in varying the lengths of message bits and codewords. The study demonstrated that AEGNN can maintain an almost identical SER to traditional methods for different coding configurations. While the framework demonstrates versatility for various configurations, the limitation in applying AE architectures to higher dimensions remains. The curse of dimensionality is even an insurmountable barrier for AE architecture in high-dimensional scenarios. This suggests that while AEGNN offers flexibility for different configuration pairs within certain limits, scalability to much greater than 8 bits remains a significant challenge for current AE architectures. 

For \textit{probabilistic shaping in VLC}, training a sampling mechanism for symbols drawn from a finite set using ML is challenging \cite{Nguyen2025VLC, aref2022end}. There are difficulties in estimating the gradients of cost functions related to the probabilities of constellations.

\subsubsection{FSO Systems}
An end-to-end AE-based transceiver has been proposed specifically for FSO communication systems \cite{zhang2023autoencoder}. The research aims to address the challenges posed by atmospheric conditions in UAV-to-ground FSO links, where atmospheric turbulence can cause intensity fluctuations and beam wandering. Notably, the proposed model demonstrated a remarkable 12 dB performance gain over traditional pulse position modulation (PPM) transmitters, especially in the presence of atmospheric distortions such as scintillation and beam spread. However, maintaining \textit{coding efficiency} while increasing transmission power directly impacts the overall effectiveness of transmission, which is an essential requirement for reliable long-distance optical links. Additionally, there is a challenge in optimizing transmission schemes due to a \textit{trade-off between symbol differentiation and error floor}. In addition, \textit{performance degradation in strong turbulence} is still an open challenge. 

On the other hand, Cao et al. \cite{cao2023end} investigated the use of convolutional AEs for improving FSO communication through end-to-end AE learning. Their work focused on addressing complex problems of practical FSO deployments due to factors like sensitivity, pointing inaccuracies, and atmospheric turbulence. The proposed model demonstrated improved performance compared to conventional ML detection or estimation methods when CSI is unknown. While the proposed system shows improved resilience against turbulence compared to conventional systems, its BER performance still degrades with increasing turbulence intensity. This highlights that despite the improvements, the system remains vulnerable to strong atmospheric effects, which emphasizes the need for effective modeling and mitigation strategies.

\subsubsection{Underwater Optical Communication Systems}
Underwater optical communication presents a unique set of challenges, most notably \textit{severe channel attenuation} caused by scattering and absorption effects in the aquatic environment \cite{zhai2020design}. These impairments significantly degrade signal strength over distance. In addition, conventional transceiver technologies often fail to mitigate these losses. Consequently, there is a pressing need for innovation in both \textit{hardware design and signal processing approaches}. Achieving high-bandwidth communication rates in such conditions remains particularly difficult, further constraining the deployment of real-time and data-intensive underwater applications.

In response to these limitations, recent work has proposed an \textit{AE-based model tailored to underwater optical communication}. This data-driven architecture has demonstrated the potential to enhance transceiver performance, particularly under severe channel attenuation conditions \cite{zhai2020design}. Importantly, the AE structure exhibits optimal performance in scenarios involving high symbol complexity, where traditional modulation and detection schemes often struggle. By jointly learning an end-to-end mapping between transmitted and received signals, the AE effectively adapts to the statistical characteristics of the underwater channel. 

However, the proposed AE architecture does not fully meet the requirements for \textit{non-negative and peak-limited optical communication}. These physical constraints necessitate modifications to the standard AE output to be compatible with optical transmission hardware, which is a crucial consideration for practical implementation.

\subsection{Semantic Communication}
Semantic communication focuses on transmitting the meaning of information, which introduces challenges beyond traditional reliable bit transmission, such as quantifying meaning, dealing with background knowledge differences, and handling semantic noise, as listed in Fig.~\ref{fig_challenge_4}.

\begin{figure}[ht!]
    \definecolor{leo-blue}{HTML}{297360}
\definecolor{geo-orange}{HTML}{796b6d}
\tikzset{box/.style={rectangle, draw=black, rounded corners=2pt, thick},
    param/.style={box, font=\large, minimum width=4cm, minimum height=0.6cm, text width=7.5cm, align=left, drop shadow},
    header/.style={box, font=\Large, text=black, minimum width=5cm, minimum height=0.8cm, drop shadow},
    icon/.style={circle, draw, thin, minimum size=6mm}}

\centering
\resizebox{1.0\columnwidth}{!}{
\begin{tikzpicture}[y=-0.8cm]

\node[header, fill=gray!30] at (0,-0.4) {Semantic Communication};
\node[header, text=white, fill=leo-blue] at (4cm,1) {Solutions};
\node[header, text=white, fill=geo-orange] at (-4cm,1) {Challenges};

\node[param, fill=geo-orange!30] at (-4,2.6) {\textbf{Defining and Quantifying Meaning:} define, model, and quantify meaning, high recovery rates};

\node[param, fill=geo-orange!30] at (-4,4.7) {\textbf{Semantic Noise and Quality Metrics:} lack of objective metrics to model
and evaluate semantic noise};
\node[param, fill=geo-orange!30] at (-4,7.1) {\textbf{Background Knowledge Alignment:} common BK between
communication nodes, convey AE-based systems to intended meaning};

\node[param, fill=leo-blue!30] at (4,2.6) {AE-based technique for learning conceptual spaces , ... active challenge area };

\node[param, fill=leo-blue!30] at (4,4.7) {AE-based image quality metric ... active challenge area};

\node[param, fill=leo-blue!30] at (4,7.1) {AE-based semantic communication framework, ... active challenge area};


\draw[] (-0.15,2.6) -- (0.15,2.6);
\draw[] (-0.15,4.7) -- (0.15,4.7);
\draw[] (-0.15,7.1) -- (0.15,7.1);

\end{tikzpicture}
}
    \caption{Semantic communication challenges and solutions}
    \label{fig_challenge_4}
\end{figure}

\subsubsection{Defining and Quantifying Meaning}
The question of how to formally \textit{define, model, and quantify meaning} remains one of the most persistent challenges in artificial intelligence and cognitive science. Despite current research, there is no consensus on a unified framework for modeling meaning in a way that is both computationally tractable and semantically expressive. Existing approaches often struggle with interpretability and fail to provide a clear understanding of the learned representations, which can limit their utility in applications that require human-aligned semantic reasoning. The absence of a standardized model for capturing conceptual meaning impedes progress in areas such as explainable AI, semantic search, and cross-modal reasoning.

To address these gaps, recent work has proposed an \textit{AE-based technique for learning conceptual spaces} which is a framework initially inspired by cognitive theories of concept representation \cite{wheeler2024autoencoder}. This approach enables the unsupervised learning of interpretable, semantically structured domains from raw input data, guided by high-level property labels. 
While the proposed framework learns a domain from raw data, the challenge of obtaining semantic property labels for training remains difficult. The study primarily considered linear dimensions and Euclidean semantic distortion, thereby a need for exploration into more general approaches is required. Furthermore, the lack of interpretability in deep learning systems was a motivation, but the interpretability of the learned dimensions in complex scenarios beyond the tested datasets could still be challenging. Crucially, achieving \textit{high recovery rates} in scenarios with minimal changes in data, such as time series data, is a significant problem \cite{Oh2024data}. This suggests that AEs may struggle to capture and reproduce subtle semantic nuances or small changes in meaning that do not drastically alter the underlying data structure.

\subsubsection{Semantic Noise and Quality Metrics}
In the context of image transmission, particularly in emerging 6G communication systems, the lack of \textit{objective metrics to model} and \textit{evaluate semantic noise} poses a significant challenge. Unlike traditional distortions such as compression artefacts or Gaussian noise, semantic noise pertains to degradations that alter the meaning or recognisability of visual content. Existing noise removal techniques are largely ineffective in addressing these deeper, meaning-altering perturbations, as they typically focus on pixel-level fidelity rather than semantic integrity \cite{samarathunga2024autoencoder}. Currently, there is no widely accepted quality metric capable of capturing and quantifying semantic noise. Conventional full-reference and no-reference quality metrics often do not account for the semantic content of images, instead relying on structural or perceptual similarity measures that do not reflect human understanding of image quality \cite{samarathunga2024autoencoder}. 

To address this gap, an \textit{AE-based image quality metric (AEQM)} has been proposed to explicitly model and quantify semantic noise \cite{samarathunga2024autoencoder}. A convolutional AE architecture is utilized to handle the semantic content of the images. To quantify semantic noise, the model calculates the MSE between the latent vector spaces of undistorted and distorted images. The proposed model demonstrated a high correlation (88\%) with subjective quality assessments. However, the AEQM requires considerable computational power due to its complex architecture, which is particularity less efficient than simpler metrics like PSNR. In addition, while showing high correlation, the unpredictability of no-reference quality metrics in general could still pose challenges. 

\subsubsection{Background Knowledge Alignment}
In scenarios involving multiple nodes (like relay channels), there is a challenge in achieving common \textit{background knowledge (BK)} between communication nodes \cite{luo2022autoencoder}. Traditional relay schemes are often inadequate for semantic transmission, as they primarily focus on symbol-level forwarding and lack the mechanisms to account for contextual disparities between the source and destination \cite{luo2022autoencoder}. This limitation becomes more pronounced in scenarios requiring the accurate transmission of semantic meaning, where the destination interpretation heavily relies on pre-existing knowledge that may not align with the source's context \cite{luo2022autoencoder}. 

To address these issues, an \textit{AE-based semantic communication framework} has been proposed, specifically designed to operate over wireless relay channels \cite{luo2022autoencoder}. This framework improves semantic understanding by jointly learning to encode and decode messages in a way that is robust to variations in background knowledge. A key component of this approach is the semantic forward (SF) protocol, which enables relay nodes to forward semantically meaningful representations rather than raw symbols, which can thereby bridge knowledge gaps between communicating parties. Although AE-based systems can encode and decode sentences from a semantic dimension, it is a challenge for AE-based systems to \textit{convey the intended meaning} accurately when aligning the interpretation of this semantic information when BK varies between transmitter and receiver.

Table~\ref{SemanticComm} lists the state-of-the-art comparison of semantic communication and hybrid quantum-classical communication. 

\begin{table*}[ht!]
\caption{Comparison of AE-based models in semantic and quantum communication}
\label{SemanticComm}
\resizebox{\textwidth}{!}{%
\centering
\begin{tabular}{p{0.4cm} p{4cm} p{2.5cm} p{2cm} p{3.5cm} p{3.5cm}}
\toprule
\textbf{Ref.} & \textbf{Main Contribution} & \textbf{Model(s) Used} & \textbf{Dataset/Data Type} & \textbf{Best Performance} & \textbf{Limitation} \\ \midrule
\cite{wheeler2024autoencoder} & Quantify meaning & Modified AE (dense, conv) & CelebA & 99\% bit reduction&Reliance on high-level property labels \\
\midrule
\cite{samarathunga2024autoencoder} & Model semantic noise, objective quality metrics & Convolutional AE&COCO 2017&88\% correlation with subjective quality&Computational expense \\
\midrule
\cite{luo2022autoencoder} & Address background knowledge discrepancy & Transformer-based&Sentence-level semantic info&BLEU score close to 0.8&Ensuring shared background knowledge\\ \midrule
\cite{tabi2022hybrid} & Hybrid quantum-classical approach for robust communication &hybrid quantum-classical AE using QNN & messages sampled uniformly,AWGN & Accuracy increased 40\% with single, 55\% with double re-uploading, 20\% gain with Weighted double re-uploading & Inference time (100ms) exceeds real-time requirements\\ \midrule
\cite{tabi2025quantum} & Adaptability and robustness of QNNs in representing complex classical problems &hybrid quantum-classical AE using QNN &  standard encoded radio signals, AWGN and Rayleigh fading & Superior performance with reduced parameters and maintained accuracy, particularly with quantum encoders & optimization for near-term quantum devices, considering complex modulations computational \\ \midrule
\bottomrule
\end{tabular}
}
\end{table*}

\subsection{Hybrid Quantum-Classical Communication}
This emerging field combines quantum and classical techniques, which introduces unique challenges related to integrating these different paradigms, managing noise in quantum systems, and optimizing performance on near-term quantum hardware, as listed in Fig.~\ref{fig_challenge_5}.

\begin{figure}[ht!]
    \definecolor{leo-blue}{HTML}{008b8b}
\definecolor{geo-orange}{HTML}{e20222}              
\tikzset{box/.style={rectangle, draw=black, rounded corners=2pt, thick},
    param/.style={box, font=\large, minimum width=4cm, minimum height=0.6cm, text width=7.5cm, align=left, drop shadow},
    header/.style={box, font=\Large, text=black, minimum width=5cm, minimum height=0.8cm, drop shadow},
    icon/.style={circle, draw, thin, minimum size=6mm}}

\centering
\resizebox{1.0\columnwidth}{!}{
\begin{tikzpicture}[y=-0.8cm]

\node[header, fill=gray!30] at (0,-0.2) {Hybrid Quantum-Classical Communication};
\node[header, text=white, fill=leo-blue] at (4cm,1) {Solutions};
\node[header, text=white, fill=geo-orange] at (-4cm,1) {Challenges};

\node[param, fill=geo-orange!30] at (-4,3) {\textbf{Quantum Dataset and Encoding:} high-dimensional quantum datasets, adapting to different encoding schemes and scaling models };

\node[param, fill=geo-orange!30] at (-4,5.4) {\textbf{System Integration and Optimization:} high-performance near real-time, hyperparameter, adapting QNN architecture};

\node[param, fill=leo-blue!30] at (4,3) {Hybrid quantum-classical AE architecture ... active challenge area };

\node[param, fill=leo-blue!30] at (4,5.4) {Hybrid quantum-classical AE architecture ... active challenge area};



\draw[] (-0.15,3) -- (0.15,3);
\draw[] (-0.15,5.4) -- (0.15,5.4);

\end{tikzpicture}
}
    \caption{Hybrid quantum-classical communication challenges and solutions}
    \label{fig_challenge_5}
\end{figure}

\subsubsection{Quantum Dataset and Encoding}
The integration of quantum machine learning techniques into communication systems introduces a new set of challenges in relation to \textit{quantum hardware limitations} and \textit{performance variability}. In experimental settings, factors such as fluctuating photon numbers, variable bin sizes, and measurement noise complicate the application of conventional dimensionality reduction tools like t-SNE. Notably, the sensitivity of t-SNE to hyperparameters often results in overlapping clusters, which reduces interpretability in high-dimensional quantum datasets \cite{Mahesh2025Quantum}. Furthermore, the performance of quantum models degrades significantly when the quantum efficiency drops below 0.2, while saturation effects in classification accuracy are observed as bin sizes exceed optimal thresholds.

Although \textit{VAE models} demonstrate robustness in such noisy quantum environments, they must contend with inherent losses and noise associated with real-world quantum experimental setups \cite{Mahesh2025Quantum}. As such, there is a growing need to optimize these models for near-term quantum devices, which are constrained by limited qubit counts, decoherence, and gate infidelities \cite{tabi2025quantum}. Furthermore, \textit{adapting to different encoding schemes and scaling models} effectively across diverse quantum architectures remain active areas of exploration. In response, recent work has introduced \textit{hybrid quantum-classical AE frameworks} designed for end-to-end radio communication, effectively leveraging the representational capacity of QNNs to model channel behaviors and reconstruct signals with enhanced efficiency \cite{tabi2022hybrid}. These systems demonstrate the viability of QNNs in channel autoencoder architectures, which offer potential gains in training speed, inference latency, and robustness against quantum noise \cite{tabi2025quantum}. Moreover, optimization of quantum circuit designs has been shown to reduce training overhead and improve runtime performance. This highlights a clear path forward for deploying QNNs in near-term quantum-enabled communication platforms.

To further support the development of QNNs for practical applications, more extensive studies are required to evaluate their performance under noisy channel conditions and to design architectures that meet stringent latency and resource constraints \cite{tabi2025quantum}.

\subsubsection{System Integration and Optimization}
Despite recent progress in quantum machine learning (QML) for communication systems, several challenges remain unaddressed in the context of system-level integration and real-time performance. In particular, \textit{high-performance near real-time applications} represent an area where further empirical investigation is required to assess practical viability under operational constraints \cite{tabi2022hybrid}. Performance under complex channel noise conditions also requires a more rigorous study, especially as QNN models scale to broader deployment environments. Another persistent challenge lies in \textit{hyperparameter tuning} during end-to-end system training, which plays a critical role in minimizing estimation errors and improving overall system generalizability \cite{tabi2022hybrid}. Furthermore, \textit{adapting QNN architectures} to diverse and noisy channel conditions presents both theoretical and implementation-level complexities, particularly when integrating data re-uploading mechanisms and tailoring core circuit structures to specific application demands \cite{tabi2025quantum}. The adaptability of these models to \textit{general encoding schemes} (beyond standard modulation formats such as 4-QAM and 16-QAM) remains an important avenue for future research.

To address these issues, \textit{a hybrid quantum-classical AE architecture} has been proposed for end-to-end radio communication, in which a quantum decoder is paired with a classical encoder \cite{tabi2022hybrid}. This approach allows for the incorporation of quantum computation expressive power while mitigating the limitations of current quantum hardware. Notably, a generalized data re-uploading strategy has been introduced for qubit-based circuits to enable the system to satisfy stringent inference time constraints without compromising representational fidelity. Initial findings suggest that these hybrid architectures are promising candidates for communication over noisy channels. 

Rathi et al. \cite{rathi2024quantum} highlighted the utility of parametrized quantum circuits to optimize channel codes in classical, entanglement-assisted, and quantum communication scenarios. Strong performance is achieved with qubit-based frameworks in various quantum noise channels such as bit-flip, phase-flip, and depolarizing models, with learned capacities approaching theoretical limits. They used Hilbert space manipulation with loss functions tailored to classical and quantum communication capacities (cross-entropy for classical tasks, trace distance for quantum settings).

However, the performance of the hybrid AE in more complex channel noise scenarios requires additional study to validate its superiority over purely classical methods. Additionally, the is open challenge when adapting the model to higher numbers of qubits, as this requires adjustments to the QNN building blocks and necessitates a new choice of measurement observables, which can complicate the implementation.

\subsection{Other Specific Challenges}
Additional challenges associated with existing solutions are illustrated in Fig.~\ref{fig_challenge_6}. These pertain more broadly to the design of communication systems as a whole.

\begin{figure}[ht!]
    \definecolor{leo-blue}{HTML}{0272af}
\definecolor{geo-orange}{HTML}{af021b}
\tikzset{box/.style={rectangle, draw=black, rounded corners=2pt, thick},
    param/.style={box, font=\large, minimum width=4cm, minimum height=0.6cm, text width=7.5cm, align=left, drop shadow},
    header/.style={box, font=\Large, text=black, minimum width=5cm, minimum height=0.8cm, drop shadow},
    icon/.style={circle, draw, thin, minimum size=6mm}}

\centering
\resizebox{1.0\columnwidth}{!}{
\begin{tikzpicture}[y=-0.8cm]

\node[header, fill=gray!30] at (0,-0.4) {Other Specific Topics in Communication Systems};
\node[header, text=white, fill=leo-blue] at (4cm,0.8) {Solutions};
\node[header, text=white, fill=geo-orange] at (-4cm,0.8) {Challenges};

\node[param, fill=geo-orange!30] at (-4,2.4) {\textbf{Complexity and Computational Constraints:} computation and storage demands in resource-constraint HW};

\node[param, fill=geo-orange!30] at (-4,4.3) {\textbf{Training and Convergence:} high-dimensional data, ensure convergence};
\node[param, fill=geo-orange!30] at (-4,5.9) {\textbf{Data Representation and Rate:} one-hot low spectral efficiency};
\node[param, fill=geo-orange!30] at (-4,7.5) {\textbf{Data Scarcity:} Collecting/Classifying RF signals, large labeled datasets};

\node[param, fill=leo-blue!30] at (4,2.4) {Tensor decomposition, shallow network architectures, low-precision compression, binary NN, hybrid architecture ... };

\node[param, fill=leo-blue!30] at (4,4.3) {Segmented GAN architecture, Community-based AE, Two-step training};

\node[param, fill=leo-blue!30] at (4,5.9) {Generalized data representation (GDR), ... active challenge area};
\node[param, fill=leo-blue!30] at (4,7.5) {VQ-VAE, deep denoising AE, semi-supervised classification, ... open challenge};

\draw[] (-0.15,2.4) -- (0.15,2.4);
\draw[] (-0.15,4.3) -- (0.15,4.3);
\draw[] (-0.15,6) -- (0.15,6);
\draw[] (-0.15,7.5) -- (0.15,7.5);

\end{tikzpicture}
}
    \caption{Other specific challenges and solutions}
    \label{fig_challenge_6}
\end{figure}

\subsubsection{Complexity and Computational Constraints}
\label{Comp_review}
Deploying data-driven NN transceivers in practical communication systems is challenged by significant computational and storage demands, especially on resource-constrained hardware \cite{ney2022hybrid}. The inherent complexity of AE-based architectures adversely impacts system scalability and real-time performance \cite{huynh2023performance}. Training and inference in artificial neural networks (ANNs), particularly large-scale models, require substantial computational resources, limiting their applicability in low-power or embedded environments \cite{ney2022hybrid}. Additionally, the large parameter space of ANNs further restricts deployment due to storage limitations \cite{ney2022hybrid}.

To address these challenges, several solutions have been proposed. \textit{Tensor decomposition techniques} (such as CP, Tucker, TT, and TR) provide an effective means to reduce both computational complexity and memory requirements without compromising performance \cite{ney2022hybrid}. These techniques result in performance degradation (0.75dB to 1.3dB bit error rate degradation) that is directly related to the compression technique facilitated by tensor decomposition. In multi-user systems, \textit{shallow network architectures} have been demonstrated to achieve satisfactory results while maintaining low complexity \cite{huynh2023performance}. Furthermore, \textit{low-precision compression methods}, such as one-bit quantization, enable efficient end-to-end learning by reducing computational burden \cite{che2022trainable}. \textit{Binary neural networks (BNNs)} represent a promising direction, offering near-equivalent performance to full-precision models with significantly reduced complexity \cite{che2022trainable}. Although using low-precision compression methods and related architectures such as BNNs are presented as beneficial for complexity reduction, the challenge of storage and computation complexities can remain significant for end-to-end learning communication systems; this is less a limitation of the low-precision method as a technique for reduction. A \textit{hybrid architecture} has also been proposed that combines the adaptability of AEs with the efficiency of conventional demapping, striking a balance between performance and computational feasibility \cite{ney2022hybrid}. The hybrid approaches can achieve optimized performance while maintaining hardware efficiency and addressing complexity \cite{ney2022hybrid}. 

\subsubsection{Training and Convergence}
Training NN for communication systems presents several challenges, particularly in the presence of \textit{high-dimensional data} and \textit{channel uncertainty}. As network size increases, especially with large input dimensions (denoted by $n$), the training process becomes susceptible to unreliable convergence and entrapment in local minima, which undermines the reliability of performance \cite{letizia2025deep}. Additionally, training models based on inaccurate or incomplete channel representations can lead to suboptimal generalization and noticeable performance degradation when deployed on actual communication channels \cite{che2022trainable}. \textit{Ensuring convergence} to an optimal solution across a diverse range of channel conditions remains a fundamental difficulty, exacerbated by the significantly increased complexity associated with training large-scale neural network architectures \cite{aoudia2021end}.

To address these challenges, several advanced solutions have been proposed. A \textit{segmented GAN architecture} incorporating copula theory has been introduced to address challenges in training NNs, particularly when dealing with high-dimensional data and unknown channels \cite{letizia2025deep}. A \textit{community-based AE architecture} enables joint optimization of transmitter and receiver functions to mitigate the limitations of prior decoupled designs \cite{asif2020ofdm}. It is also used to tackle the challenge of ensuring that the learning procedure converges to an optimal solution for every channel. \textit{Practical end-to-end learning frameworks} have also been developed, such as trainable point-to-point systems where both transmitter and receiver neural networks are optimized using real-world wireless channel data, which can improve both adaptability and robustness \cite{cammerer2020trainable}. Furthermore, a \textit{two-step training paradigm} based on transfer learning has been proposed to overcome difficulties associated with real-channel variability, which offers enhanced convergence and generalization across heterogeneous channel conditions \cite{dorner2017deep}. 

\subsubsection{Data Representation and Rate}
Efficient data representation is a pivotal factor in achieving high data rates in deep learning AE-based communication systems. Traditional encoding schemes, such as \textit{one-hot vectors}, exhibit inherently low spectral efficiency due to their sparse structure, which limits the amount of information that can be transmitted per symbol \cite{chen2019generalized}. This inefficiency necessitates the exploration of alternative representation methods capable of enhancing throughput without incurring excessive computational costs. Furthermore, scaling these representations to accommodate large message dimensions becomes increasingly impractical, as it imposes significant demands on memory and processing resources \cite{rajapaksha2020low}. Consequently, the limitations of current AE-based frameworks underscore the pressing need for innovative transmission schemes that can reconcile high data rates with practical implementation constraints \cite{chen2019generalized}. 

To address these challenges, recent efforts have introduced new transmission paradigms specifically designed to increase data rates in neural communication systems, such as \textit{generalized data representation (GDR)} \cite{chen2019generalized}. GDR is effective in representing messages with large dimensions and has low complexity.

\subsubsection{Data Scarcity}
Data scarcity presents significant challenges in communication systems, particularly concerning training data for AE. Collecting extensive RF signal datasets is difficult, especially under diverse operational conditions and low SNR environments \cite{nirmal2021deep}. This difficulty leads to constraints for traditional signal classification methods due to limited training data \cite{kompella2024augmenting}. Small sample sizes may lead to overfitting issues when training models \cite{letizia2021capacity}. This highlights a need for better handling of limited data. Challenges also exist in classifying wireless RF signals in complex environments with limited labeled training data, and progress on environment-independent AE deep learning models remains slow, which requires large labeled datasets for training. Achieving efficient communication with limited training data is also a discussed problem.

AE and their variants are explored as potential solutions. A VQ-VAE is proposed to augment training data for classifying wireless RF signals in complex environments with limited labeled training data \cite{kompella2024augmenting}. Deep Denoising AE (DAEs) are highlighted as effective in scenarios with limited training data, which reportedly outperform existing solutions for channel modeling and precoding \cite{zhang2022svd}. AEs are also demonstrated as feasible for joint communication and sensing of multiple targets, with a novel counting encoding method proposed that is particularly suitable for systems with a limited number of available snapshots \cite{muth2023autoencoder}. A semi-supervised classification method combining SHAE and a semi-supervised linear discriminant is proposed to enhance generalization performance with limited labeled data \cite{liu2020data}. A data-driven approach using AEs can maintain performance even when perfect channel information is unavailable \cite{le2022ris}.

Furthermore, the discussion around adapting to channel characteristics with measured data \cite{xu2019performance}, optimizing training efficiency \cite{li2025semantic}, and applying learning strategies without continuous on-device learning \cite{karanov2025towards} implies the potential benefit of leveraging knowledge gained from one scenario (where data might be more abundant) to another (where data might be scarce). This aligns with the general concept of transfer learning, where a model trained on a related task or dataset is adapted for a new task with potentially less data. This particular aspect necessitates further deep investigation.

However, general challenges remain for AE, such as difficulties training with high code dimensions, which can lead to local minima, and the instability of some mutual information estimators with larger values, although the direct link between these general issues and limited data is not explicitly detailed \cite{letizia2021capacity}. The performance of AEs is also observed to be influenced by training procedures, architecture design, and hyper-parameter tuning, factors that can be complicated by limited data \cite{vijayakumari2025survey, che2022trainable}. The high complexity of end-to-end learning communication systems using AEs hinders further development, which requires significant computational resources, a challenge potentially exacerbated with complex limited datasets. The counting encoding method, used for limited snapshots, may lead to a lower detection rate compared to one-hot encoding. Performance may degrade significantly when domain regularities are not applicable, which could be an issue if limited datasets do not capture diverse real-world scenarios \cite{yu2017autoencoders, muth2023autoencoder}. This concludes the need for better handling of limited data by investigating the other existing methods (which could include some AE applications) as proposed in \cite{alzubaidi2023survey}.

\subsubsection{Emerging Hardware Technologies}
AE-based approaches are increasingly aligning with next-generation communication hardware technologies, particularly neuromorphic computing and photonic neural networks. This convergence represents a paradigm shift towards energy-efficient, high-speed, and intelligent communication systems that can meet the demands of 6G and beyond. The neuromorphic joint source-channel coding (NeuroJSCC) framework represents a pioneering all-spike solution that combines neuromorphic sensing using event-driven cameras (DVS), spiking neural networks (SNNs) for encoding/decoding
Impulse Radio (IR) for wireless transmission \cite{skatchkovsky2020end}. This system achieves remarkable energy efficiency with consumption proportional to spike activity, enabling picojoule-per-spike operation. However, End-to-end training of all components, including the SNNs and channel model, can be computationally demanding, where training separate models at different SNR levels is expensive. In addition, the complexity of the network structure might lead to over parametrization and overfitting, especially when increasing transmission rates does not necessarily lead to better performance. 

A photonic encoder-decoder (PED) based on an all-optical variational autoencoder is a remarkable system using linear diffractive processors, photonic VAEs for high-throughput image transmission and WDM-based broadcast-and-weight architectures \cite{chen2023photonic}. This is a clever implementation, since the system leverages optical signals to perform encoding, encryption, compression, and decoding without converting signals into electronic form and back (i.e., no analog-digital conversions). These frameworks demonstrate the potential for tera-operations per second (TOPS) throughput, far exceeding conventional digital signal processing capabilities. However, the experimental setup underlines potential challenges. The reliance on precise alignment of optical components in the adaptive training section suggests that even minor misalignments may affect performance. Although the adaptive tuning strategy is beneficial for handling variations in the system, the need for retraining and fine-tuning, particularly with a limited dataset, may limit rapid deployment and adaptability in dynamically changing transmission conditions. 

The development of neuromorphic and photonic technologies is still in its early stages, with ongoing research needed to address material challenges and improve the maturity level of these technologies \cite{mehonic2024roadmap}. Despite these challenges, the potential benefits of integrating AE-based approaches with next-generation communication hardware are substantial, offering a transformative impact on the field of artificial intelligence and communication systems.

\section{(Non-) and Differentiable Channel and Components}
\begin{figure}[htbp]
    \centering
    \includegraphics[width=1.0\linewidth]{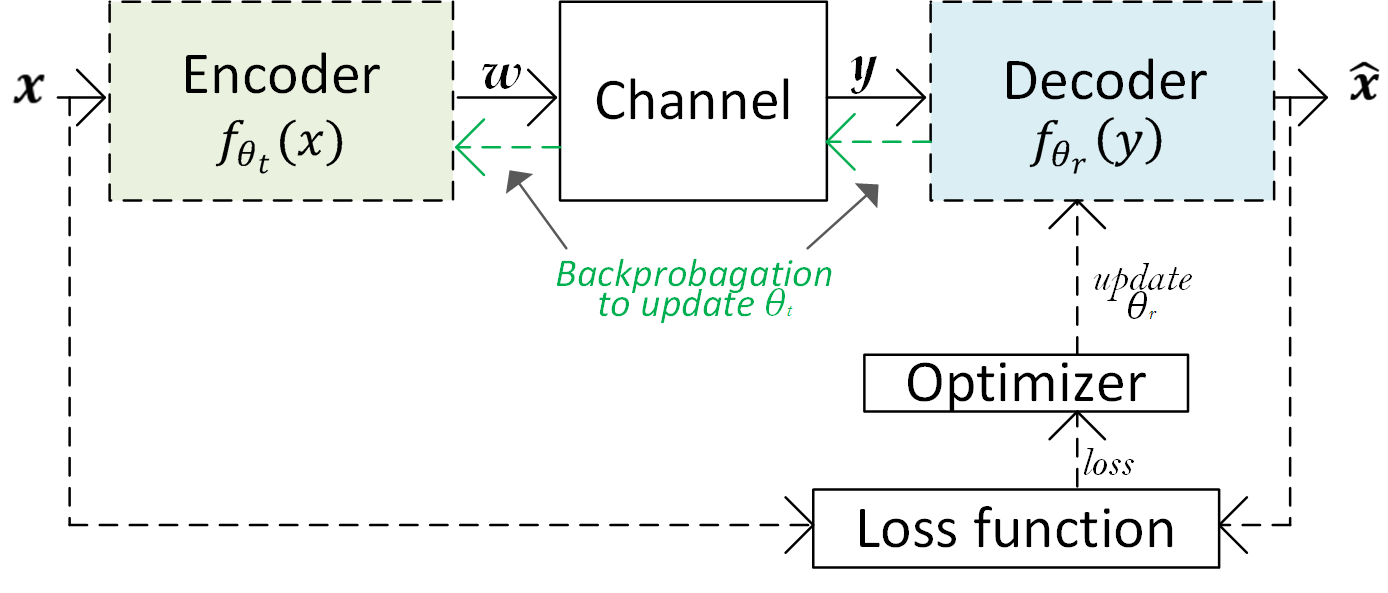}
    \caption{Backpropagating gradients for training the encoder to achieve AE end-to-end training}
    \label{fig_diff}
\end{figure}

A differentiable channel refers to a channel model that can be represented mathematically and whose behavior can be differentiated with respect to its input. Specifically, it means that the channel output (received signal) can be expressed as a differentiable function of its input (transmitted signal). This property is crucial for training end-to-end neural network-based AE learning, where a single neural network represents the entire communication system (transmitter, channel, and receiver). The AE aims to learn an efficient representation of transmitted information, and having a differentiable channel allows gradients to flow through the entire system during training, as shown in Fig.~\ref{fig_diff}.

\subsection{Determining Differentiability}
Whether a channel is differentiable depends on the specific characteristics of the channel model. a differentiable function (channel) typically has a tangent line at every point along its graph. If the tangent exists, the function is continuous, and its derivative exists, making it differentiable \cite{ActiveCalculas}. Here are some considerations:
\paragraph{Mathematical formulation} If the channel can be described by a mathematical function, such as linear channel, AWGN, or Rayleigh fading, it is likely to be differentiable. For example, the AWGN channel can be expressed as $y=x+n$, where $y$ is the received signal, $x$ is the transmitted signal, and $n$ is the noise. This channel is differentiable because its behavior can be expressed as a continuous function.
\paragraph{Discrete Effects} Non-differentiability often arises from discrete effects, such as quantization and modulation. If the channel introduces discrete operations, such as a threshold, it may not be differentiable. In such cases, alternative training methods such as reinforcement learning are needed.
\paragraph{Noise Models} Noise models matter. AWGN is differentiable due to its Gaussian noise, but other noise models, such as impulsive noise, may not be. 
\paragraph{Physical limitations} Real-world channels have physical constraints, such as power limits or bandwidth limits. These constraints affect differentiability. For instance, clipping can make a channel non-differentiable.

\begin{figure*}[h!]
\centering
\begin{tikzpicture}[
    node distance=1.2cm,
    every node/.style={draw, rectangle, rounded corners, minimum width=3cm, draw=colorgreen, drop shadow},
    every path/.style={draw, ->, >=open triangle 45}]

    \node (solutions) [align=center, fill=colorgreen!30] {\baselineskip=12pt Non-differentiable\\ Channel Solutions};
    \node (alt_train) [align=center, fill=colorgreen!20, left of=solutions, node distance=5cm] {Alternating\\training algorithm};
    \node (learn_stoch) [align=center, fill=colorgreen!20, above of=alt_train] {Learning stochastic\\ channel model};
    \node (two_phase) [below of=alt_train, fill=colorgreen!20] {Two-phase training strategy};
    \node (rl_base) [align=center, fill=colorgreen!20, below of=two_phase] {Reinforcement\\learning-based approaches};
    \node (genetic_algos) [right of=solutions, fill=colorgreen!20, node distance=5cm] {Genetic algorithms};
    \node (grad_free) [align=center, fill=colorgreen!20, above of=genetic_algos] {Gradient-free\\ training method};
    \node (diff_dsp) [align=center, fill=colorgreen!20,below of=genetic_algos] {Differentiable\\ DSP algorithms};
    \node (stoch_pert) [align=center, fill=colorgreen!20,below of=diff_dsp] {Stochastic Perturbation\\ Technique};
    \node (CGAN) [align=center, fill=colorgreen!20,below of=solutions, yshift=-1.4cm] {Conditional GAN};

    \draw (solutions) to (alt_train);
    \draw (solutions) to[bend left=-15](learn_stoch);
    \draw (solutions) to[bend left=15] (two_phase);
    \draw (solutions) to[bend left=30] (rl_base);
    \draw (solutions) to[bend left=15](grad_free);
    \draw (solutions) to (genetic_algos);
    \draw (solutions) to[bend right=15] (diff_dsp);
    \draw (solutions) to[bend right=30] (stoch_pert);
    \draw (solutions) to (CGAN);

\end{tikzpicture}
    \caption{Solutions for Non-differentiable Channel}
    \label{fig_non-diff}
\end{figure*}
\subsection{Solution Approaches for Non-differentiable}
\label{Non_diff_solution}
A key challenge in end-to-end AE learning for communication systems is the need for a differentiable channel model to enable transmitter training through backpropagation \cite{dorner2017deep}. End-to-end training becomes feasible by treating the channel as a differentiable component. However, accurate differentiable channel models are often unavailable in practical scenarios, leading to performance degradation when deployed. If the channel is non-differentiable, due to nonlinearities or discrete effect, alternative approaches are needed, such as: 

\paragraph{Alternating training algorithm} The idea behind this is to optimize the transmitter and receiver independently. The iterative optimization process involves alternating between optimizing the receiver with a fixed transmitter and subsequently optimizing the transmitter with a fixed receiver. This approach aims to enhance the overall system performance through this iterative refinement. This can be achieved through an alternating optimization approach, where the receiver is trained using supervised learning while the transmitter is optimized using reinforcement learning \cite{Aoudia2018, Goutay2019}.
\paragraph{Differentiable DSP algorithms} An end-to-end differentiable AE channel require all components, including DSP algorithms, to be differentiable to enable gradient-based optimization. Traditional DSP algorithms often incorporate non-differentiable operations, hindering their direct integration into the differentiable channel. Therefore the authors of \cite{Rode2023} proposed to modify blind phase search (BPS) algorithm and make it differentiable to include it in the end-to-end constellation shaping.
\paragraph{Learning stochastic channel model} This approach involves learning a surrogate channel model, either through supervised learning \cite{Wang2020} or adversarial methods \cite{Shea2019}. The key idea is to learn a differentiable generative model of the channel in the form of a GAN, which can then be used to train the AE, and to approximate end-to-end channel responses. The learned model is subsequently employed to train the transceiver. This enables direct optimization of the modulation and encoding components within an end-to-end AE model \cite{jafarkhani2024modulation}. The accuracy of this learned model is crucial for the overall system performance. In \cite{Yuzhe2023}, a channel-sensitive AE (CSAE) incorporates a pre-trained conditional GAN (CGAN) to model optical fiber communication systems with varying impairments.
\paragraph{Two-phase training strategy \cite{dorner2017deep}} With this approach, first train the AE using a channel model designed closely to the real channel such as a stochastic channel model. Upon deploying the trained model for real-wold transmission, performance is contingent on the model's accuracy. To mitigate this, based on measurement data, a find-tuning phase focusing exclusively on the receiver part of the AE, which leads to a sub-optimal training.   
\paragraph{Reinforcement learning-based approaches} The idea is to train an neural network (NN)-based transmitter using policy gradients while employing a non-differentiable receiver that treats detection as a clustering problem. 
\paragraph{Stochastic perturbation techniques} A gradient free optimization techniques called simultaneous perturbation stochastic optimization has been proposed in \cite{Raj2018}. It is a novel, single-stage approach to train deep learning-based communication systems directly in real-world channel environments. By circumventing the need for explicit channel modeling, this way leverages stochastic approximation to estimate gradients, enabling end-to-end optimization. 
\paragraph{Gradient-free training method} It is a derivative-free optimization method to eliminate the need for a differentiable channel model in AE training. This enables simultaneous optimization of the encoder and decoder for arbitrary, even non-numerical, channel environments. The authors of \cite{Jovanovic2021} proposed the use of the cubature kalman filter (CKF) which offers gradient-free training while maintaining accuracy, thereby facilitating online adaptation.

\paragraph{Genetic algorithms (GAs)} While traditional end-to-end learning relies heavily on backpropagation, which requires the system to be differentiable, genetic algorithms offer an alternative optimization method that does not rely on gradient information \cite{Yao1999, Yan2013, Brito2016, Vural2017}. The advantage of using GAs are as (1) \textit{optimization without gradients:} GAs optimize a population of solutions through evolutionary processes such as selection, crossover, and mutation. This optimization process does not require the computation of gradients and can be applied to non-differentiable functions. Thus, if the channel in the communication system is non-differentiable, due to involving quantization, discrete noise, or other non-continuous processes, traditional gradient-based methods fail to optimize the AE effectively. GAs, can still optimize the system by treating it as a black-box and relying on fitness evaluations rather than gradient information, (2) \textit{Fitness evaluation:} In the context of an AE, GAs can evaluate the fitness of each individual in the population based on how well the AE reconstructs the input data after it has passed through the channel. The fitness function can be designed to measure the reconstruction error or any other relevant performance metric, and (3) \textit{Population diversity:} GAs maintain a population of potential solutions, which can help to explore the solution space more thoroughly compared to gradient descent methods that could get stuck in local minima. This can be particularly beneficial in complex, non-differentiable landscapes.

\paragraph{Conditional GAN} The conceptual structure involves the use of a conditional GAN to represent the conditional distribution of a channel \cite{ye2018channel}. This allows channel effects to be learned directly from the data rather than relying on expert knowledge or assumed mathematical models. The learned conditional GAN acts as a surrogate for the real channel, which allows gradients to pass through to the transmitter, thus enabling the back-propagation of gradients to the transmitter even when the real channel is unknown, which is facilitating end-to-end learning. For time-varying channels, received pilot data is also included as part of the conditioning information. This enables the conditional GAN to generate data corresponding to specific instantaneous channel conditions.

\section{Lessons Learned}
Based on the exhaustive review described above, a compilation of recommended methodologies has been identified to ensure robust design of AE and reliable deployment in practical settings.
\paragraph{Regularization Strategies} The physical constraints such as power limits and computational efficiency such as sparsity must be balanced. In wireless communication, it shall use constraint in the encoder to enforce physical limits such as power normalization. Dropout is less common; instead, GAN-based adversarial training helps model channel variations. In addition, batch normalization is widely utilized instead of dropout. In optical communication, it has been seen that sparse AEs can reduce complexity for long-haul dispersion compensation, and VAE-based regularization (KL divergence) improves robustness to phase noise.
\paragraph{Loss Function Choices} for Regression Tasks (symbol recovery), MSE dominates but fails at low SNR; Huber loss is more robust. For Classification Tasks (modulation detection),  MI-based losses excel in capacity-approaching codes, while CCE is preferred for modulation tasks. Hybrid losses have also been incorporated such as combine MSE + PAPR penalty for OFDM waveform design. 
\paragraph{Hyperparameter Tuning Techniques} Like activation functions (ELU/Tanh) and learning rates (cyclic) are critical for convergence in noisy environments. LeakyReLU/ELU is preferred over ReLU for gradient flow in fading channels, while Tanh can mitigate exploding gradients in optical systems. Furthermore, cyclic LR can avoid vanishing gradients. In terms of learning rate, cyclic learning rates help escape local minima in end-to-end training.
\paragraph{Deployment Reliability} In the context of hardware constraint deployment, quantization-aware training such as 1-bit AE can reduce the FPGA resource. In addition, hybrid model-based AE designs facilitate FPGA implementation. Transfer learning can fine-tune AEs on real-world data and helps in channel mismatch mitigation.

\section{Future Research Areas}
\label{sec6} 
Future research can further advance the application of AEs in communication systems. Therefore, there are numerous unresolved challenges and inherent limitations when it comes to using AEs in communication systems. It is imperative that these issues undergo comprehensive investigation to uncover novel solutions or refine existing methodologies that address these challenges.

\subsubsection{Advanced AE architectures} 
\label{VI_A}
Investigate more advanced AE architectures, such as Bayesian AEs \cite{kingma2013auto}, diffusion AEs \cite{ho2020denoising}, contractive AEs \cite{rifai2011contractive}, denoising AEs \cite{vincent2008extracting}, semi-supervised AEs \cite{kingma2014semi}, and masked AEs \cite{he2022masked}. Combining and modifying these architectures could lead to improved performance and new applications in communication systems. In addition, it is necessary to make \textit{quantitative comparisons} of the system performance between different AE architectures. This includes evaluating complexity, efficiency, and effectiveness in various communication scenarios, as well as comparing hardware implementations such as FPGA.

\subsubsection{Hybrid models} 
\label{VI_B}
It develops hybrid models that combine the strengths of both data-driven and model-based approaches. This could involve integrating traditional mathematical models with deep learning techniques to enhance the robustness and adaptability of communication systems. The researcher can also investigate the use of quantum circuits in OFDM and MIMO systems. Furthermore, integrating quantum CNNs into existing CNN-based solutions could be a promising direction to handle large block sizes \cite{zhang2024hybrid}. 

\subsubsection{Complexity and Computational Constraints} 
\label{VI_C}
AE architectures often lead to significant computational and storage demands, which hinders scalability and real-time performance, especially on resource-constrained hardware such as FPGAs \cite{ney2022hybrid, goutay2022applications}. This includes developing low-complexity AE architectures and exploring techniques such as low-precision compression and binary neural networks. Training and inference for large AE models require substantial resources. Even with techniques like tensor decomposition or low-precision methods, complexity can remain a challenge. This requires a carefully adjusted AE structure \cite{uhlemann2020deep}. Further research can focus on the challenges of real-world deployment, such as the need for extensive training data, the risk of overfitting, and the complexity of practical implementations. This includes developing robust training strategies, data augmentation techniques, and methods for handling noisy or corrupted data.

\subsubsection{Data and Training Issues}
\label{VI_D}
\textit{Data Scarcity:} collecting sufficient and diverse training data, particularly under varying channel conditions or low SNR, is difficult. Limited data can lead to overfitting and challenges in training models that generalize well. \textit{Training Difficulty and Convergence:} Training large or complex AE networks can be susceptible to unreliable convergence and entrapment in local minima \cite{letizia2025deep}. Ensuring the learning procedure converges to an optimal solution across diverse channel conditions is challenging \cite{karanov2018end}. Hyperparameter tuning is often meticulous and critical for performance \cite{cammerer2020trainable, stark2019joint}. Challenges exist related to \textit{overfit the training data}, which can lead to poor performance with new unseen data. To solve this, a powerful regularization technique is to equip each layer with batch normalization. This normalization significantly accelerates training and effectively mitigates overfitting \cite{ioffe2015batch, alzubaidi2021review}. \textit{Channel Model Dependency and Mismatch:} Training models based on inaccurate or incomplete channel representations can lead to suboptimal generalization and performance degradation when deployed on actual channels \cite{vijayakumari2025survey}. Achieving adaptability across varying channel conditions not perfectly matched during training remains a persistent challenge \cite{lee2020autoencoder}. 

\subsubsection{Non-differentiable channels and DSP components}
\label{VI_E}
Some approaches require differentiable channel models for back-propagation, which can hinder development unless workarounds are employed \cite{xu2019performance, aoudia2019model}. Therefore, it is required to explore alternative approaches for end-to-end learning in real-world scenarios where the channel and transceiver are non-differentiable or the channel is unknown. This could involve reinforcement learning, stochastic perturbation techniques, gradient-free training methods, and genetic algorithms. In addition, differentiable DSP techniques, such as equalization, CPE, and LPN compensation are required. This includes integrating differentiable DSP algorithms into AE models for end-to-end optimization.

\subsubsection{Hardware and Implementation Issues}
\label{VI_F}
It is crucial to incorporate \textit{physical constraints}, such as channel capacity, power budget, and bandwidth limitations to ensure practical feasibility. However, excessive imposition of constraints can potentially hinder reconstruction performance due to noise in the received data. Therefore, carefully balancing the trade-off between reconstruction loss and physical constraints is vital to achieve optimal system optimization. Hardware effects, such as \textit{pulse shaping} and \textit{quantization}, are often not fully accounted for in AE/VAE models, requiring significant effort to implement them as neural network layers \cite{dorner2017deep}. Furthermore, \textit{Continuous transmissions} pose challenges related to \textit{timing synchronization} and \textit{sampling frequency offset}, which must be integrated into the training process. Synchronization in single-carrier AE systems is difficult \cite{dorner2017deep, cammerer2018end}.

\subsubsection{Scalability and High Dimensions} 
\label{VI_I}
AE architectures can face the curse of dimensionality, which leads to exponential growth in complexity and training difficulties when dealing with a large number of antennas (MIMO) or high-dimensional data representations. Scaling representations to accommodate large message dimensions can be impractical due to memory and processing demands \cite{o2017introduction, aoudia2019model}.

\subsubsection{Interpretability and Trust} 
A \textit{lack of transparency and trust} exists in deep learning AE. Understanding why AEs outperform traditional coding technologies remains an open challenge. The complexity of neural networks makes it difficult to provide clear explanations, hindering user trust \cite{tang2022meta, song2023autoencoders}.The \textit{interpretability} of the AE learning could limit its performance, because existing AE/VAE systems can be perceived as \textit{black-box} solutions, the complexity of interpreting the performance and behavior of AE/VAE systems is a significant challenges with potential pitfalls in understanding the learned schemes \cite{song2022benchmarking, goutay2022applications, uhlemann2020deep}.

\subsubsection{System-Level and Practical Considerations} 
End-to-end AE designs often \textit{lack modularity}, which requires separate designs for different source types or scenarios, and there is a lack of coding standards for learned methods. Also there is a need for \textit{rate adaptation}, as current AEs often have a fixed data rate \cite{song2022benchmarking}. In addition, deep Learning AE models must consistently outperform existing solutions in accuracy, efficiency, and robustness for industry integration. The integration of DL into \textit{industry standards} is not imminent due to these challenges \cite{letizia2025deep}.

\subsubsection{Federated Learning for Distributed AE Training} 
Researchers can investigate different federated learning algorithms and optimization techniques to enhance the efficiency and convergence speed of distributed AE training \cite{yuan2024decentralized}. Exploring the use of federated learning in various communication scenarios, such as decentralized communication systems, could provide valuable insights for practical applications. Additionally, addressing challenges such as data heterogeneity, device availability, and communication constraints in federated learning for autoencoders could be important research directions.

\subsubsection{Edge-AI Integration for Low-Latency Systems} 
Future research can focus on optimizing the AE models for deployment on edge devices with limited computational resources \cite{liu2022intrusion}. Developing lightweight and efficient autoencoder architectures that can run on edge-AI hardware while maintaining high performance could be a key area of interest. Additionally, exploring the integration of edge-AI with other emerging technologies, such as 5G or 6G networks, could further enhance the capabilities of low-latency communication systems.

\section{Conclusion}
\label{sec7}
The integration of AEs into communication systems represents a paradigm shift from traditional model-based approaches to data-driven methodologies. This survey highlighted the potential of AEs to overcome the inherent limitations of communication systems including mathematical models to  provide more accurate and adaptable solutions for next-generation communication systems. The review also highlighted the extensive application of AEs in various domains, including wireless, optical, semantic, and quantum communications, and provided a critical analysis discussion. Despite the progress, challenges persist, such as handling non-differentiable channels, computational complexity, convergence, data representation and data training including data scarcity. Moreover, this paper provides an examination of the computational complexity calculation associated with AE-based systems using FLOPS. This insight is crucial to guide the practical implementation of AEs in resource-constrained environments. Future research should focus on addressing these challenges, exploring advanced AE architectures, and developing hybrid models that combine the strengths of both data-driven and model-based approaches. In doing so, the field can move closer to realizing the full potential of AEs in creating efficient, reliable, and high-performance communication systems.

\bibliographystyle{IEEEtran}
\bibliography{references}

\begin{thebibliography}{100}
\providecommand{\url}[1]{#1}
\csname url@samestyle\endcsname
\providecommand{\newblock}{\relax}
\providecommand{\bibinfo}[2]{#2}
\providecommand{\BIBentrySTDinterwordspacing}{\spaceskip=0pt\relax}
\providecommand{\BIBentryALTinterwordstretchfactor}{4}
\providecommand{\BIBentryALTinterwordspacing}{\spaceskip=\fontdimen2\font plus
\BIBentryALTinterwordstretchfactor\fontdimen3\font minus \fontdimen4\font\relax}
\providecommand{\BIBforeignlanguage}[2]{{%
\expandafter\ifx\csname l@#1\endcsname\relax
\typeout{** WARNING: IEEEtran.bst: No hyphenation pattern has been}%
\typeout{** loaded for the language `#1'. Using the pattern for}%
\typeout{** the default language instead.}%
\else
\language=\csname l@#1\endcsname
\fi
#2}}
\providecommand{\BIBdecl}{\relax}
\BIBdecl

\bibitem{saleh2019fundamentals}
B.~E.~A. Saleh and M.~C. Teich, \emph{Fundamentals of Photonics}.\hskip 1em plus 0.5em minus 0.4em\relax John Wiley \& Sons, 2019.

\bibitem{senior2009optical}
J.~M. Senior and M.~Y. Jamro, \emph{Optical fiber communications: principles and practice}.\hskip 1em plus 0.5em minus 0.4em\relax Pearson Education, 2009.

\bibitem{keiser2021fiber}
G.~Keiser, \emph{Fiber Optic Communications}.\hskip 1em plus 0.5em minus 0.4em\relax Springer, 2021.

\bibitem{sun2014mimo}
S.~Sun, T.~S. Rappaport, R.~W. Heath, A.~Nix, and S.~Rangan, ``Mimo for millimeter-wave wireless communications: Beamforming, spatial multiplexing, or both?'' \emph{IEEE Communications Magazine}, vol.~52, no.~12, pp. 110--121, 2014.

\bibitem{kaminow2010optical}
I.~Kaminow, T.~Li, and A.~E. Willner, \emph{Optical fiber telecommunications VB: systems and networks}.\hskip 1em plus 0.5em minus 0.4em\relax Elsevier, 2010.

\bibitem{mukherjee2006optical}
B.~Mukherjee, \emph{Optical WDM networks}.\hskip 1em plus 0.5em minus 0.4em\relax Springer Science \& Business Media, 2006.

\bibitem{marzetta2016fundamentals}
T.~L. Marzetta, E.~G. Larsson, H.~Yang, and H.~Q. Ngo, \emph{Fundamentals of massive MIMO}.\hskip 1em plus 0.5em minus 0.4em\relax Cambridge University Press, 2016.

\bibitem{tse2005fundamentals}
D.~Tse and P.~Viswanath, \emph{Fundamentals of wireless communication}.\hskip 1em plus 0.5em minus 0.4em\relax Cambridge university press, 2005.

\bibitem{heath2018foundations}
R.~W. Heath~Jr and A.~Lozano, \emph{Foundations of MIMO communication}.\hskip 1em plus 0.5em minus 0.4em\relax Cambridge University Press, 2018.

\bibitem{marcuse2013theory}
D.~Marcuse, \emph{Theory of dielectric optical waveguides}.\hskip 1em plus 0.5em minus 0.4em\relax Elsevier, 2013.

\bibitem{snyder1983optical}
A.~Snyder, ``Optical waveguide theory,'' 1983.

\bibitem{david2012microwave}
M.~P. David, ``Microwave engineering, ; joihn wiley \& sons,'' \emph{Inc.: Hoboken, NJ, USA}, 2012.

\bibitem{o2017introduction}
T.~O’shea and J.~Hoydis, ``An introduction to deep learning for the physical layer,'' \emph{IEEE Transactions on Cognitive Communications and Networking}, vol.~3, no.~4, pp. 563--575, 2017.

\bibitem{cammerer2020trainable}
S.~Cammerer, F.~A. Aoudia, S.~D{\"o}rner, M.~Stark, J.~Hoydis, and S.~Ten~Brink, ``Trainable communication systems: Concepts and prototype,'' \emph{IEEE Transactions on Communications}, vol.~68, no.~9, pp. 5489--5503, 2020.

\bibitem{raj2020design}
V.~Raj and S.~Kalyani, ``Design of communication systems using deep learning: A variational inference perspective,'' \emph{IEEE Transactions on Cognitive Communications and Networking}, vol.~6, no.~4, pp. 1320--1334, 2020.

\bibitem{yu2017autoencoders}
S.~Yu, M.~Emigh, E.~Santana, and J.~C. Pr{\'\i}ncipe, ``Autoencoders trained with relevant information: Blending shannon and wiener's perspectives,'' in \emph{2017 IEEE International Conference on Acoustics, Speech and Signal Processing (ICASSP)}.\hskip 1em plus 0.5em minus 0.4em\relax IEEE, 2017, pp. 6115--6119.

\bibitem{song2022benchmarking}
J.~Song, C.~H{\"a}ger, J.~Schr{\"o}der, T.~J. O’Shea, E.~Agrell, and H.~Wymeersch, ``Benchmarking and interpreting end-to-end learning of mimo and multi-user communication,'' \emph{IEEE Transactions on Wireless Communications}, vol.~21, no.~9, pp. 7287--7298, 2022.

\bibitem{zhai2020design}
Z.~Zhai, ``Design of an autoencoder-based underwater optical communication transceiver in attenuation channel,'' in \emph{Proceedings of the 2020 International Conference on Computing, Networks and Internet of Things}, 2020, pp. 151--157.

\bibitem{li2025semantic}
M.~Li, K.~Shen, and S.~Cui, ``A semantic approach to successive interference cancellation for multiple access networks,'' \emph{IEEE Internet of Things Journal}, 2025.

\bibitem{lee2020autoencoder}
H.~Lee, C.~Eom, H.~Lee, and C.~Lee, ``Autoencoder based communication system using multi-dimensional constellations,'' in \emph{2020 35th International Technical Conference on Circuits/Systems, Computers and Communications (ITC-CSCC)}.\hskip 1em plus 0.5em minus 0.4em\relax IEEE, 2020, pp. 481--484.

\bibitem{huynh2023performance}
T.~N.~T. Huynh, T.~K. Hoang, and T.-N. Dao, ``Performance analysis of an autoencoder-based communication system with numerous transmitter-receiver pairs,'' in \emph{2023 12th International Conference on Control, Automation and Information Sciences (ICCAIS)}.\hskip 1em plus 0.5em minus 0.4em\relax IEEE, 2023, pp. 541--546.

\bibitem{ney2022hybrid}
J.~Ney, B.~Hammoud, and N.~Wehn, ``A hybrid approach combining ann-based and conventional demapping in communication for efficient fpga-implementation,'' in \emph{2022 IEEE International Parallel and Distributed Processing Symposium Workshops (IPDPSW)}.\hskip 1em plus 0.5em minus 0.4em\relax IEEE, 2022, pp. 92--95.

\bibitem{zhao2023end}
W.~Zhao and S.~Hu, ``End-to-end auto-encoder system for deep residual shrinkage network for awgn channels,'' \emph{Journal of Computer and Communications}, vol.~11, no.~5, pp. 161--176, 2023.

\bibitem{vijayakumari2025survey}
K.~Vijayakumari and K.~Anusudha, ``Survey of hybrid deep learning autoencoders for enhanced visible light communication systems,'' \emph{Procedia Computer Science}, vol. 252, pp. 100--107, 2025.

\bibitem{karanov2020end2}
B.~Karanov, P.~Bayvel, and L.~Schmalen, ``End-to-end learning in optical fiber communications: Concept and transceiver design,'' in \emph{2020 European Conference on Optical Communications (ECOC)}.\hskip 1em plus 0.5em minus 0.4em\relax IEEE, 2020, pp. 1--4.

\bibitem{dorner2017deep}
S.~D{\"o}rner, S.~Cammerer, J.~Hoydis, and S.~Ten~Brink, ``Deep learning based communication over the air,'' \emph{IEEE Journal of Selected Topics in Signal Processing}, vol.~12, no.~1, pp. 132--143, 2017.

\bibitem{cammerer2018end}
S.~Cammerer, S.~D{\"o}rner, J.~Hoydis, and S.~ten Brink, ``End-to-end learning for physical layer communications,'' in \emph{International Zurich Seminar on Information and Communication (IZS 2018). Proceedings}.\hskip 1em plus 0.5em minus 0.4em\relax ETH Zurich, 2018, pp. 51--52.

\bibitem{wang2017deep}
T.~Wang, C.-K. Wen, H.~Wang, F.~Gao, T.~Jiang, and S.~Jin, ``Deep learning for wireless physical layer: Opportunities and challenges,'' \emph{China Communications}, vol.~14, no.~11, pp. 92--111, 2017.

\bibitem{letizia2025deep}
N.~A. Letizia, ``Deep learning models for physical layer communications,'' PhD thesis, University in Klagenfurt, Faculty of Technical Sciences, Feb 2025.

\bibitem{wu2020deep}
D.~Wu, M.~Nekovee, and Y.~Wang, ``Deep learning-based autoencoder for m-user wireless interference channel physical layer design,'' \emph{IEEE Access}, vol.~8, pp. 174\,679--174\,691, 2020.

\bibitem{song2023autoencoders}
J.~Song, \emph{Autoencoders for Physical-Layer Communications: Approaches and Applications}.\hskip 1em plus 0.5em minus 0.4em\relax Chalmers Tekniska Hogskola (Sweden), 2023.

\bibitem{Erpek2019}
T.~Erpek, T.~J. O’Shea, Y.~E. Sagduyu, Y.~Shi, and T.~C. Clancy, ``Deep learning for wireless communications,'' in \emph{Development and Analysis of Deep Learning Architectures}, Nov 2019, pp. 223--266.

\bibitem{lau2022machine}
A.~P.~T. Lau and F.~N. Khan, \emph{Machine Learning for Future Fiber-Optic Communication Systems}.\hskip 1em plus 0.5em minus 0.4em\relax Academic Press, 2022.

\bibitem{alnaseri2024Noisy}
O.~Alnaseri and Y.~Himeur, ``End-to-end deep learning in phase noisy coherent optical link,'' in \emph{2024 IEEE Future Networks World Forum (FNWF)}, 2024, pp. 719--724.

\bibitem{goutay2021end}
M.~Goutay, F.~A. Aoudia, J.~Hoydis, and J.-M. Gorce, ``End-to-end learning of ofdm waveforms with papr and aclr constraints,'' in \emph{2021 IEEE Globecom Workshops (GC Wkshps)}.\hskip 1em plus 0.5em minus 0.4em\relax IEEE, 2021, pp. 1--6.

\bibitem{mohamed2022lstm}
A.~Mohamed, A.~S.~T. Eldien, M.~M. Fouda, and R.~S. Saad, ``Lstm-autoencoder deep learning technique for papr reduction in visible light communication,'' \emph{IEEE Access}, vol.~10, pp. 113\,028--113\,034, 2022.

\bibitem{aoudia2018end}
F.~A. Aoudia and J.~Hoydis, ``End-to-end learning of communications systems without a channel model,'' in \emph{2018 52nd Asilomar Conference on Signals, Systems, and Computers}.\hskip 1em plus 0.5em minus 0.4em\relax IEEE, 2018, pp. 298--303.

\bibitem{asif2020ofdm}
K.~M. Asif and A.~Trivedi, ``Ofdm ensemble autoencoder using cnn and spsa for end-to-end learning communication systems,'' in \emph{2020 IEEE 4th Conference on Information \& Communication Technology (CICT)}.\hskip 1em plus 0.5em minus 0.4em\relax IEEE, 2020, pp. 1--6.

\bibitem{sakketou2019invariance}
F.~Sakketou and N.~Ampazis, ``On the invariance of the selu activation function on algorithm and hyperparameter selection in neural network recommenders,'' in \emph{Artificial Intelligence Applications and Innovations: 15th IFIP WG 12.5 International Conference, AIAI 2019, Hersonissos, Crete, Greece, May 24--26, 2019, Proceedings 15}.\hskip 1em plus 0.5em minus 0.4em\relax Springer, 2019, pp. 673--685.

\bibitem{dulhare2020machine}
U.~N. Dulhare, K.~Ahmad, and K.~A.~B. Ahmad, \emph{Machine learning and big data: concepts, algorithms, tools and applications}.\hskip 1em plus 0.5em minus 0.4em\relax John Wiley \& sons, 2020.

\bibitem{roberts2022principles}
D.~A. Roberts, S.~Yaida, and B.~Hanin, \emph{The principles of deep learning theory}.\hskip 1em plus 0.5em minus 0.4em\relax Cambridge University Press Cambridge, MA, USA, 2022, vol.~46.

\bibitem{montavon2012neural}
G.~Montavon, G.~Orr, and K.-R. M{\"u}ller, \emph{Neural networks: tricks of the trade}.\hskip 1em plus 0.5em minus 0.4em\relax springer, 2012, vol. 7700.

\bibitem{Goodfellow2016}
I.~Goodfellow, Y.~Bengio, and A.~Courville, \emph{Deep Learning}.\hskip 1em plus 0.5em minus 0.4em\relax Cambridge, Massachusetts: The MIT Press, 2016.

\bibitem{Bishop2006}
C.~M. Bishop, \emph{Pattern Recognition and Machine Learning}.\hskip 1em plus 0.5em minus 0.4em\relax Springer, 2006.

\bibitem{Krizhevsky2012}
A.~Krizhevsky, I.~Sutskever, and G.~E. Hinton, ``Imagenet classification with deep convolutional neural networks,'' in \emph{Advances in Neural Information Processing Systems}, vol.~25, 2012.

\bibitem{LeCun2015}
Y.~LeCun, Y.~Bengio, and G.~Hinton, ``Deep learning,'' \emph{Nature}, vol. 521, no. 7553, pp. 436--444, May 2015.

\bibitem{morocho2020learning}
M.~E. Morocho-Cayamcela, J.~N. Njoku, J.~Park, and W.~Lim, ``Learning to communicate with autoencoders: Rethinking wireless systems with deep learning,'' in \emph{2020 International Conference on Artificial Intelligence in Information and Communication (ICAIIC)}.\hskip 1em plus 0.5em minus 0.4em\relax IEEE, 2020, pp. 308--311.

\bibitem{duque2024autoencoders}
V.~Duque, J.~Lewandowsky, M.~Adrat, T.~Hardenbicker, and P.~Jax, ``Autoencoders for signal enhancement in communication systems,'' in \emph{2024 International Conference on Military Communication and Information Systems (ICMCIS)}.\hskip 1em plus 0.5em minus 0.4em\relax IEEE, 2024, pp. 01--09.

\bibitem{aoudia2022waveform}
F.~A. Aoudia and J.~Hoydis, ``Waveform learning for next-generation wireless communication systems,'' \emph{IEEE Transactions on Communications}, vol.~70, no.~6, pp. 3804--3817, 2022.

\bibitem{alzubaidi2021review}
L.~Alzubaidi, J.~Zhang, A.~J. Humaidi, A.~Al-Dujaili, Y.~Duan, O.~Al-Shamma, J.~Santamar{\'\i}a, M.~A. Fadhel, M.~Al-Amidie, and L.~Farhan, ``Review of deep learning: concepts, cnn architectures, challenges, applications, future directions,'' \emph{Journal of big Data}, vol.~8, pp. 1--74, 2021.

\bibitem{berahmand2024autoencoders}
K.~Berahmand, F.~Daneshfar, E.~S. Salehi, Y.~Li, and Y.~Xu, ``Autoencoders and their applications in machine learning: a survey,'' \emph{Artificial Intelligence Review}, vol.~57, no.~2, p.~28, 2024.

\bibitem{b13}
F.~Buchali, F.~Steiner, G.~Böcherer, L.~Schmalen, P.~Schulte, and W.~Idler, ``Rate adaptation and reach increase by probabilistically shaped 64-qam: An experimental demonstration,'' \emph{Journal of Lightwave Technology}, vol.~34, no.~7, pp. 1599--1609, 2016.

\bibitem{stark2019joint}
M.~Stark, F.~A. Aoudia, and J.~Hoydis, ``Joint learning of geometric and probabilistic constellation shaping,'' in \emph{2019 IEEE Globecom Workshops (GC Wkshps)}.\hskip 1em plus 0.5em minus 0.4em\relax IEEE, 2019, pp. 1--6.

\bibitem{b15}
F.~A. Aoudia and J.~Hoydis, ``Joint learning of probabilistic and geometric shaping for coded modulation systems,'' in \emph{GLOBECOM 2020 - 2020 IEEE Global Communications Conference}, Taipei, Taiwan, 2020, pp. 1--6.

\bibitem{b16}
T.~Fehenberger, A.~Alvarado, G.~Böcherer, and N.~Hanik, ``On probabilistic shaping of quadrature amplitude modulation for the nonlinear fiber channel,'' \emph{Journal of Lightwave Technology}, vol.~34, no.~21, pp. 5063--5073, Nov 2016.

\bibitem{b17}
M.~P. Yankov \emph{et~al.}, ``Constellation shaping for wdm systems using 256qam/1024qam with probabilistic optimization,'' \emph{Journal of Lightwave Technology}, vol.~34, no.~22, pp. 5146--5156, Nov 2016.

\bibitem{aref2022end}
V.~Aref and M.~Chagnon, ``End-to-end learning of joint geometric and probabilistic constellation shaping,'' in \emph{2022 Optical Fiber Communications Conference and Exhibition (OFC)}.\hskip 1em plus 0.5em minus 0.4em\relax IEEE, 2022, pp. 1--3.

\bibitem{neskorniuk2022model}
V.~Neskorniuk, A.~Carnio, D.~Marsella, S.~K. Turitsyn, J.~E. Prilepsky, and V.~Aref, ``Model-based deep learning of joint probabilistic and geometric shaping for optical communication,'' in \emph{CLEO: Science and Innovations}.\hskip 1em plus 0.5em minus 0.4em\relax Optica Publishing Group, 2022, pp. SW4E--5.

\bibitem{neskorniuk2022memory}
V.~Neskorniuk \emph{et~al.}, ``Memory-aware end-to-end learning of channel distortions in optical coherent communications,'' \emph{Optics Express}, vol.~31, no.~1, pp. 1--20, 2022.

\bibitem{hunger2005floating}
R.~Hunger, \emph{Floating Point Operations in Matrix-vector Calculus}.\hskip 1em plus 0.5em minus 0.4em\relax Munich University of Technology, Inst. for Circuit Theory and Signal Processing, 2005.

\bibitem{zou2021channel}
C.~Zou, F.~Yang, J.~Song, and Z.~Han, ``Channel autoencoder for wireless communication: State of the art, challenges, and trends,'' \emph{IEEE Communications Magazine}, vol.~59, no.~5, pp. 136--142, 2021.

\bibitem{che2022trainable}
B.~Che, X.~Li, Z.~Chen, and Q.~He, ``Trainable communication systems based on the binary neural network,'' \emph{Frontiers in Communications and Networks}, vol.~3, p. 878170, 2022.

\bibitem{xu2019performance}
J.~Xu, W.~Chen, B.~Ai, R.~He, Y.~Li, J.~Wang, T.~Juhana, and A.~Kurniawan, ``Performance evaluation of autoencoder for coding and modulation in wireless communications,'' in \emph{2019 11th International Conference on Wireless Communications and Signal Processing (WCSP)}.\hskip 1em plus 0.5em minus 0.4em\relax IEEE, 2019, pp. 1--6.

\bibitem{aoudia2019model}
F.~A. Aoudia and J.~Hoydis, ``Model-free training of end-to-end communication systems,'' \emph{IEEE Journal on Selected Areas in Communications}, vol.~37, no.~11, pp. 2503--2516, 2019.

\bibitem{davaslioglu2022autoencoder}
K.~Davaslioglu, T.~Erpek, and Y.~E. Sagduyu, ``Autoencoder communications with optimized interference suppression for nextg ran,'' in \emph{2022 IEEE Future Networks World Forum (FNWF)}.\hskip 1em plus 0.5em minus 0.4em\relax IEEE, 2022, pp. 164--169.

\bibitem{choubey2022autoencoder}
N.~Choubey, A.~Trivedi, and V.~S. Kushwah, ``Autoencoder for end-to-end learning communication system based on noma,'' in \emph{2022 IEEE 6th Conference on Information and Communication Technology (CICT)}.\hskip 1em plus 0.5em minus 0.4em\relax IEEE, 2022, pp. 1--5.

\bibitem{jiang2019turbo}
Y.~Jiang, H.~Kim, H.~Asnani, S.~Kannan, S.~Oh, and P.~Viswanath, ``Turbo autoencoder: Deep learning based channel codes for point-to-point communication channels,'' \emph{Advances in neural information processing systems}, vol.~32, 2019.

\bibitem{jiang2020joint}
H.~A. S. K. S.~O. Yihan~Jiang, Hyeji~Kim and P.~Viswanath, ``Joint channel coding and modulation via deep learning,'' in \emph{2020 IEEE 21st International Workshop on Signal Processing Advances in Wireless Communications (SPAWC)}.\hskip 1em plus 0.5em minus 0.4em\relax IEEE, 2020, pp. 1--5.

\bibitem{balevi2020autoencoder}
E.~Balevi and J.~G. Andrews, ``Autoencoder-based error correction coding for one-bit quantization,'' \emph{IEEE Transactions on Communications}, vol.~68, no.~6, pp. 3440--3451, 2020.

\bibitem{van2020deep}
T.~Van~Luong, Y.~Ko, N.~A. Vien, M.~Matthaiou, and H.~Q. Ngo, ``Deep energy autoencoder for noncoherent multicarrier mu-simo systems,'' \emph{IEEE Transactions on Wireless Communications}, vol.~19, no.~6, pp. 3952--3962, 2020.

\bibitem{caciularu2020unsupervised}
A.~Caciularu and D.~Burshtein, ``Unsupervised linear and nonlinear channel equalization and decoding using variational autoencoders,'' \emph{IEEE Transactions on Cognitive Communications and Networking}, vol.~6, no.~3, pp. 1003--1018, 2020.

\bibitem{zhang2021design}
Y.~Zhang, H.~Wu, and M.~Coates, ``On the design of channel coding autoencoders with arbitrary rates for isi channels,'' \emph{IEEE Wireless Communications Letters}, vol.~11, no.~2, pp. 426--430, 2021.

\bibitem{saidutta2021joint}
Y.~M. Saidutta, A.~Abdi, and F.~Fekri, ``Joint source-channel coding over additive noise analog channels using mixture of variational autoencoders,'' \emph{IEEE Journal on Selected Areas in Communications}, vol.~39, no.~7, pp. 2000--2013, 2021.

\bibitem{letizia2021capacity}
N.~A. Letizia and A.~M. Tonello, ``Capacity-driven autoencoders for communications,'' \emph{IEEE Open Journal of the Communications Society}, vol.~2, pp. 1366--1378, 2021.

\bibitem{khan2019robust}
F.~N. Khan and A.~P.~T. Lau, ``Robust and efficient data transmission over noisy communication channels using stacked and denoising autoencoders,'' \emph{China Communications}, vol.~16, no.~8, pp. 72--82, 2019.

\bibitem{ke2021real}
Z.~Ke and H.~Vikalo, ``Real-time radio technology and modulation classification via an lstm auto-encoder,'' \emph{IEEE Transactions on Wireless Communications}, vol.~21, no.~1, pp. 370--382, 2021.

\bibitem{ali2017automatic}
A.~Ali and F.~Yangyu, ``Automatic modulation classification using deep learning based on sparse autoencoders with nonnegativity constraints,'' \emph{IEEE signal processing letters}, vol.~24, no.~11, pp. 1626--1630, 2017.

\bibitem{zhu2017modulation}
X.~Zhu and T.~Fujii, ``A modulation classification method in cognitive radios system using stacked denoising sparse autoencoder,'' in \emph{2017 IEEE Radio and Wireless Symposium (RWS)}.\hskip 1em plus 0.5em minus 0.4em\relax IEEE, 2017, pp. 218--220.

\bibitem{rajapaksha2020low}
N.~R. Nuwanthika~Rajapaksha and M.~Latva-aho, ``Low complexity autoencoder based end-to-end learning of coded communications systems,'' in \emph{2020 IEEE 91st Vehicular Technology Conference (VTC2020-Spring)}.\hskip 1em plus 0.5em minus 0.4em\relax IEEE, 2020, pp. 1--7.

\bibitem{rajapaksha1911low}
N.~Rajapaksha, N.~Rajatheva, and M.~Latva-aho, ``Low complexity autoencoder based end-to-end learning of coded communications systems,'' in \emph{2020 IEEE 91st Vehicular Technology Conference (VTC2020-Spring)}.\hskip 1em plus 0.5em minus 0.4em\relax IEEE, 2020, pp. 1--7.

\bibitem{dorner2021bit}
S.~D{\"o}rner, S.~Rottacker, M.~Gauger, and S.~ten Brink, ``Bit-wise autoencoder for multiple antenna systems,'' in \emph{2021 17th International Symposium on Wireless Communication Systems (ISWCS)}.\hskip 1em plus 0.5em minus 0.4em\relax IEEE, 2021, pp. 1--5.

\bibitem{bui2023deep}
T.~T.~T. Bui, X.~N. Tran, and A.~H. Phan, ``Deep learning based mimo systems using open-loop autoencoder,'' \emph{AEU-International Journal of Electronics and Communications}, vol. 168, p. 154712, 2023.

\bibitem{zhao2021variational}
T.~Zhao and F.~Li, ``Variational-autoencoder signal detection for mimo-ofdm-im,'' \emph{Digital Signal Processing}, vol. 118, p. 103230, 2021.

\bibitem{tao2020autoencoder}
J.~Tao, J.~Chen, J.~Xing, S.~Fu, and J.~Xie, ``Autoencoder neural network based intelligent hybrid beamforming design for mmwave massive mimo systems,'' \emph{IEEE Transactions on Cognitive Communications and Networking}, vol.~6, no.~3, pp. 1019--1030, 2020.

\bibitem{cherif2023autoencoder}
M.~Cherif, A.~Arfaoui, and R.~Bouallegue, ``Autoencoder-based deep learning for massive multiple-input multiple-output uplink under high-power amplifier non-linearities,'' \emph{IET Communications}, vol.~17, no.~2, pp. 162--170, 2023.

\bibitem{sahay2023defending}
R.~Sahay, M.~Zhang, D.~J. Love, and C.~G. Brinton, ``Defending adversarial attacks on deep learning-based power allocation in massive mimo using denoising autoencoders,'' \emph{IEEE Transactions on Cognitive Communications and Networking}, vol.~9, no.~4, pp. 913--926, 2023.

\bibitem{le2022ris}
H.~A. Le, T.~Van~Chien, W.~Choi \emph{et~al.}, ``Ris-assisted mimo communication systems: Model-based versus autoencoder approaches,'' in \emph{2022 IEEE 33rd Annual International Symposium on Personal, Indoor and Mobile Radio Communications (PIMRC)}.\hskip 1em plus 0.5em minus 0.4em\relax IEEE, 2022, pp. 1--6.

\bibitem{jang2019deep}
Y.~Jang, G.~Kong, M.~Jung, S.~Choi, and I.-M. Kim, ``Deep autoencoder based csi feedback with feedback errors and feedback delay in fdd massive mimo systems,'' \emph{IEEE Wireless Communications Letters}, vol.~8, no.~3, pp. 833--836, 2019.

\bibitem{hussien2023prvnet}
M.~Hussien, ``Prvnet: Variational autoencoders for massive mimo csi feedback,'' \emph{Authorea Preprints}, 2023.

\bibitem{ravula2021deep}
S.~Ravula and S.~Jain, ``Deep autoencoder-based massive mimo csi feedback with quantization and entropy coding,'' in \emph{2021 IEEE Global Communications Conference (GLOBECOM)}.\hskip 1em plus 0.5em minus 0.4em\relax IEEE, 2021, pp. 1--6.

\bibitem{zhang2023deep}
X.~Zhang and M.~Vaezi, ``Deep autoencoder-based z-interference channels with perfect and imperfect csi,'' \emph{IEEE Transactions on Communications}, 2023.

\bibitem{shin2024autoencoder}
J.~Shin and X.~Jin, ``Autoencoder--based mimo cooperative communications with quantize--forward relaying,'' \emph{IEEE Transactions on Vehicular Technology}, 2024.

\bibitem{Hangyang2023massive}
H.~Shan, X.~Chen, H.~Yin, L.~Chen, and G.~Wei, ``Joint sparse autoencoder based massive mimo csi feedback,'' \emph{IEEE Communications Letters}, vol.~27, no.~4, pp. 1150--1154, 2023.

\bibitem{aoudia2021end}
F.~A. Aoudia and J.~Hoydis, ``End-to-end learning for ofdm: From neural receivers to pilotless communication,'' \emph{IEEE Transactions on Wireless Communications}, vol.~21, no.~2, pp. 1049--1063, 2021.

\bibitem{felix2018ofdm}
A.~Felix, S.~Cammerer, S.~D{\"o}rner, J.~Hoydis, and S.~Ten~Brink, ``Ofdm-autoencoder for end-to-end learning of communications systems,'' in \emph{2018 IEEE 19th International Workshop on Signal Processing Advances in Wireless Communications (SPAWC)}.\hskip 1em plus 0.5em minus 0.4em\relax IEEE, 2018, pp. 1--5.

\bibitem{marasinghe2024constellation}
D.~Marasinghe, L.~H. Nguyen, J.~Mohammadi, Y.~Chen, T.~Wild, and N.~Rajatheva, ``Constellation shaping under phase noise impairment for sub-thz communications,'' in \emph{ICC 2024-IEEE International Conference on Communications}.\hskip 1em plus 0.5em minus 0.4em\relax IEEE, 2024, pp. 3833--3838.

\bibitem{xu2022turbo}
C.~Xu, T.~Van~Luong, L.~Xiang, S.~Sugiura, R.~G. Maunder, L.-L. Yang, and L.~Hanzo, ``Turbo detection aided autoencoder for multicarrier wireless systems: Integrating deep learning into channel coded systems,'' \emph{IEEE Transactions on Cognitive Communications and Networking}, vol.~8, no.~2, pp. 600--614, 2022.

\bibitem{alnaseri2025papr}
O.~Alnaseri, I.~R. Al-Saedi, Y.~Himeur, and H.~Li, ``Deep learning autoencoders for reducing papr in coherent optical systems,'' in \emph{2025 IEEE International Conference on Communications (ICC)}, 2025.

\bibitem{Geiger2025Joint}
B.~Geiger, F.~Liu, S.~Lu, A.~Rode, and L.~Schmalen, ``Joint optimization of geometric and probabilistic constellation shaping for ofdm-isac systems,'' in \emph{2025 IEEE 5th International Symposium on Joint Communications and Sensing (JCS)}, 2025, pp. 1--6.

\bibitem{chen2019generalized}
X.~Chen, J.~Cheng, Z.~Zhang, L.~Wu, and J.~Dang, ``A generalized data representation and training-performance analysis for deep learning based communication systems,'' in \emph{2019 IEEE 90th Vehicular Technology Conference (VTC2019-Fall)}.\hskip 1em plus 0.5em minus 0.4em\relax IEEE, 2019, pp. 1--5.

\bibitem{muth2023autoencoder}
C.~Muth and L.~Schmalen, ``Autoencoder-based joint communication and sensing of multiple targets,'' in \emph{WSA \& SCC 2023; 26th International ITG Workshop on Smart Antennas and 13th Conference on Systems, Communications, and Coding}.\hskip 1em plus 0.5em minus 0.4em\relax VDE, 2023, pp. 1--6.

\bibitem{nirmal2021deep}
I.~Nirmal, A.~Khamis, M.~Hassan, W.~Hu, and X.~Zhu, ``Deep learning for radio-based human sensing: Recent advances and future directions,'' \emph{IEEE Communications Surveys \& Tutorials}, vol.~23, no.~2, pp. 995--1019, 2021.

\bibitem{kompella2024augmenting}
S.~K. Kompella, K.~Davaslioglu, Y.~E. Sagduyu, and S.~Kompella, ``Augmenting training data with vector-quantized variational autoencoder for classifying rf signals,'' in \emph{MILCOM 2024-2024 IEEE Military Communications Conference (MILCOM)}.\hskip 1em plus 0.5em minus 0.4em\relax IEEE, 2024, pp. 1--6.

\bibitem{abdalzaher2021deep}
M.~S. Abdalzaher, M.~Elwekeil, T.~Wang, and S.~Zhang, ``A deep autoencoder trust model for mitigating jamming attack in iot assisted by cognitive radio,'' \emph{IEEE Systems Journal}, vol.~16, no.~3, pp. 3635--3645, 2021.

\bibitem{liu2020data}
M.~Liu, G.~Liao, N.~Zhao, H.~Song, and F.~Gong, ``Data-driven deep learning for signal classification in industrial cognitive radio networks,'' \emph{IEEE Transactions on Industrial Informatics}, vol.~17, no.~5, pp. 3412--3421, 2020.

\bibitem{teganya2021deep}
Y.~Teganya and D.~Romero, ``Deep completion autoencoders for radio map estimation,'' \emph{IEEE Transactions on Wireless Communications}, vol.~21, no.~3, pp. 1710--1724, 2021.

\bibitem{almazrouei2019using}
E.~Almazrouei, G.~Gianini, C.~Mio, N.~Almoosa, and E.~Damiani, ``Using autoencoders for radio signal denoising,'' in \emph{Proceedings of the 15th ACM International Symposium on QoS and Security for Wireless and Mobile Networks}, 2019, pp. 11--17.

\bibitem{karanov2018end}
B.~Karanov, M.~Chagnon, F.~Thouin, T.~A. Eriksson, H.~B{\"u}low, D.~Lavery, P.~Bayvel, and L.~Schmalen, ``End-to-end deep learning of optical fiber communications,'' \emph{Journal of Lightwave Technology}, vol.~36, no.~20, pp. 4843--4855, 2018.

\bibitem{karanov2020optical}
B.~Karanov, M.~Chagnon, V.~Aref, D.~Lavery, P.~Bayvel, and L.~Schmalen, ``Optical fiber communication systems based on end-to-end deep learning,'' in \emph{2020 IEEE Photonics Conference (IPC)}.\hskip 1em plus 0.5em minus 0.4em\relax IEEE, 2020, pp. 1--2.

\bibitem{karanov2025towards}
B.~Karanov, ``Towards robust end-to-end neural network-based transceivers for short reach fiber links,'' \emph{Optical Fiber Technology}, vol.~90, p. 104069, 2025.

\bibitem{lauinger2023improving}
V.~Lauinger, F.~Buchali, and L.~Schmalen, ``Improving the bootstrap of blind equalizers with variational autoencoders,'' in \emph{Optical Fiber Communication Conference}.\hskip 1em plus 0.5em minus 0.4em\relax Optica Publishing Group, 2023, pp. M2F--4.

\bibitem{uhlemann2020deep}
T.~Uhlemann, S.~Cammerer, A.~Span, S.~D{\"o}rner, and S.~ten Brink, ``Deep-learning autoencoder for coherent and nonlinear optical communication,'' in \emph{Photonic Networks; 21th ITG-Symposium}.\hskip 1em plus 0.5em minus 0.4em\relax VDE, 2020, pp. 1--8.

\bibitem{karanov2020end}
B.~Karanov, V.~Oliari, M.~Chagnon, G.~Liga, A.~Alvarado, V.~Aref, D.~Lavery, P.~Bayvel, and L.~Schmalen, ``End-to-end learning in optical fiber communications: Experimental demonstration and future trends,'' in \emph{2020 European Conference on Optical Communications (ECOC)}.\hskip 1em plus 0.5em minus 0.4em\relax IEEE, 2020, pp. 1--4.

\bibitem{Rode2023}
A.~Rode, B.~Geiger, S.~Chimmalgi, and L.~Schmalen, ``End-to-end optimization of constellation shaping for wiener phase noise channels with a differentiable blind phase search,'' \emph{Journal of Lightwave Technology}, vol.~41, no.~12, pp. 3849--3859, Jun 2023.

\bibitem{lauinger2022blind}
F.~B. Vincent~Lauinger and L.~Schmalen, ``Blind equalization and channel estimation in coherent optical communications using variational autoencoders,'' \emph{IEEE Journal on Selected Areas in Communications}, vol.~40, no.~9, pp. 2529--2539, 2022.

\bibitem{song2023blind}
J.~Song, V.~Lauinger, C.~H{\"a}ger, J.~Schr{\"o}der, A.~G. i~Amat, L.~Schmalen, and H.~Wymeersch, ``Blind frequency-domain equalization using vector-quantized variational autoencoders,'' in \emph{49th European Conference on Optical Communications (ECOC 2023)}, vol. 2023.\hskip 1em plus 0.5em minus 0.4em\relax IET, 2023, pp. 1222--1225.

\bibitem{jones2018deep}
R.~T. Jones, T.~A. Eriksson, M.~P. Yankov, and D.~Zibar, ``Deep learning of geometric constellation shaping including fiber nonlinearities,'' in \emph{2018 European Conference on Optical Communication (ECOC)}.\hskip 1em plus 0.5em minus 0.4em\relax IEEE, 2018, pp. 1--3.

\bibitem{jones2019end}
R.~T. Jones, M.~P. Yankov, and D.~Zibar, ``End-to-end learning for gmi optimized geometric constellation shape,'' in \emph{45th European Conference on Optical Communication (ECOC 2019)}.\hskip 1em plus 0.5em minus 0.4em\relax IET, 2019, pp. 1--4.

\bibitem{rode2022geometric}
A.~Rode, B.~Geiger, and L.~Schmalen, ``Geometric constellation shaping for phase-noise channels using a differentiable blind phase search,'' in \emph{2022 Optical Fiber Communications Conference and Exhibition (OFC)}.\hskip 1em plus 0.5em minus 0.4em\relax IEEE, 2022, pp. 1--3.

\bibitem{rode2023optimized}
A.~Rode, W.~A. Gebrehiwot, S.~Chimmalgi, and L.~Schmalen, ``Optimized geometric constellation shaping for wiener phase noise channels with viterbi-viterbi carrier phase estimation,'' in \emph{49th European Conference on Optical Communications (ECOC 2023)}, vol. 2023.\hskip 1em plus 0.5em minus 0.4em\relax IET, 2023, pp. 1457--1460.

\bibitem{rode2025machine}
A.~Rode, M.~Farsi, V.~Lauinger, M.~Karlsson, E.~Agrell, L.~Schmalen, and C.~H{\"a}ger, ``Machine learning opportunities for integrated polarization sensing and communication in optical fibers,'' \emph{Optical Fiber Technology}, vol.~90, p. 104047, 2025.

\bibitem{chimmalgi2025end}
S.~Chimmalgi, L.~Schmalen, and V.~Aref, ``End-to-end learning of probabilistic constellation shaping through importance sampling,'' \emph{IEEE Photonics Technology Letters}, 2025.

\bibitem{kim2022autoencoding}
J.~Kim, H.~Lee, and S.-H. Park, ``Autoencoding graph neural networks for scalable transceiver design,'' in \emph{2022 IEEE 96th Vehicular Technology Conference (VTC2022-Fall)}.\hskip 1em plus 0.5em minus 0.4em\relax IEEE, 2022, pp. 1--2.

\bibitem{liu2020autoencoder}
X.~Liu, Z.~Wei, A.~Pepe, Z.~Wang, and H.~Fu, ``Autoencoder for optical wireless communication system in atmospheric turbulence,'' in \emph{2020 Opto-Electronics and Communications Conference (OECC)}.\hskip 1em plus 0.5em minus 0.4em\relax IEEE, 2020, pp. 1--3.

\bibitem{soltani2018autoencoder}
M.~Soltani, W.~Fatnassi, A.~Aboutaleb, Z.~Rezki, A.~Bhuyan, and P.~Titus, ``Autoencoder-based optical wireless communications systems,'' in \emph{2018 IEEE Globecom Workshops (GC Wkshps)}.\hskip 1em plus 0.5em minus 0.4em\relax IEEE, 2018, pp. 1--6.

\bibitem{si2020model}
L.-H. Si-Ma, Z.-R. Zhu, and H.-Y. Yu, ``Model-aware end-to-end learning for siso optical wireless communication over poisson channel,'' \emph{IEEE Photonics Journal}, vol.~12, no.~6, pp. 1--15, 2020.

\bibitem{zhang2023autoencoder}
Q.~Zhang, G.~Chen, B.~Liu, X.~Zhi, S.~Zhan, J.~Zhang, N.~Jiang, B.~Cao, and Z.~Li, ``An autoencoder-based transceiver for uav-to-ground free space optical communication,'' in \emph{2023 Opto-Electronics and Communications Conference (OECC)}.\hskip 1em plus 0.5em minus 0.4em\relax IEEE, 2023, pp. 1--3.

\bibitem{cao2023end}
M.~Cao, R.~Wang, Y.~Zhang, H.~Deng, L.~Zhou, and H.~Wang, ``An end-to-end autoencoder for fso system under unknown csi scenarios,'' in \emph{2023 21st International Conference on Optical Communications and Networks (ICOCN)}.\hskip 1em plus 0.5em minus 0.4em\relax IEEE, 2023, pp. 1--3.

\bibitem{Nguyen2025VLC}
T.~K. Nguyen, T.~V. Pham, H.~D. Le, C.~T. Nguyen, and A.~T. Pham, ``Multi-user visible light communications with probabilistic constellation shaping and precoding,'' \emph{IEEE Transactions on Communications}, pp. 1--1, 2025.

\bibitem{wheeler2024autoencoder}
D.~Wheeler and B.~Natarajan, ``Autoencoder-based domain learning for semantic communication with conceptual spaces,'' in \emph{2024 Wireless Telecommunications Symposium (WTS)}.\hskip 1em plus 0.5em minus 0.4em\relax IEEE, 2024, pp. 1--6.

\bibitem{Oh2024data}
J.~Oh, Y.~Choi, C.~Oh, and W.~Na, ``Autoencoder-based data compression model experiment for semantic communication,'' in \emph{2024 International Conference on Information Networking (ICOIN)}, 2024, pp. 553--556.

\bibitem{samarathunga2024autoencoder}
P.~Samarathunga, T.~Fernando, V.~Gowrisetty, T.~Atulugama, and A.~Fernando, ``Autoencoder based image quality metric for modelling semantic noise in semantic communications,'' \emph{Electronics Letters}, vol.~60, no.~4, p. e13115, 2024.

\bibitem{luo2022autoencoder}
X.~Luo, B.~Yin, Z.~Chen, B.~Xia, and J.~Wang, ``Autoencoder-based semantic communication systems with relay channels,'' in \emph{2022 IEEE International Conference on Communications Workshops (ICC Workshops)}.\hskip 1em plus 0.5em minus 0.4em\relax IEEE, 2022, pp. 711--716.

\bibitem{tabi2022hybrid}
Z.~Tabi, B.~Bak{\'o}, D.~T. Nagy, P.~Vaderna, Z.~Kallus, P.~H{\'a}ga, and Z.~Zimbor{\'a}s, ``Hybrid quantum-classical autoencoders for end-to-end radio communication,'' in \emph{2022 IEEE/ACM 7th Symposium on Edge Computing (SEC)}.\hskip 1em plus 0.5em minus 0.4em\relax IEEE, 2022, pp. 468--473.

\bibitem{tabi2025quantum}
Z.~I. Tabi, B.~Bak{\'o}, D.~T. Nagy, P.~Vaderna, Z.~Kallus, P.~H{\'a}ga, and Z.~Zimbor{\'a}s, ``Quantum-classical autoencoder architectures for end-to-end radio communication,'' \emph{IEEE Access}, 2025.

\bibitem{Mahesh2025Quantum}
M.~Bhupati, A.~Mall, A.~Kumar, and P.~K. Jha, ``Deep learning-based variational autoencoder for classification of quantum and classical states of light,'' \emph{Advanced Physics Research}, vol.~4, no.~2, p. 2400089, 2025.

\bibitem{rathi2024quantum}
L.~Rathi, S.~DiAdamo, and A.~Shabani, ``Quantum autoencoders for learning quantum channel codes,'' in \emph{2024 16th International Conference on COMmunication Systems \& NETworkS (COMSNETS)}.\hskip 1em plus 0.5em minus 0.4em\relax IEEE, 2024, pp. 988--993.

\bibitem{zhang2022svd}
X.~Zhang, M.~Vaezi, and T.~J. O’Shea, ``Svd-embedded deep autoencoder for mimo communications,'' in \emph{ICC 2022-IEEE International Conference on Communications}.\hskip 1em plus 0.5em minus 0.4em\relax IEEE, 2022, pp. 5190--5195.

\bibitem{alzubaidi2023survey}
L.~Alzubaidi, J.~Bai, A.~Al-Sabaawi, J.~Santamar{\'\i}a, A.~S. Albahri, B.~S.~N. Al-Dabbagh, M.~A. Fadhel, M.~Manoufali, J.~Zhang, A.~H. Al-Timemy \emph{et~al.}, ``A survey on deep learning tools dealing with data scarcity: definitions, challenges, solutions, tips, and applications,'' \emph{Journal of Big Data}, vol.~10, no.~1, p.~46, 2023.

\bibitem{skatchkovsky2020end}
N.~Skatchkovsky, H.~Jang, and O.~Simeone, ``End-to-end learning of neuromorphic wireless systems for low-power edge artificial intelligence,'' in \emph{2020 54th Asilomar Conference on Signals, Systems, and Computers}.\hskip 1em plus 0.5em minus 0.4em\relax IEEE, 2020, pp. 166--173.

\bibitem{chen2023photonic}
Y.~Chen, T.~Zhou, J.~Wu, H.~Qiao, X.~Lin, L.~Fang, and Q.~Dai, ``Photonic unsupervised learning variational autoencoder for high-throughput and low-latency image transmission,'' \emph{Science Advances}, vol.~9, no.~7, p. eadf8437, 2023.

\bibitem{mehonic2024roadmap}
A.~Mehonic, D.~Ielmini, K.~Roy, O.~Mutlu, S.~Kvatinsky, T.~Serrano-Gotarredona, B.~Linares-Barranco, S.~Spiga, S.~Savel’ev, A.~G. Balanov \emph{et~al.}, ``Roadmap to neuromorphic computing with emerging technologies,'' \emph{APL Materials}, vol.~12, no.~10, 2024.

\bibitem{ActiveCalculas}
M.~Boelkins, ``Ac active calculus,'' available: https://activecalculus.org/single/book-1.html.

\bibitem{Aoudia2018}
F.~A. Aoudia and J.~Hoydis, ``End-to-end learning of communications systems without a channel model,'' in \emph{2018 52nd Asilomar Conference on Signals, Systems, and Computers}, Pacific Grove, CA, USA, 2018, pp. 298--303.

\bibitem{Goutay2019}
M.~Goutay, F.~A. Aoudia, and J.~Hoydis, ``Deep reinforcement learning autoencoder with noisy feedback,'' in \emph{2019 International Symposium on Modeling and Optimization in Mobile, Ad Hoc, and Wireless Networks (WiOPT)}, Avignon, France, 2019.

\bibitem{Wang2020}
D.~W. et~al., ``Data-driven optical fiber channel modeling: A deep learning approach,'' \emph{Journal of Lightwave Technology}, vol.~38, no.~17, pp. 4730--4743, Sept 2020.

\bibitem{Shea2019}
T.~J. O’Shea, T.~Roy, and N.~West, ``Approximating the void: Learning stochastic channel models from observation with variational generative adversarial networks,'' in \emph{2019 International Conference on Computing, Networking and Communications (ICNC)}, Honolulu, HI, USA, 2019, pp. 681--686.

\bibitem{jafarkhani2024modulation}
H.~Jafarkhani, H.~Maleki, and M.~Vaezi, ``Modulation and coding for noma and rsma,'' \emph{Proceedings of the IEEE}, 2024.

\bibitem{Yuzhe2023}
Y.~Li, H.~Chang, R.~Gao, Q.~Zhang, F.~Tian, H.~Yao, Q.~Tian, Y.~Wang, X.~Xin, F.~Wang, and L.~Rao, ``End-to-end deep learning of joint geometric probabilistic shaping using a channel-sensitive autoencoder,'' \emph{Electronics}, vol.~12, no.~20, p. 4234, Jan 2023.

\bibitem{Raj2018}
V.~Raj and S.~Kalyani, ``Backpropagating through the air: Deep learning at physical layer without channel models,'' \emph{IEEE Communications Letters}, vol.~22, no.~11, pp. 2278--2281, Nov 2018.

\bibitem{Jovanovic2021}
O.~Jovanovic, M.~P. Yankov, F.~D. Ros, and D.~Zibar, ``Gradient-free training of autoencoders for non-differentiable communication channels,'' \emph{Journal of Lightwave Technology}, vol.~39, no.~20, pp. 6381--6391, Oct 2021.

\bibitem{Yao1999}
X.~Yao, ``Evolving artificial neural networks,'' \emph{Proceedings of the IEEE}, vol.~87, no.~9, pp. 1423--1447, Sep 1999.

\bibitem{Yan2013}
K.~Yan, F.~Yang, C.~Pan, J.~Song, F.~Ren, and J.~Li, ``Genetic algorithm aided gray-apsk constellation optimization,'' in \emph{2013 9th International Wireless Communications and Mobile Computing Conference (IWCMC)}, Sardinia, Italy, 2013, pp. 1705--1709.

\bibitem{Brito2016}
N.~S.~D. Brito and S.~M.~A. Cruz, ``Genetic algorithms in wireless communications: A survey,'' \emph{IEEE Communications Surveys \& Tutorials}, vol.~18, no.~1, pp. 96--125, 2016, first Quarter.

\bibitem{Vural2017}
S.~A.~B. Vural and E.~A. Aydin, ``Evolutionary optimization in communication networks: A survey,'' \emph{IEEE Transactions on Evolutionary Computation}, vol.~21, no.~2, pp. 166--184, Apr 2017.

\bibitem{ye2018channel}
H.~Ye, G.~Y. Li, B.-H.~F. Juang, and K.~Sivanesan, ``Channel agnostic end-to-end learning based communication systems with conditional gan,'' in \emph{2018 IEEE Globecom Workshops (GC Wkshps)}.\hskip 1em plus 0.5em minus 0.4em\relax IEEE, 2018, pp. 1--5.

\bibitem{kingma2013auto}
D.~P. Kingma, M.~Welling \emph{et~al.}, ``Auto-encoding variational bayes,'' 2013.

\bibitem{ho2020denoising}
J.~Ho, A.~Jain, and P.~Abbeel, ``Denoising diffusion probabilistic models,'' \emph{Advances in neural information processing systems}, vol.~33, pp. 6840--6851, 2020.

\bibitem{rifai2011contractive}
S.~Rifai, P.~Vincent, X.~Muller, X.~Glorot, and Y.~Bengio, ``Contractive auto-encoders: Explicit invariance during feature extraction,'' in \emph{Proceedings of the 28th international conference on international conference on machine learning}, 2011, pp. 833--840.

\bibitem{vincent2008extracting}
P.~Vincent, H.~Larochelle, Y.~Bengio, and P.-A. Manzagol, ``Extracting and composing robust features with denoising autoencoders,'' in \emph{Proceedings of the 25th international conference on Machine learning}, 2008, pp. 1096--1103.

\bibitem{kingma2014semi}
D.~P. Kingma, S.~Mohamed, D.~Jimenez~Rezende, and M.~Welling, ``Semi-supervised learning with deep generative models,'' \emph{Advances in neural information processing systems}, vol.~27, 2014.

\bibitem{he2022masked}
K.~He, X.~Chen, S.~Xie, Y.~Li, P.~Doll{\'a}r, and R.~Girshick, ``Masked autoencoders are scalable vision learners,'' in \emph{Proceedings of the IEEE/CVF conference on computer vision and pattern recognition}, 2022, pp. 16\,000--16\,009.

\bibitem{zhang2024hybrid}
B.~Zhang, G.~Zheng, and N.~Van~Huynh, ``A hybrid quantum-classical autoencoder framework for end-to-end communication systems,'' \emph{IEEE Wireless Communications Letters}, 2024.

\bibitem{goutay2022applications}
M.~Goutay, ``Applications of deep learning to the design of enhanced wireless communication systems,'' PhD thesis, UNIVERSITY OF LYON, School of Electronics, Electrotechnics and Automation, January 2022.

\bibitem{ioffe2015batch}
S.~Ioffe and C.~Szegedy, ``Batch normalization: Accelerating deep network training by reducing internal covariate shift,'' in \emph{International conference on machine learning}, 2015, pp. 448--456.

\bibitem{tang2022meta}
H.~Tang, ``Meta-modeling for autoencoder-based end-to-end communications systems,'' in \emph{2022 3rd International Conference on Electronic Communication and Artificial Intelligence (IWECAI)}.\hskip 1em plus 0.5em minus 0.4em\relax IEEE, 2022, pp. 472--476.

\bibitem{yuan2024decentralized}
L.~Yuan, Z.~Wang, L.~Sun, P.~S. Yu, and C.~G. Brinton, ``Decentralized federated learning: A survey and perspective,'' \emph{IEEE Internet of Things Journal}, vol.~11, no.~21, pp. 34\,617--34\,638, 2024.

\bibitem{liu2022intrusion}
C.~Liu, R.~Antypenko, I.~Sushko, and O.~Zakharchenko, ``Intrusion detection system after data augmentation schemes based on the vae and cvae,'' \emph{IEEE Transactions on Reliability}, vol.~71, no.~2, pp. 1000--1010, 2022.

\end{thebibliography}

\end{document}